\documentclass[preprint2]{aastex6}
\usepackage{multirow}

\begin{document}

\title{Evolution of mass and velocity filed in comic web: comparison between baryonic and dark matter}

\author{Weishan Zhu$^{1}$ and  Long-Long Feng$^{1,2}$}
\affil{$^{1}$Institute of Astronomy and Space Science, School of Physics and Astronomy, \\
Sun Yat-Sen University, Guangzhou 510275, China\\
$^{2}$Purple Mountain Observatory, CAS, Nanjing, 210008, China}
\begin{abstract}
We investigate the evolution of cosmic web since $z=5$ in grid based cosmological hydrodynamical simulations, focusing on the mass and velocity field of both baryonic and cold dark matter. The tidal tensor of density is used as the main method for web identification, with $\lambda_{th}=0.2-1.2$. The evolution trends in baryonic and dark matter are similar, while moderate differences are observed. Sheets appeared early and their large scale pattern may have been set up by $z=3$. In term of mass, filaments superseded sheets as the primary collapsing structures at $z\sim2-3$. Tenuous filaments assembled with each other to form prominent ones at $z<~2$. In accordance with the construction of the frame of sheets, the cosmic divergence velocity, $v_{div}$, had been well developed above 2-3 Mpc by z=3. Afterwards, curl velocity, $v_{curl}$, grown dramatically along with the rising of filaments, become comparable to $v_{div}$, for $<2-3 Mpc$ at $z=0$. The scaling of $v_{curl}$ can be described by the hierarchical turbulence model. The alignment between vorticity and eigenvectors of shear tensor in baryonic matter field resembles dark matter, and is even moderately stronger between $\vec{\omega}$ and $\vec{e}_1$, and $\vec{e}_3$. Compared with dark matter, mildly less baryonic matter is found residing in filaments and clusters, and its vorticity has been developed more significantly below $2-3 Mpc$. These differences may be underestimated due to the limited resolution and lack of star formation in our simulation. The impact of the change of dominant structures in over-dense regions at $z\sim2-3$ on galaxy formation and evolution is shortly discussed.
\end{abstract}

\section{Introduction}
The spatial distribution of cosmic matter on large scale is in a pattern named as cosmic web, consisting of voids, sheets/walls, filaments and clusters/knots, which was firstly predicted by theoretical study and have been confirmed by galaxy redshift survey(e.g. de Lapparent et al. 1986; Colless et al. 2003; Tegmark et al. 2004; Mehmet et al. 2014). The formation and evolution of cosmic web have evoked numerous investigation in the past several decades(see van de Weygaert \& Bond 2008). Using a linear Lagrangian model, Zel'dovich(1970) firstly pointed out that the growth of density perturbation would depend on the deformation tensor, which is connected to the tidal shear field. The Zel'dovich approximation predicted the formation of pancake and filamentary large scale structures in a sequential order due to the anisotropic gravitational collapse in the linear and mildly nonlinear regimes. 

The homogeneous ellipsoidal collapse model(Icke 1973; White \& Silk 1979) predicted the appearance and evolution of flattened and elongated structures in the quasi-linear regime. Integrated with the effects of external tidal field(Eisenstein \& Loeb 1995; Bond \& Myers 1996), the ellipsoidal model has became a useful tool to describe the distribution of virialized objects within the bottom-up structure formation scenario. Based on the peak patch formalism (Bond \& Myers 1996), the anisotropic property of gravitational collapse were incorporated within the hierarchical clustering picture, and led to the heuristic cosmic web model given in Bond, Kofman \& Pogosyan(1996). According to this scenario, high density peaks in the primordial Gaussian field on large smoothing scale would evolve firstly, and then seed for the mega-parsec scale tidal shear that drives the formation of prominent filaments connecting clusters, and afterwards the sheets. The proposed evolutionary sequence is in an inverse order compared to the Zel'dovich picture, and stresses the dominance of filamentary structures instead of sheets(van de Weygaert \& Bond 2008). 

Due to the high complexity, the formation and evolution of cosmic web in the nonlinear regimes have been investigated by analysis of cosmological simulation samples recently(e.g. Aragon-Calvo et al. 2007a; Hahn et al. 2007a; Aragon-Calvo, van de Weygaert \& Jones 2010; Cautun et al. 2014 etc.). Various numerical methods have been developed to identify morphological components of the cosmic web in both simulation samples and observed galaxy distribution(Aragon-Calvo et al. 2007a; Hahn et al. 2007; Forero-Romero et al.2009; Bond et al. 2010a; Sousbie 2011; Hoffman et al. 2012; Cautun et al. 2013). Of particular interest is to reconstruct the large scale filamentary structures in observations and simulations, and study statistical properties of their lengths and widths, and alignment with galaxies and halos (Colberg et al. 2005; Zhang et al. 2009; Sousbie et al. 2011; Tempel et al 2014; Gheller et al. 2015). With the aid of numerical simulations, the mass and volume content, the density distribution, and the spatial extent of cosmic web components are investigated in detail, as well as their redshift evolution(for details see e.g. Hahn et al. 2007b; Bond et al. 2010; Cautun et al. 2014, hereafter C14). Filaments are identified as the most prominent structures of the web and contain up to $~50\%$ of the cosmic mass since $z=2$. Many of the properties and evolution of massive filaments in simulation samples are generally in agreement with the prediction of the peak patch model. Nevertheless, clusters became a significant component only at $z< 0.5$ and tenuous filaments are found to be dominate at redshift $z>1$, which however can not well explained by the peak patch model. A more comprehensive theory for the intricate cosmic web may be needed, of which the evolution of singular structures in Zel'dovich approximation might be a key factor(Hidding, Shandarin \& van de Weygaert 2014). 

On the other hand, the evolution of cosmic web at redshifts higher than 2 is an important section to build the whole picture, by providing direct insights on properties of each component of cosmic web at early time and tracking the hierarchical assembly history of structures. A few visual impression of the filaments at $z>2$ is presented in Aragon-Calvo (2007) and Bond et al. (2010b). \textbf{C14} demonstrated that the dominancy of filaments holds up to $z=3.8$(see their Figure 24), although the margin over sheets is decreasing towards high redshift. It is reasonable to expect that the sheets might surpass the filaments, i.e., become the primary structure, at the epoch early than $z>4$ in their samples. Moreover, the bias effect between baryonic and dark matter in the nonlinear regime should be taken into account while comparing the observed cosmic web with N-body simulations. However, less efforts have been ever made to systematically investigate the evolution of baryonic matter in each large scale environment during the formation of cosmic web, except for the topic about warm and hot intergalactic medium(WHIM) at low redshifts. Cosmological hydrodynamic simulations predicted that the WHIM resides in moderate over-dense filaments and possibly walls, and host $\sim 50\%$ of the total baryons, which is still being searched by intensively observations(e.g. Cen \& Ostriker 1999; Dave et al. 2001;Bregman 2007; Fang et al. 2010; Shull et al. 2012; Tejos et al. 2016).

The velocity fields of cold dark matter and baryonic gas encode crucial information of the cosmic network, especially in the mildly and non-linear regime, and can be employed to track the formation history of cosmic structures and the mass transportation among different large scale structures. Research on the motions of matter in cosmic web could provide hints for the following important question: the impact of cosmic environment on the galaxies properties, such as the stellar mass, morphology, spins etc. Recently, the velocity fields have been investigated in simulation works looking for the origin and alignment of spins of halos with cosmic web(e.g. Aragon-Calvo et al. 2007b; Codis et al 2012; Libeskind et al 2013; Wang et al. 2013; Dubois et al. 2014), in order to interpreting the alignment of galaxies shape and distribution in different cosmic environment revealed by observation and simulations samples(e.g. Lee \& Erdogdu 2007; Zhang et al. 2009; Tempel \& Libeskind 2013; Dong et al 2014; Wang et al 2014; Zhang et al. 2015). Nevertheless, investigation on the evolution of flows of baryonic and cold dark matter during the formation of cosmic web in simulations is inadequate, in comparison with the mass content. The velocity on large scale is usually assumed to be irrotational, just as in the Zel'dovich approximation (Zel'dovich1970; Shandrin \& Zel'dovich 1989), which might become invalid in the mildly regimes. The curl velocity would be non-negligible after shell-crossing in the multi-stream regions. The properties of the irrotational and curl velocity of cosmic matter are expected to evolve along the hierarchical clustering of structures, to reflect the anisotropic gravitational collapse of the cosmic web. 

Pichon \& Bernardeau(1999) made a pioneering effort to calculate the vorticity generation in large-scale structure caustics and demonstrated that the vorticity would be significant at scales above the galaxy clusters. In our previous work(Zhu, Feng, \& Fang 2010; Zhu et al. 2013; Zhu \& Feng et al. 2015, hereafter ZF15), we found that the cosmic shocks appearing in multi-stream regions will boost the vorticity of baryonic gas by baroclinity. The vortical motions of gas are triggered and pumped up in sheets and filaments in simulations. The growth history and statistical properties of curl velocity is in accordance with the anisotropic collapse process in a bottom-up universe. In order to understand the origin of halo spins beyond the linear tidal torque theory, ambient vortical flow around dark matter halos has been studied in the past few years (e.g., Libeskind et al 2013). Libeskind, Hoffman \& Gottlober(2014) investigated the velocity shear tensor and vorticity, and their relative orientations as a function of scale and redshift, indicating the nonlinear evolution will drive the vorticity tending to perpendicular to the fastest collapsing axis. Some aspects of the vortical motions of dark matter including the growth history, statistical properties in different environment remains blur. The connection between irrotational motion and the evolution of cosmic web is also not well resolved. In addition, the galaxy spin with respect to the $e_1$ vector(normal of sheets) tends to be randomly distributed in observations, which is different from the angular momentum of halos in N-body simulations(Tempel \& Libeskind 2013). For a better understanding of the cosmic web, a detail comparison between baryonic and dark matter in regarding to the evolution and alignment of velocity shear tensor and vorticity is required, which has been partially tackled in Laigle et al. 2015.

We here present a comprehensive comparison study focusing on the evolution of volume, mass, irrotational and curl velocity of both baryonic and cold dark matter in the cosmic web since $z=5$ using fixed grid cosmological hydrodynamical simulations without star formation and feedback. The connection between matter flows and the evolution of cosmic web is also explored. This paper is organized as follows. The simulations and numerical methods are described in \S2. \S3 studies the volume and mass distribution and their evolution in cosmic web. The growth history and properties of irrotational and curl velocity of baryonic and cold dark matter during the formation and evolution of cosmic web, including magnitude, power spectrum, scaling relation are investigated in \S4. \S5 compares the alignment of velocity shear tensor and vorticity of baryonic and dark matter in each class of the cosmic web. \S6 discuss the growth of vortical motions along with the rising of filamentary structures at $z<3$, and possible impact on the formation and evolution of galaxies. We summarize our results in \S7.   

\section{Numerical Methodology}
\subsection{Simulations }
Three simulations in periodic cubical boxes of 25, 50 and $100\ h^{-1}$ Mpc were performed in the $\Lambda$CDM model with the WMAP5 normalization, in which,  the cosmological parameters were adopted such as $\Omega_{m}=0.274, \Omega_{\Lambda}=0.726,h=0.705,\sigma_{8}=0.812, \Omega_{b}=0.0456$, and $n_{s}=0.96$(Komatsu et al. 2009). We refer the simulations as L025, L050 and L100 respectively hereafter, with the digital number stands for the box size. The same random seeds and phases are used in generating the initial conditions of the three simulations. The velocity fields of baryonic matter in L025 and L100 have been investigated in ZF15. The simulations were run by the hybrid N-body/hydrodynamic cosmological code WIGEON, which employs the positivity-preserving Weighted Essentially Non-Oscillatory(WENO) finite differences scheme to solve the hydrodynamic equations, incorporating with the standard particle-mesh method as the gravitation potential solver (Feng et al. 2004; Zhu et al. 2013). All the simulations were evolved from redshift $z=99$ to $z=0$ in a $1024^3$ grid with equal number of dark matter particles. The space resolutions are $24.4 h^{-1}$ , $48.8 h^{-1}$, and $97.7 h^{-1}$ kpc respectively. The corresponding particle mass resolution are $1.30 \times 10^{6}$, $1.04 \times 10^{7}$ and  $8.32\times 10^{7}$  $M_{\odot}$. A uniform UV background is switched on at $z=11.0$. The radiative cooling and heating modules are implemented with a primordial composition $(X=0.76,Y=0.24)$ following the method in Theuns et al.(1998). The processes of star formation, AGN and their feedback are not included.

\subsection{Density and Velocity Fields Resampling and Velocity Decomposition}

To reduce the impact of discreteness, especially in under dense regions, the mass and velocity of dark matter particles are first assigned to a $512^3$ grid using the cloud-in-cell (CIC) algorithm. The resolution of CIC grid, $R_g$ is two grid units of simulations. The resampled density and velocity field on CIC grids are then smoothed with gaussian kernel of one, two, four, and eight CIC grid cell(denoted as smt01, smt02, smt04, and smt08 hereafter), using FFT.  For instance, smoothing length $R_s$ of $48.8, 97.6,  195, 390h^{-1} kpc$ are applied for L025. The effective smooth length is given by $R_{eff}=\sqrt{R_g^2+R_s^2}$. The same $512^3$ grid is also applied to resampling the density and velocity fields of gas, which are then smoothed with the same kernel as dark matter. The  peculiar velocity fields of both dark matter and baryonic matter are decomposed into three components respectively, the curl-free divergence velocity $\vec{v}_{div}$, the divergence-free curl velocity $\vec{v}_{curl}$, and the uniform velocity $\vec{v}_{unif}$ through the Helmholtz-Hodge decomposition(Sagaut \& Cambon 2008),
\begin{equation}
 \vec{v}=\vec{v}_{curl}+\vec{v}_{div}+\vec{v}_{unif}
 \end{equation}
where, $\nabla \times \vec{v}_{div}=0$ and $\nabla \cdot \vec{v}_{div} = 
\nabla \cdot \vec{v}$; $\nabla \cdot \vec{v}_{curl}=0$ and $\nabla \times \vec{v}_{curl} 
= \nabla \times \vec{v}$. The uniform component $\vec{v}_{unif}$ is both of curl-free and divergence-free, 
which is actually negligible in the simulations and will not be discussed here.

\subsection{Cosmic Web Identification}

The distributions of baryonic and dark matter at resampled grid cells are respectively classified into four categories of cosmic structures, i.e, voids, sheets/walls, filaments and clusters/knots with two identification schemes following Hanh et al.(2007) and Forero-Romero et al.(2009). The former scheme is based on the density field and the latter is on the peculiar velocity field of matter. We denote $\bold{r}$ to be the comoving coordinates and $\bold{x}=a(t)\bold{r}$ the proper coordinates, where a(t) is the cosmic expansion factor. Furthermore, the comoving peculiar velocity $\bold{v}(\bold{r},t)=\dot{\bold{r}}$ is related to the physical velocity $\bold{u}$ by $a(t)\bold{v}(\bold{r},t)=\bold{u}-\dot{a}(t)\bold{r}$. The first method(\textbf{d-web} hereafter) identifies structures on the basis of the eigenvalues $\lambda_{t,1}, \lambda_{t,2},\lambda_{t,3}$ of the tidal tensor, which is defined as the Hessian matrix of the rescaled peculiar gravitational potential $\phi$,
\begin{equation}
T_{\alpha\beta}=\frac{\partial^2\phi}{\partial r_{\alpha} \partial r_{\beta}}
\end{equation}
where  $\alpha, \beta=1,2$, and $3$ denotes for the components on three axes. The peculiar gravitation potential is rescaled by $4\pi G \bar{\rho}(t)$, and obeys $\nabla^2 \phi=\delta$, where $\bar{\rho}(t)$ is the cosmic mean density of matter and $\delta=(\rho-\bar{\rho})/\bar{\rho}$ is the overdensity field. The mean density and overdensity of baryonic matter $\bar{\rho}_b(t), \delta_b(t)$, and of dark matter $ \bar{\rho}_d(t), \delta_d(t)$ are separately used to identify webs in corresponding matter fields. The second method(\textbf{v-web} hereafter) depends on the eigenvalues $\lambda_{v,1}, \lambda_{v,2},\lambda_{v,3}$ of the rescaled velocity shear tensor, 
\begin{equation}
\Sigma_{\alpha\beta}=-\frac{1}{2H(z)}(\frac{\partial v_{\alpha}}{\partial r_{\beta}}+\frac{\partial v_{\beta}}{\partial r_{\alpha}}),
\end{equation}
where $H(z)$ is the Hubble parameter and used as the rescale factor. The derivatives of potential and velocity are performed in Fourier space. We count the number of eigenvalues above $\lambda_{th}$ at each CIC grid cell. A cell with value 3, 2, 1 or 0 is marked as cluster, filament, sheet or void respectively. The threshold value hence is important for the classification results. Forero-Romero et al(2009) pointed out that $\lambda_{th}$ could be estimated in principle, considering the association of deformation tensor, collapse time scale and the age of universe as well. A precise estimation of $\lambda_{th}$, however, needs to carefully deal with the anisotropic collapsing of structures and is not available so far. Forero-Romero et al.(2009) relaxed the assumption $\lambda_{th}=0$ used in Hanh et al.(2007), by taking account of $\lambda_{th}$ dependent on the collapse time-scale, and recommended values of  0.2-0.4 for the d-web. Hoffman et al. (2012) used $\lambda_{th} \sim 0.44$ for the v-web and a larger value, $\lambda_{th} \sim 0.7$, for d-web to provide the best match to the visual impression. Here, we investigated the threshold values of 0.2-1.2 for d-web, and $\lambda_{v, th}=0.2-0.4$ for v-web. $\lambda_{th}$ is firstly kept constant in time, and then set to varying with redshifts.

\section{The Distribution and Evolution of Mass in the Cosmic Web}

\subsection{Visual Inspection of the Cosmic Web}

\begin{figure*}[htbp]
\vspace{-1.5cm}
\gridline{
\hspace{-1.8cm}
\includegraphics[width=0.75\textwidth]{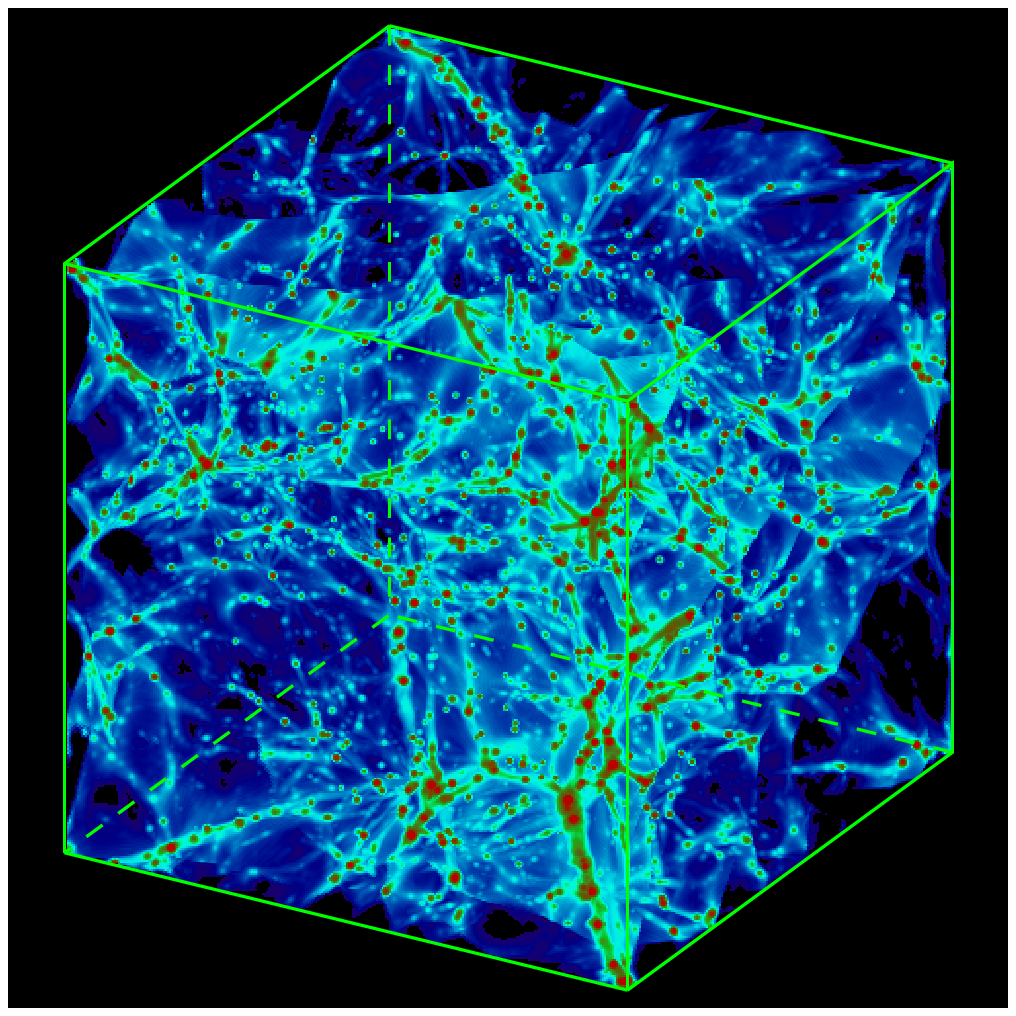}
\hspace{-5.5cm}
\includegraphics[width=0.75\textwidth]{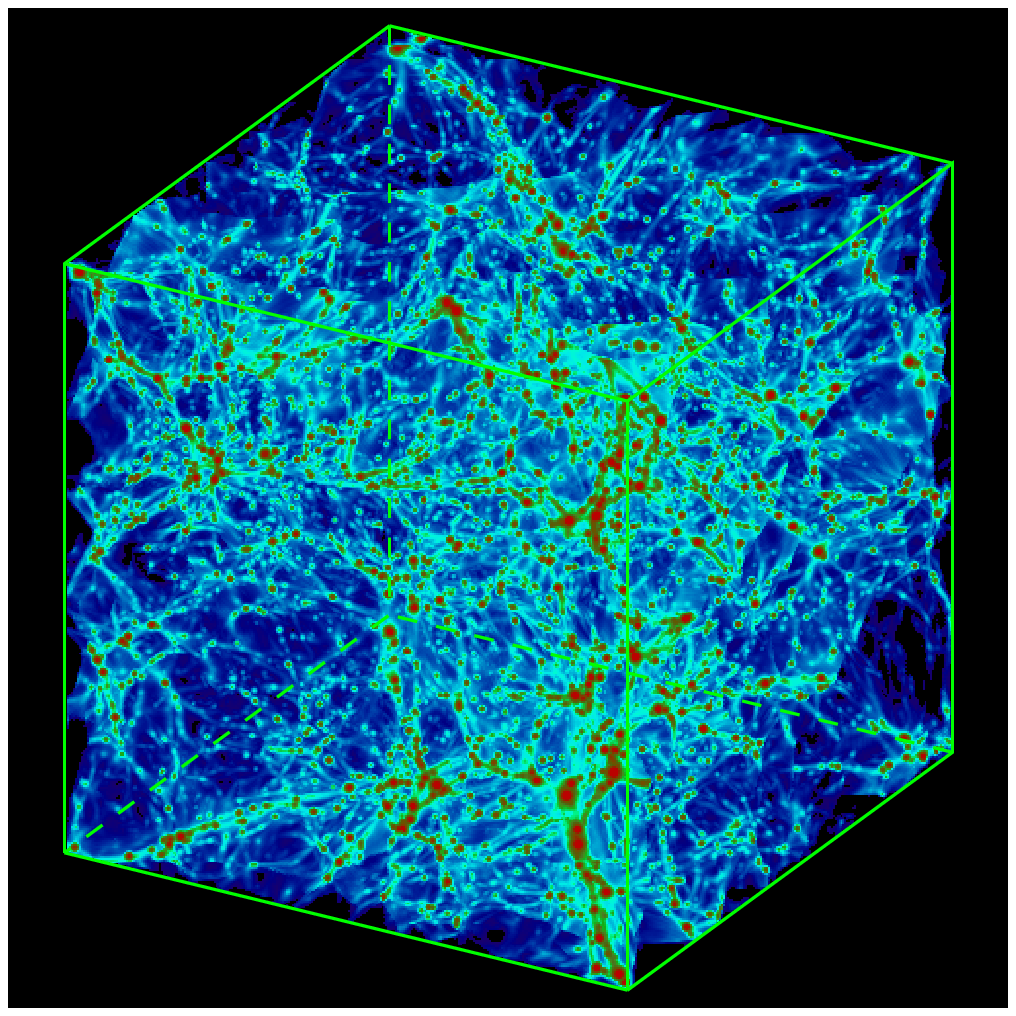}
}
\vspace{-2.2cm}
\gridline{
\hspace{-1.8cm}
\includegraphics[width=0.75\textwidth]{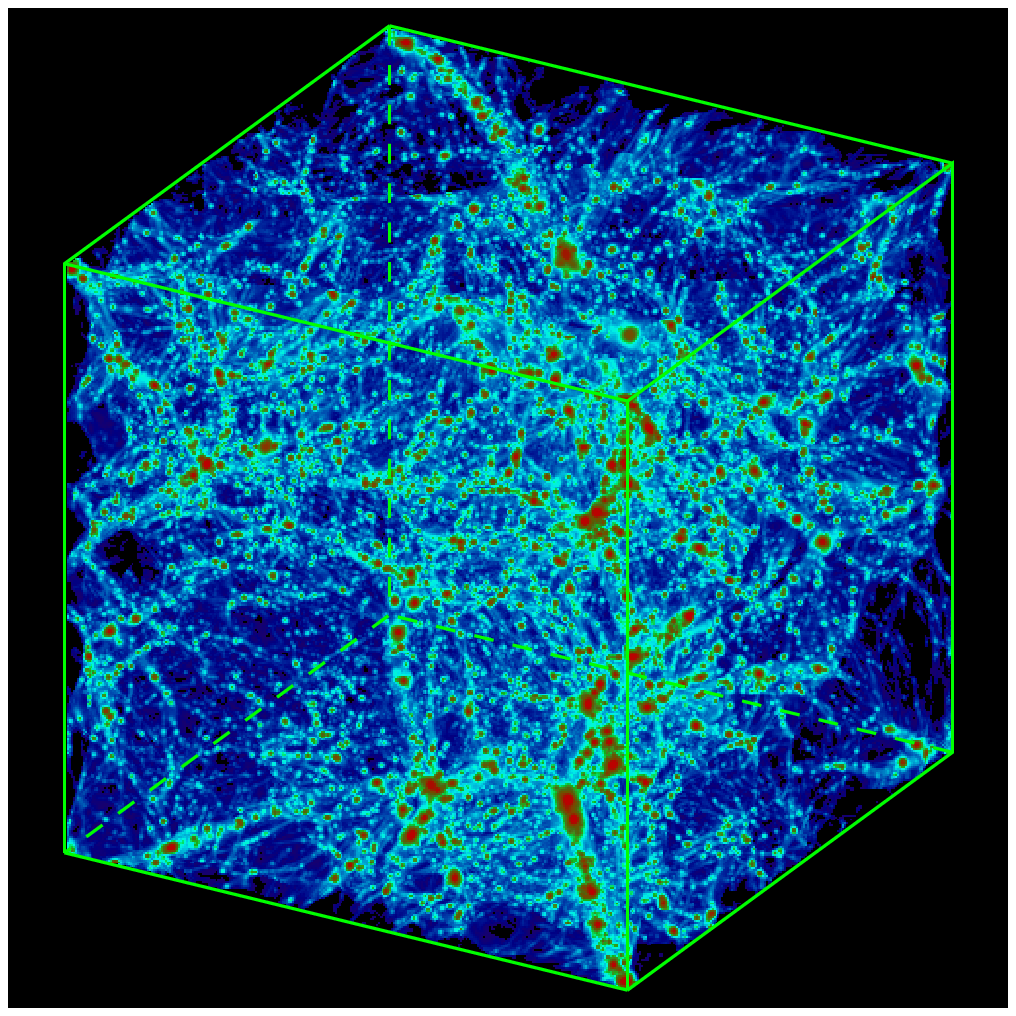}
\hspace{-5.5cm}
\includegraphics[width=0.75\textwidth]{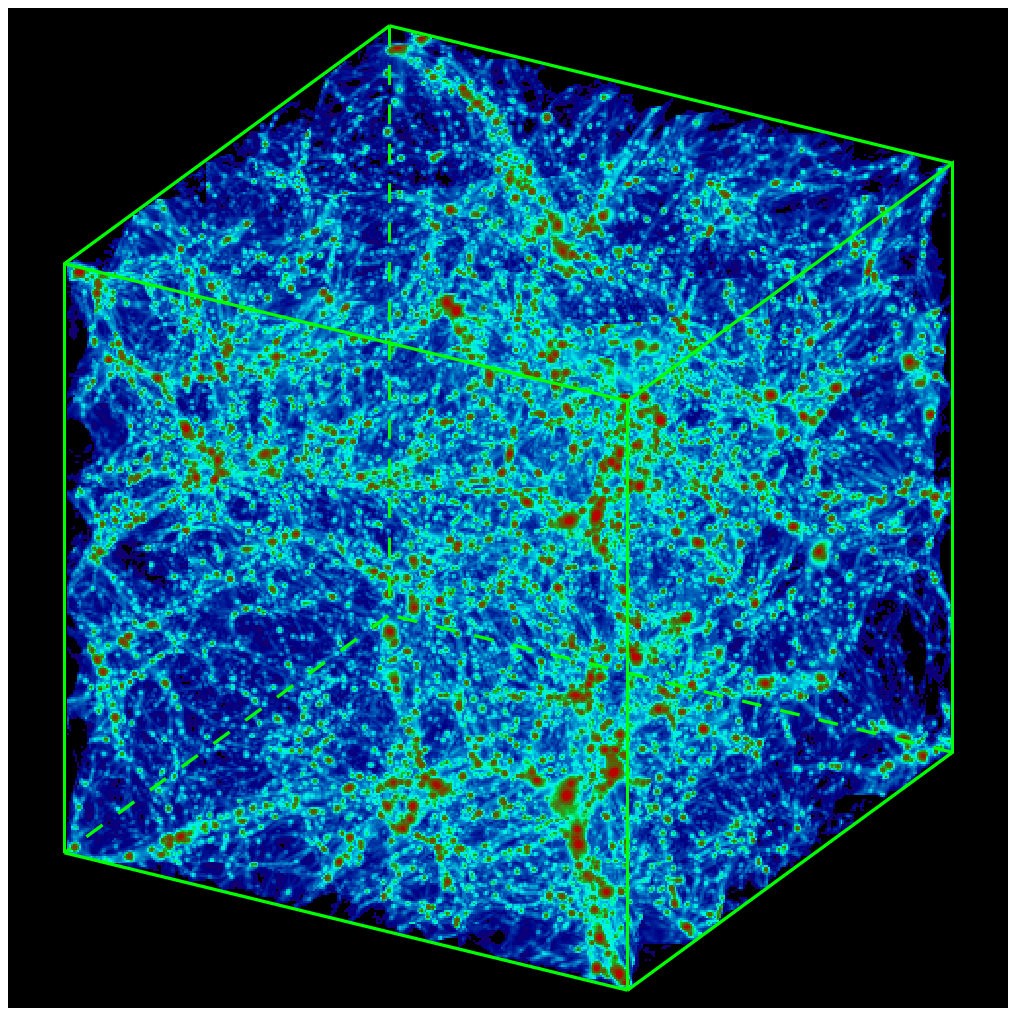}
}
\vspace{-2.2cm}
\gridline{
\hspace{-1.8cm}
\includegraphics[width=0.75\textwidth]{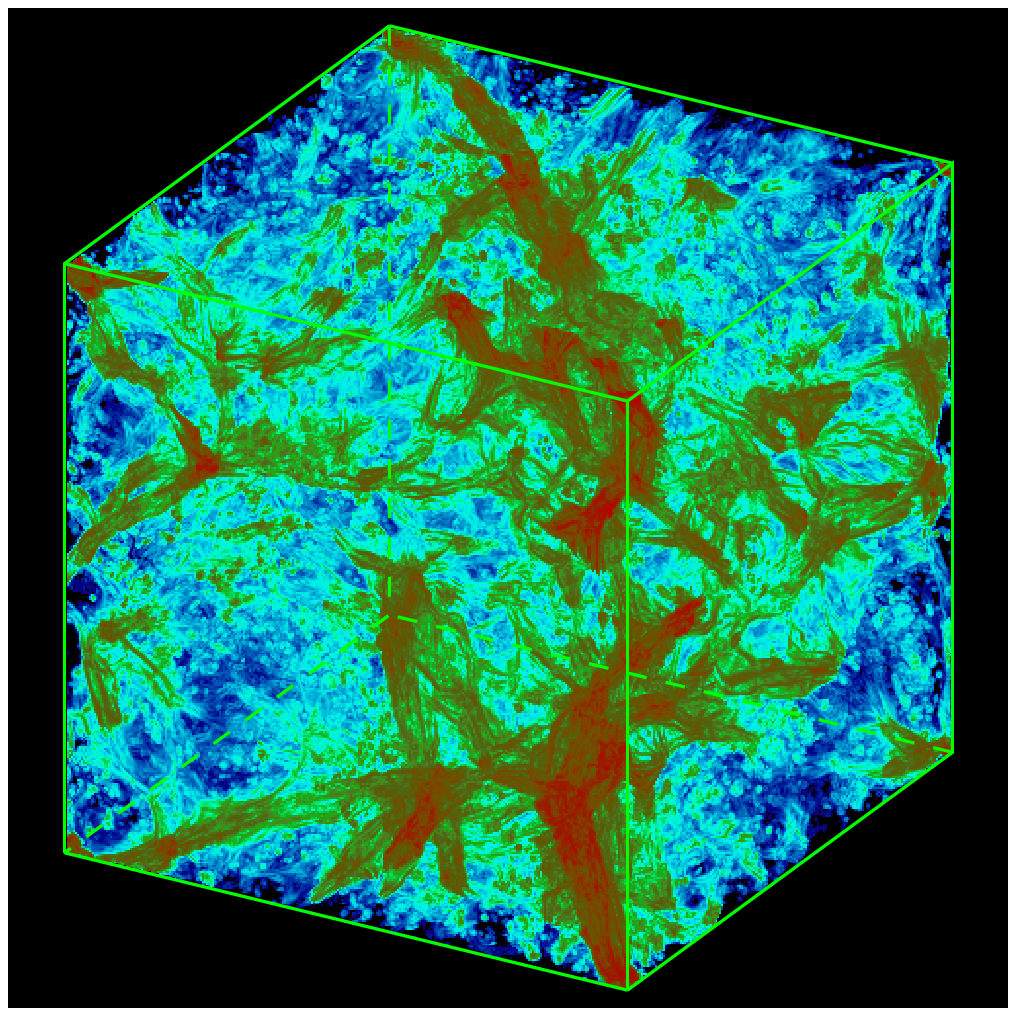}
\hspace{-5.5cm}
\includegraphics[width=0.75\textwidth]{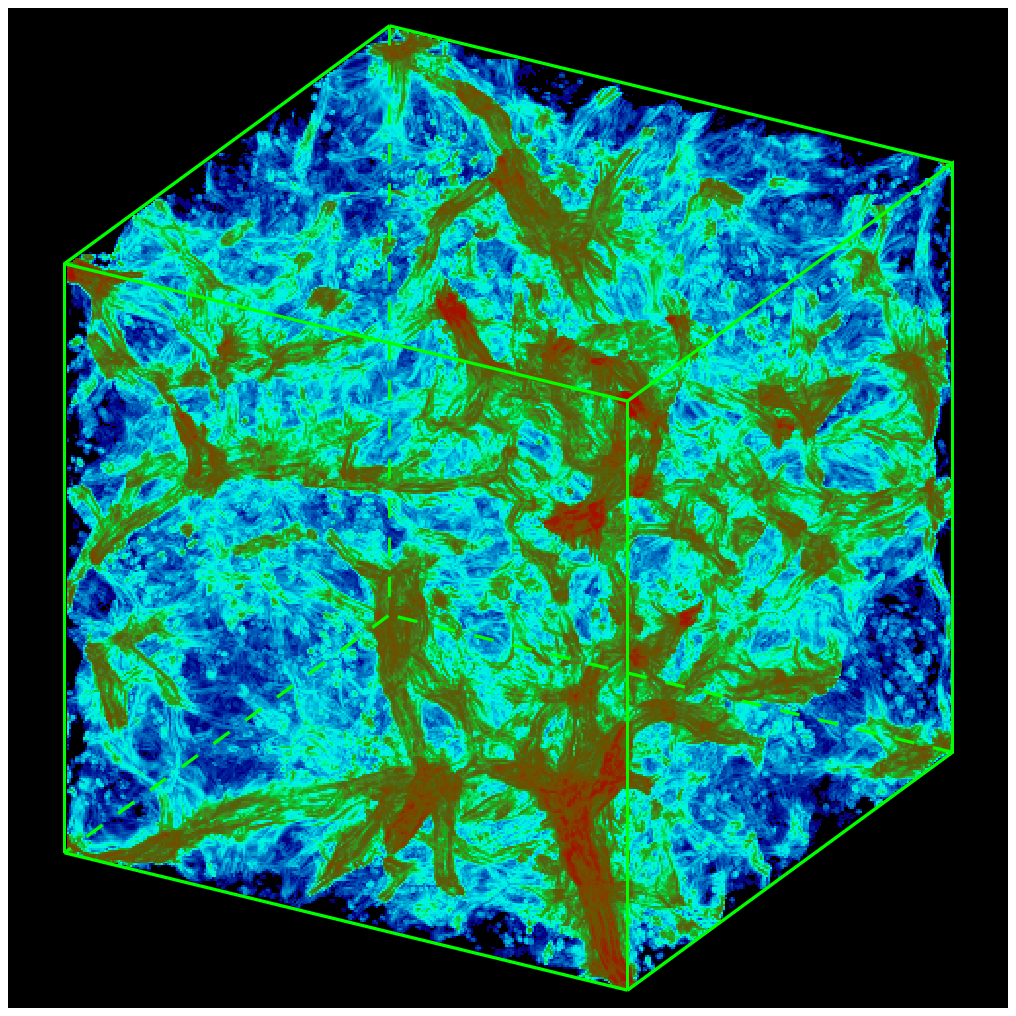}
}
\vspace{-1.2cm}
\caption{Projected three-dimensional rendering of density of baryonic matter(Top), and dark matter(Middle), and vorticity of dark matter(Bottom) in simulation L025(Left) and L050(Right).}
\label{figure1}
\end{figure*}

\begin{figure*}[htbp]
\vspace{-1.5cm}
\gridline{
\hspace{-1.8cm}
\includegraphics[width=0.75\textwidth]{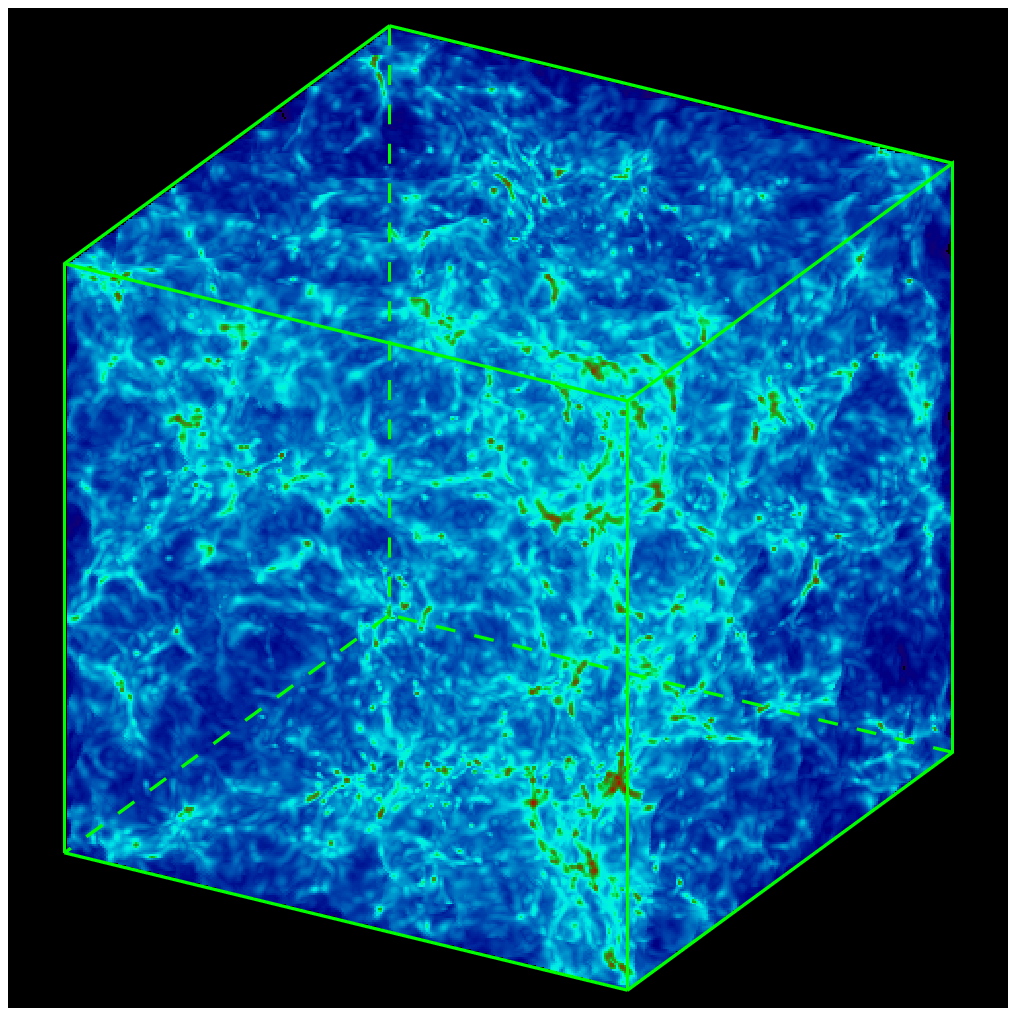}
\hspace{-5.5cm}
\includegraphics[width=0.75\textwidth]{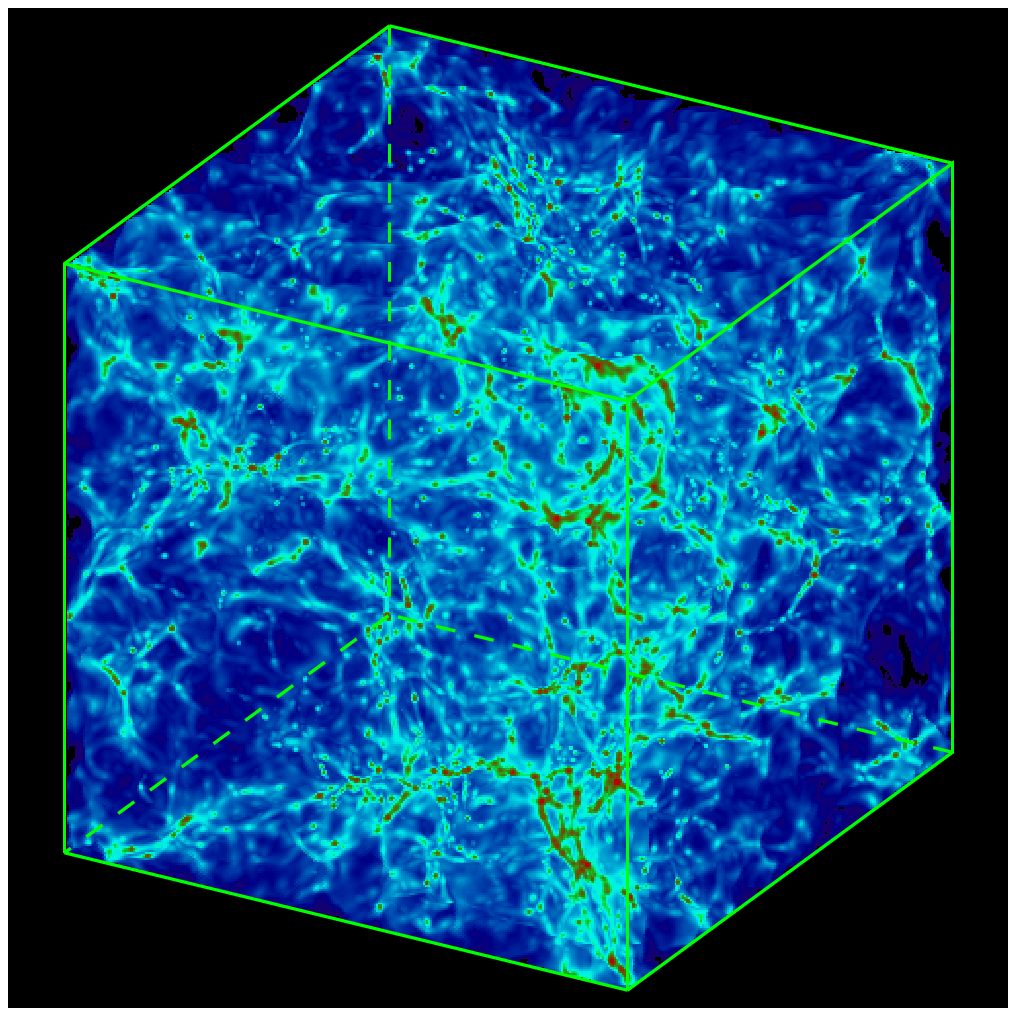}
}
\vspace{-2.2cm}
\gridline{
\hspace{-1.8cm}
\includegraphics[width=0.75\textwidth]{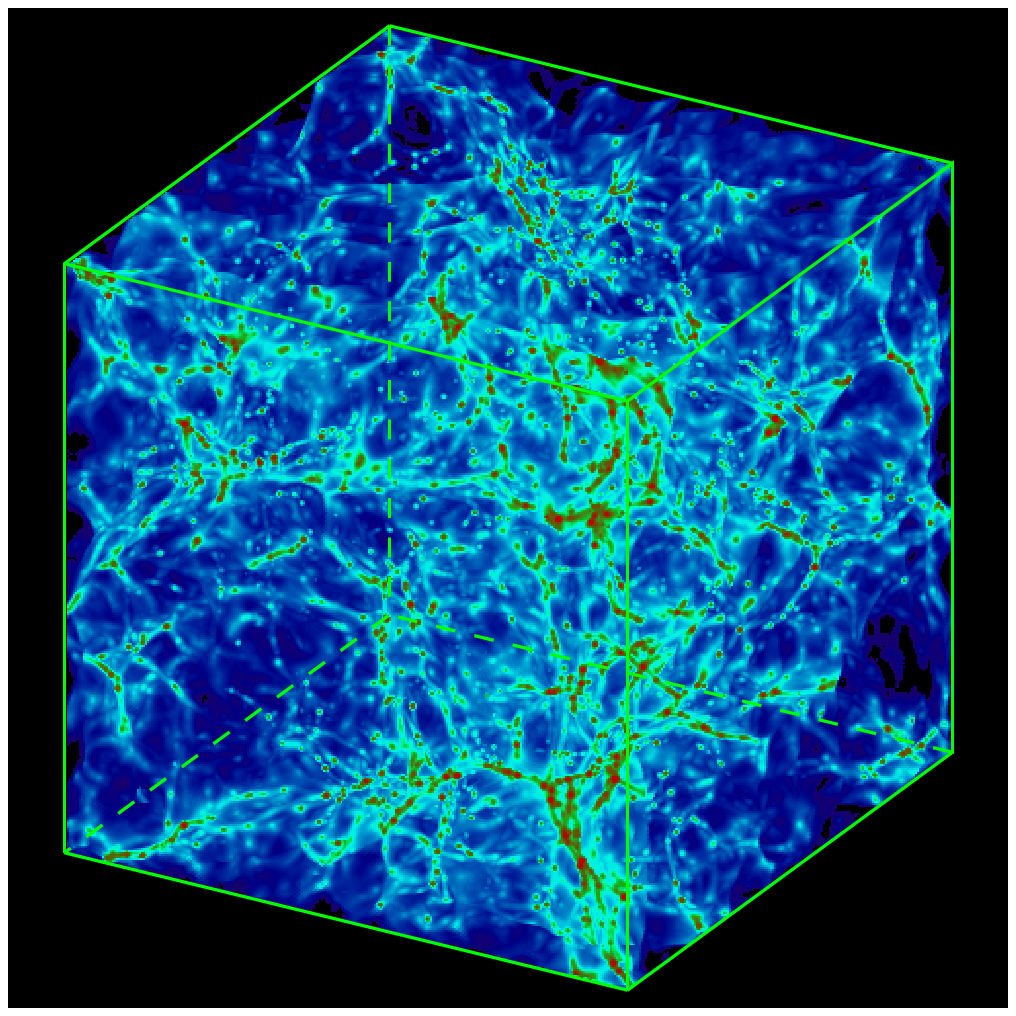}
\hspace{-5.5cm}
\includegraphics[width=0.75\textwidth]{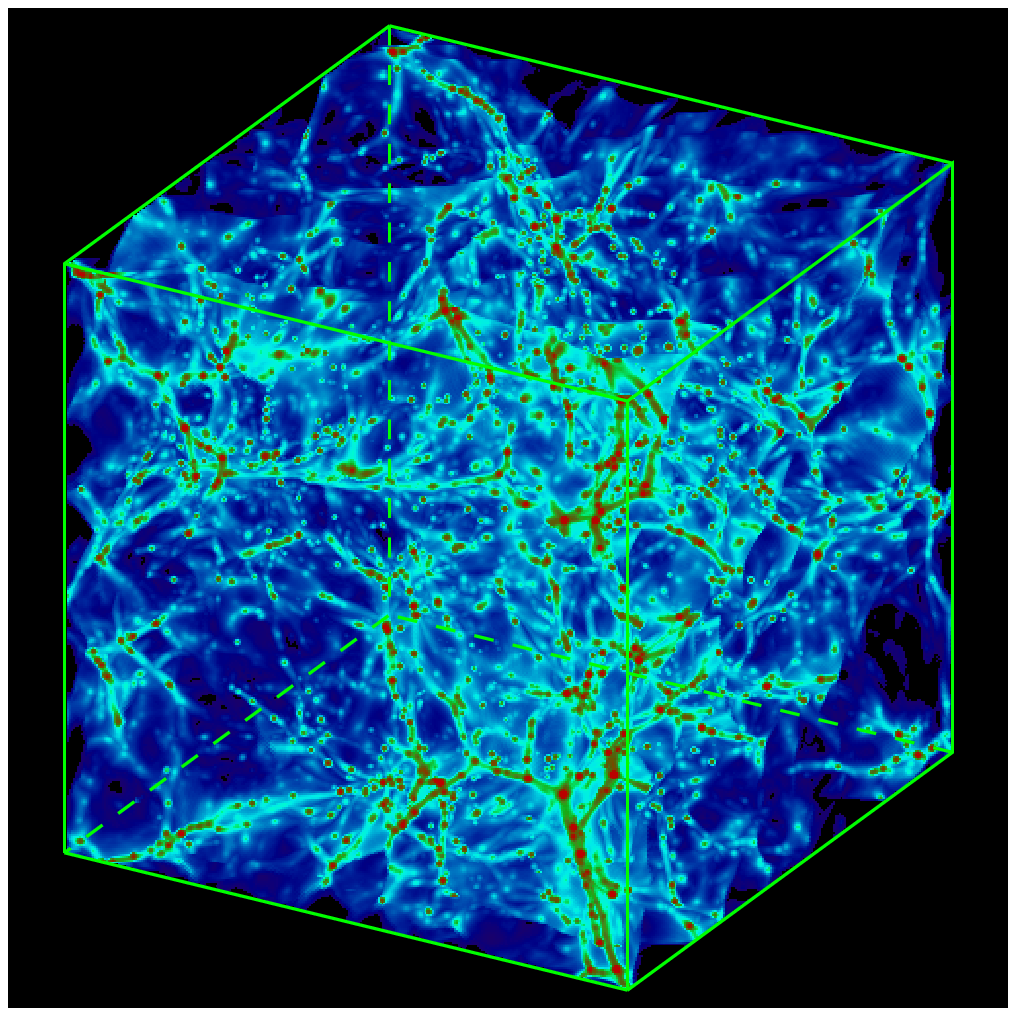}
}
\vspace{-2.2cm}
\gridline{
\hspace{-1.8cm}
\includegraphics[width=0.75\textwidth]{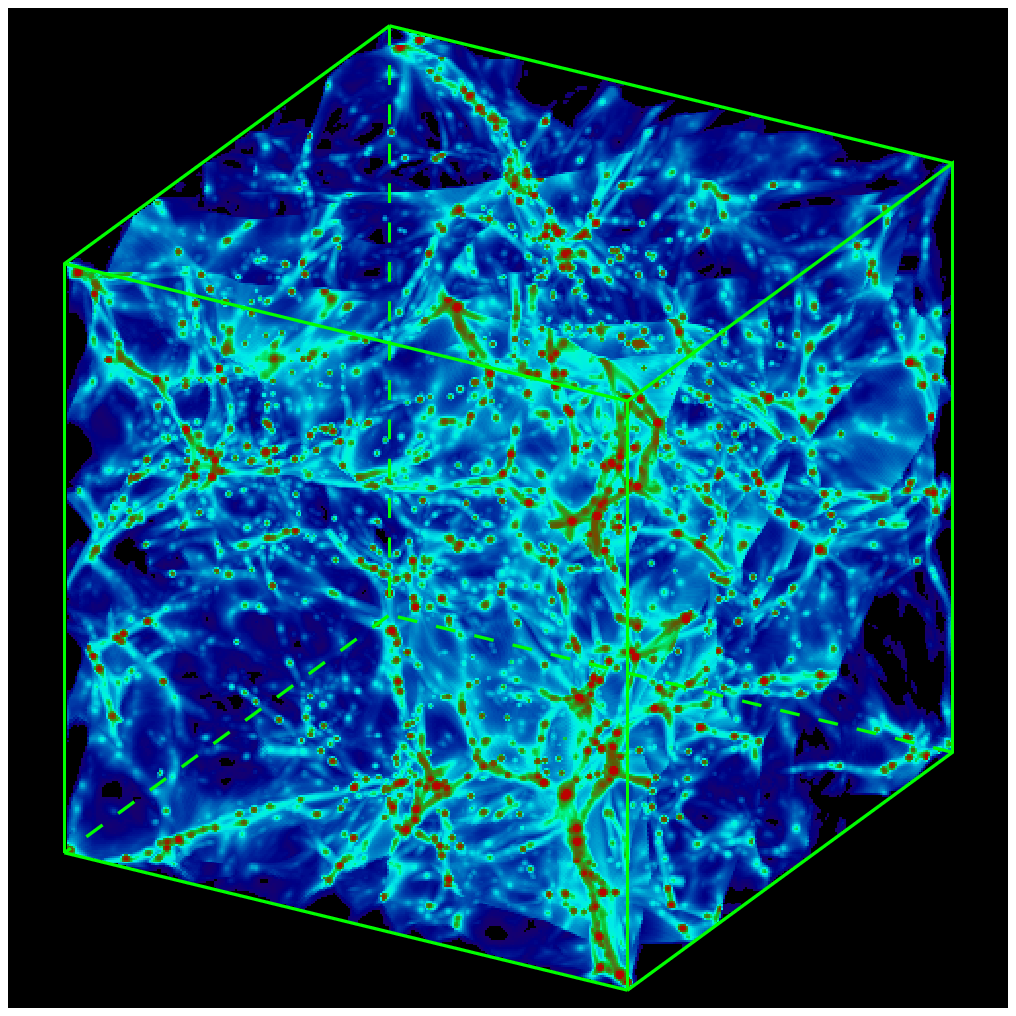}
\hspace{-5.5cm}
\includegraphics[width=0.75\textwidth]{denbafil_smt_3024.00.eps}
}
\vspace{-1.2cm}
\caption{Projected three-dimensional rendering of density filed of baryonic matter in the simulation L025 at $z=5.0, 3.0, 2.0, 1.0, 0.5, 0.0$ in turn.}
\label{figure2}
\end{figure*}

Before going to probe the volume and mass content of the cosmic web, we firstly provide the direct visual impression of mass distribution, vorticity and identified structures in this subsection. Figure 1 displays a projected three-dimensional rendering of the density distribution of baryonic and dark matter, and the vorticity of dark matter in the simulations L025 and L050 at $z=0$. A sharp picture of the cosmic web is visualized in the baryonic density field of L025(the rendering procedure is improved comparing to Figure 1 in ZF15 to feature the structures). In comparison, the large scale structures are less prominent in L050, probably because the lower spatial numerical resolution lead to lower density contrast between structures in simulation. It would take more efforts to visually identify the large scale structures in the density field of dark matter, especially in under dense regions. The structures in the vorticity field, $|\omega|= |\nabla \times v|$, of dark matter, however, is smooth and prominent. 

The vorticity of dark matter resembles the baryonic matter(see Figure 1. in Zhu \& Feng 2015) in both the spatial configuration and strength distribution. The curl motions have been developed significantly in the regions surrounding filaments and clusters, showing more extended morphologies than the density field. The evolution of the cosmic web since $z=5.0$ in L025 is demonstrated in Figure 2. Vague proto-sheets and sheets appear at early times. Small scale filaments can also be found visually at $z=5$, when most of them are tenuous. Prominent filaments emerge lately at a redshift between $z=2$ and $z=1$, mainly built from tenuous predecessors. Evident cluster regions form at around $z=0.5$, in agreement with \textbf{C14}. 

\begin{figure*}[htbp]
\vspace{-1.5cm}
\gridline{
\hspace{-1.8cm}
\includegraphics[width=0.75\textwidth]{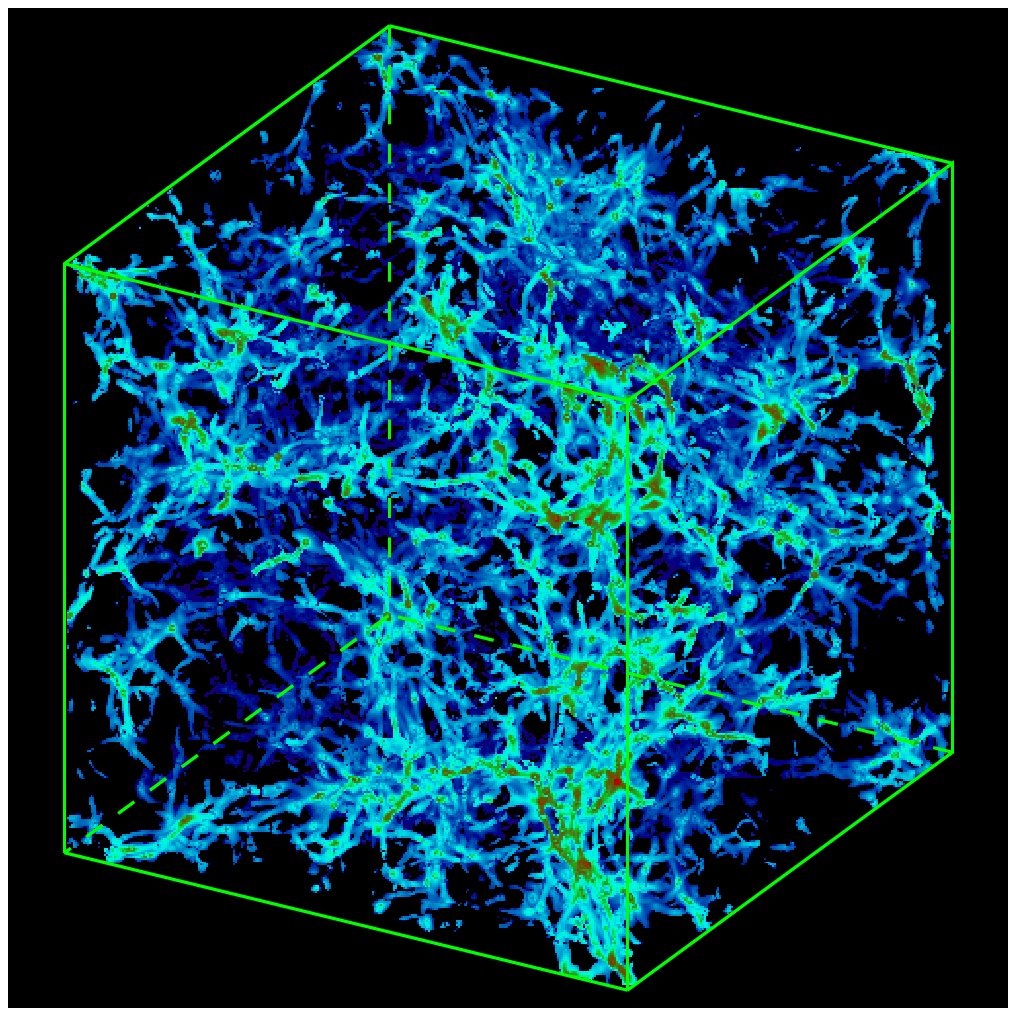}
\hspace{-5.5cm}
\includegraphics[width=0.75\textwidth]{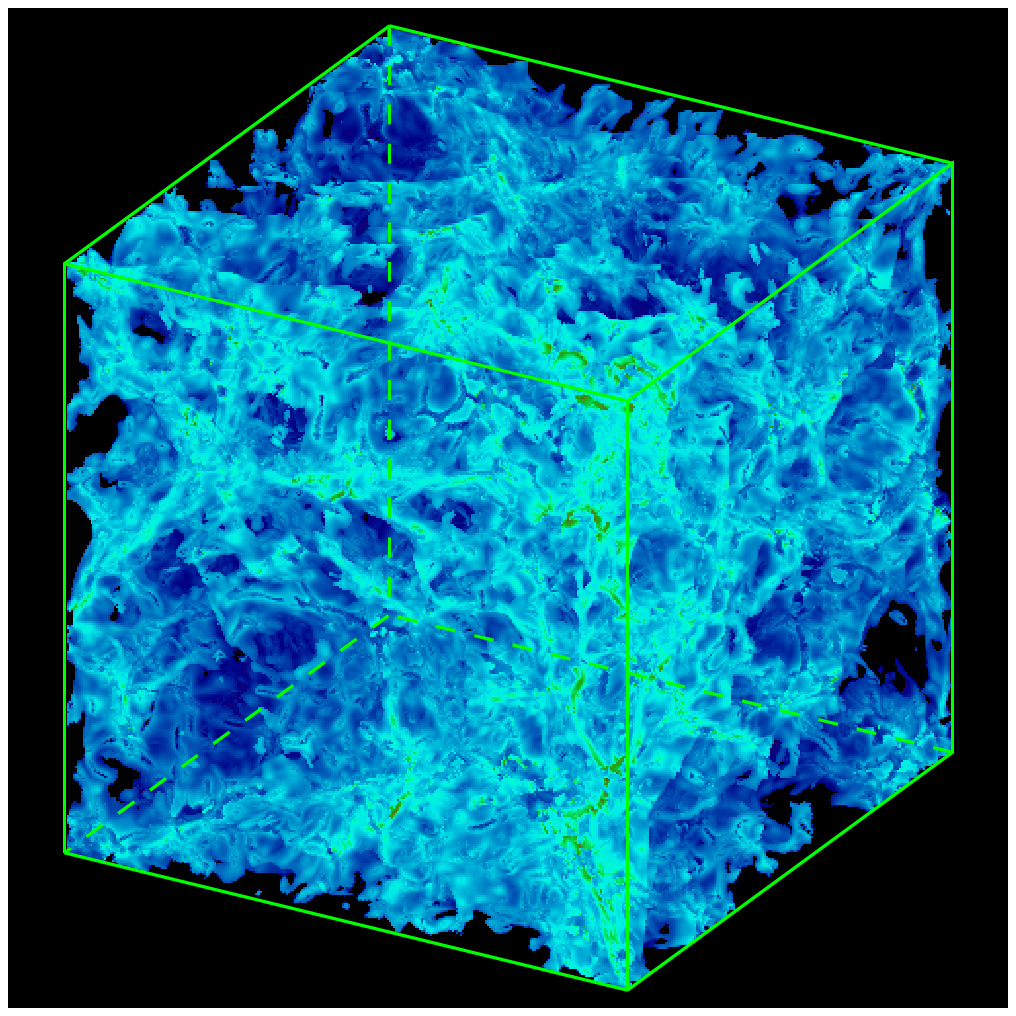}
}
\vspace{-2.2cm}
\gridline{
\hspace{-1.8cm}
\includegraphics[width=0.75\textwidth]{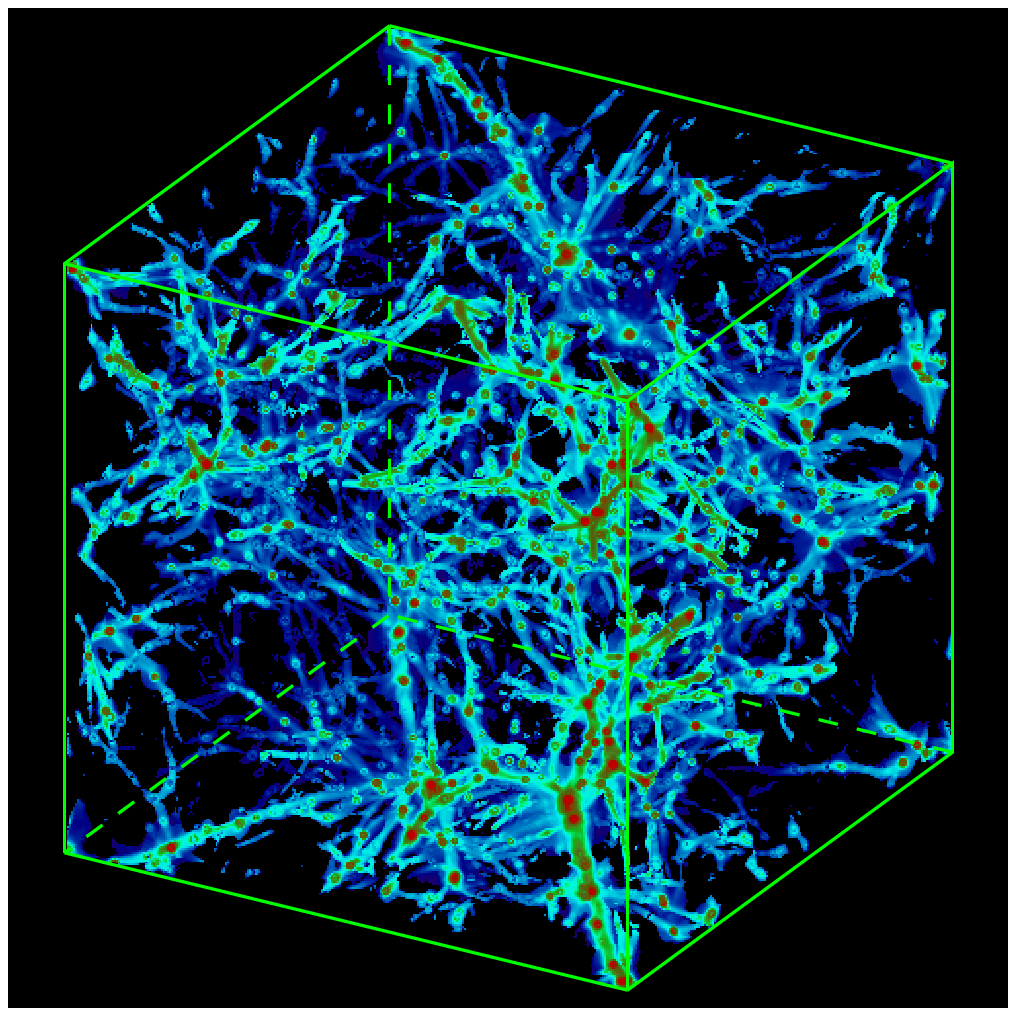}
\hspace{-5.5cm}
\includegraphics[width=0.75\textwidth]{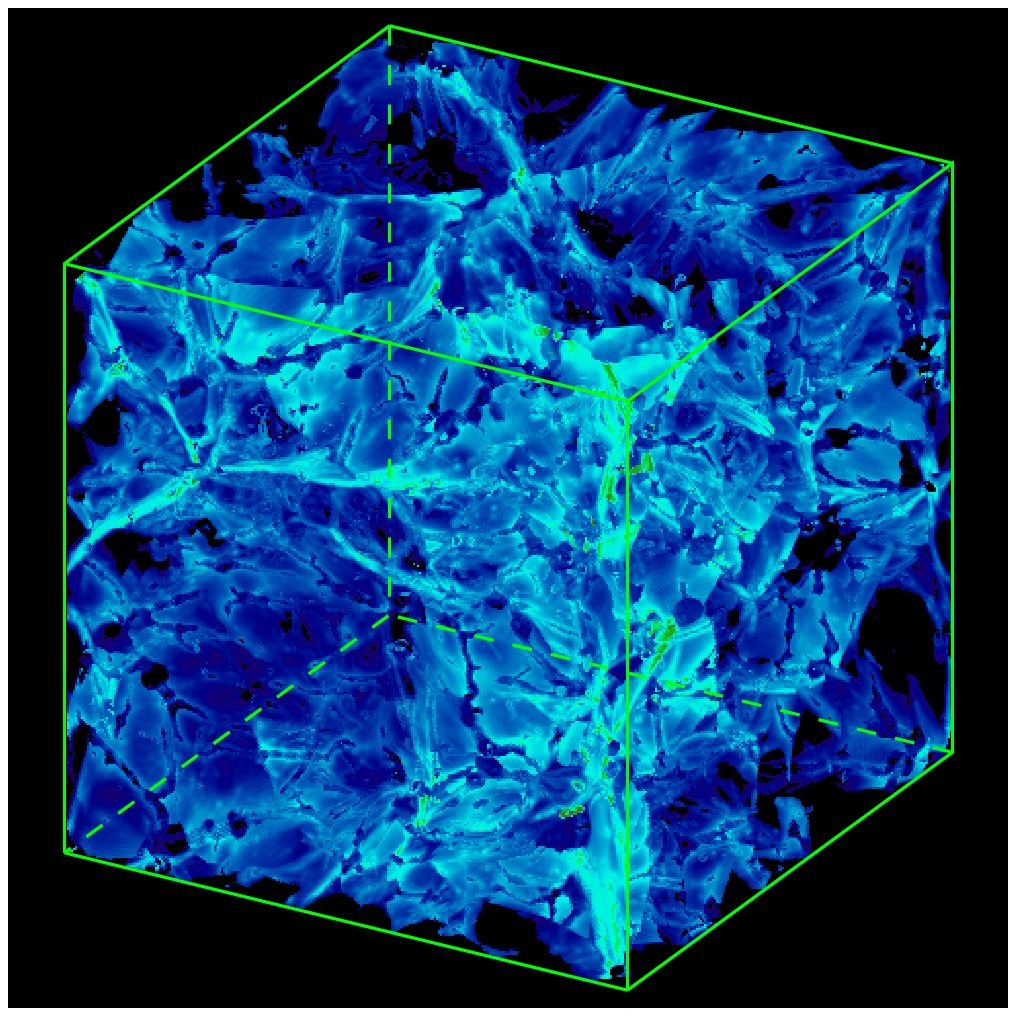}
}
\vspace{-2.2cm}
\gridline{
\hspace{-1.8cm}
\includegraphics[width=0.75\textwidth]{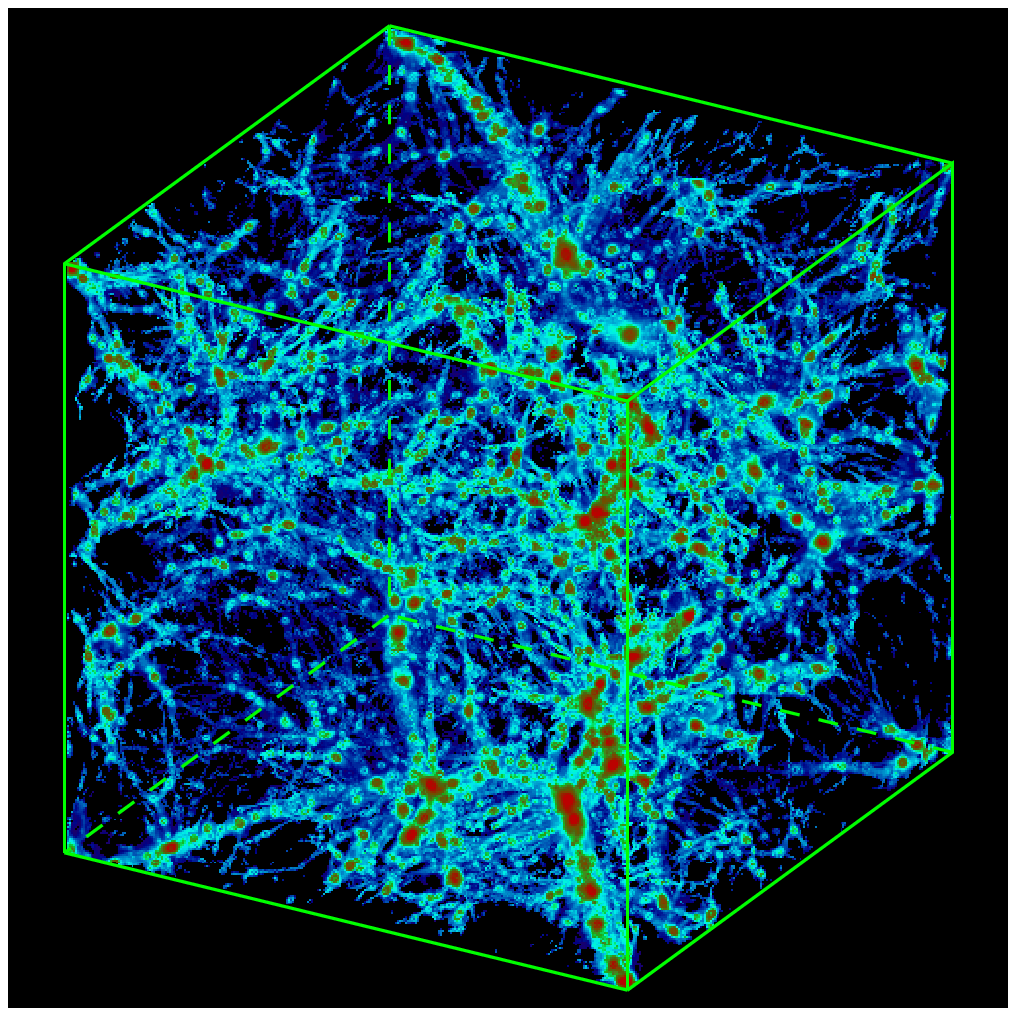}
\hspace{-5.5cm}
\includegraphics[width=0.75\textwidth]{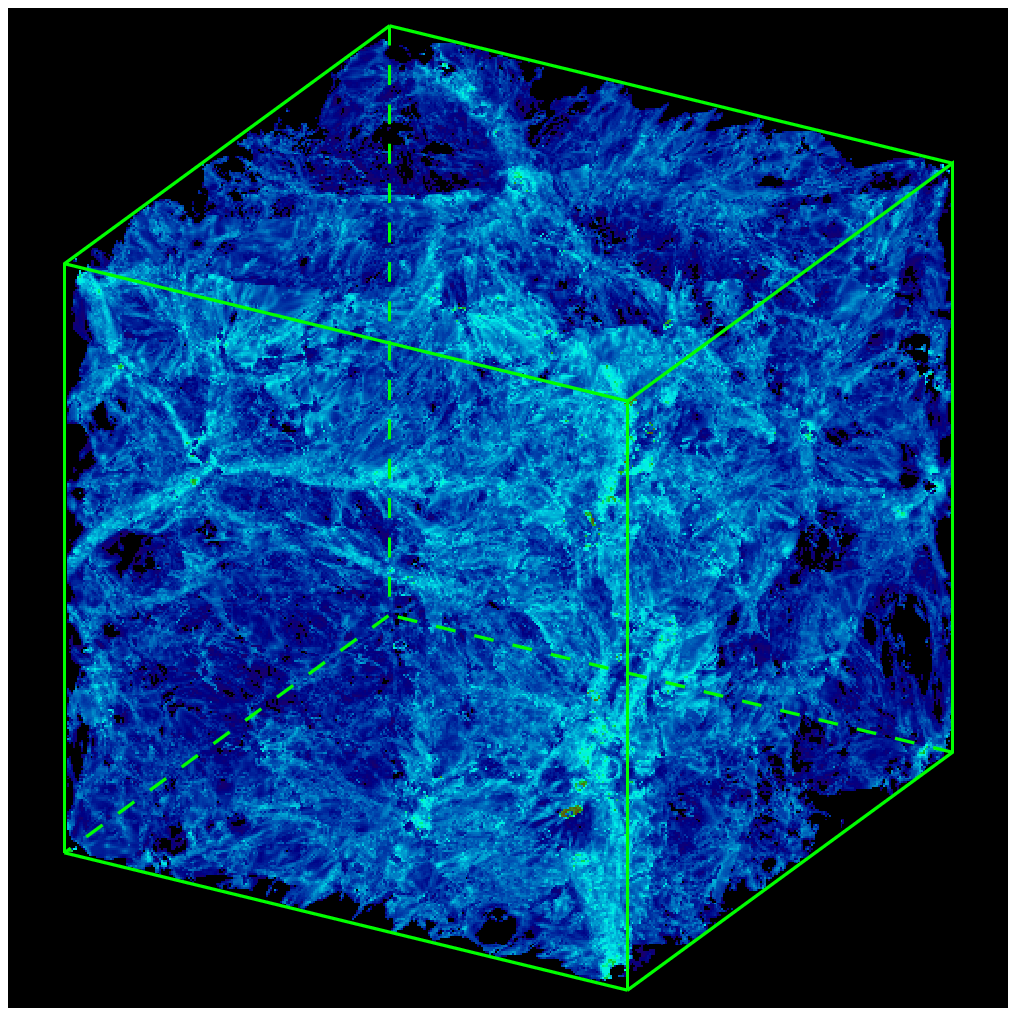}
}
\vspace{-1.2cm}
\caption{The filaments(Left) and sheets/walls(Right) identified by \textbf{d-web} for baryon at z=3(Top) and z=0(Middle), and for cdm at z=0(Bottom).}
\label{figure3}
\end{figure*}

The filaments and sheets identified by \textbf{d-web} at $z=3.0$ and $z=0.0$ in L025 with a constant $\lambda_{th}=0.6$ over redshift and $R_s=R_g$ are presented in Figure 3. The \textbf{d-web} is capable of finding out the filaments with various length and radius, as well as the sheets. The spacial distribution of filaments at $z=3$ is significant different with $z=0$ on small scale, while is similar on large scale to some extent. Larger smooth length and varying $\lambda_{th}$ at different redshifts may enhance the similarity, as in Bond et al.(2010b). The most prominent filaments are connecting to massive clusters at $z=0$. However, most of the filaments at high redshifts do not connect to any clusters. They will coalesced into prominent ones at lower redshifts, e.g., around $z=1$. The \textbf{d-web} classification has extracted the filaments from the voids and sheets embracing them, leaving empty tubes(Figure 2b. and 2d.). The sheets are fluffy without distinct boundary at high redshifts, due to low density contrast with respect to the voids. The sheets in baryonic matter evolve to a much smooth and thin shape at $z=0$. The distribution of sheets on larger scale at $z=3$ is similar as $z=0$ at certain level, which we will revisit later with slice view. The discreteness of cold dark matter particles leads to the identified filaments and sheets appearing relatively rough. The sheets are suffering more severely from the under-sampling in low density regions, even after smoothing. 

\subsection{Mass Distribution and Its Evolution}

The volume and mass content are the most explicit quantities to demonstrate the evolution of cosmic web(Hahn et al. 2007b; Bond et al. 2010; C14). Table 1 summarizes the mass fraction and mean density of each cosmic web component identified by \textbf{d-web} in our simulations at $z=0$ with $\lambda_{th}=0.6$ and $R_s=R_g$. A lower fraction of baryonic matter is found to residing in clusters and filaments, which is more significant in L025 with the smallest box size. The mass fraction of gas in filaments is less than dark matter by $8\%$ in its absolute value, while for both sheets and voids it is $\sim 3.6\%$ higher. It implies that more baryonic matter is likely to inhabit in the under-dense or mildly over-dense region at $z=0$, which may be attributed to thermal pressure effect and non-thermal turbulent motions. The box size has minor effects on the results on these two global quantities. The mean density of baryon in each environment is usually close to dark matter, i.e., $1+\delta >50$ in clusters, and around $\sim 6, 0.8, 0.2$ in filaments, sheets and voids respectively, in agreement with the results in recent simulation studies(e.g. Aragon-Calvo, van de Weygaert \& Jones 2010; C14). In L025, however, the mean density of baryon in clusters is nearly twice of dark matter. In comparison, the excursion model in Shen et al.(2007) gives an approximation of $36$ and $6$ times denser than the critical density for filaments and sheets. The median density of filaments of baryonic matter is also lower than the estimated density of WHIM, i.e., $\sim 10-30$(Dave et al. 2001; Fang et al. 2010).  It is probably because the mean density in the background, i.e., the voids, is only $0.2$ times of the cosmic mean. Meanwhile, more matter will reside in clusters and filaments, when the simulation box get smaller and the resolution is increased. 

In Figure 4., we investigate the evolution of volume and mass fraction in voids, sheets, filaments and clusters since $z=5.0$ in L025 with the d-web scheme. The smooth length is fixed as $R_s=R_g$, meanwhile three values of  $\lambda_{th}$, i.e., $0.2, 0.4, 0.6$ are probed. The overall evolutionary histories inferred from the different values of $\lambda_{th}$ are similar. The volume fraction of each component varies mildly along with time, while for the evolution of mass fraction, it is clearly more significant. The volume fraction of clusters is less than $1\%$ throughout time, and hence is not plotted in this figure. The exact volume fractions in voids and sheets, as well as the mass fraction in voids  at given redshift are sensitive to the value of $\lambda_{th}$. A lower $\lambda_{th}$ would lead to more cells are identified as parts of filaments and sheets, instead of voids. For $\lambda_{th}=0.4, 0.6$, the voids occupy the largest volume share, and then sheets and filaments in turn. However, the volume fraction of sheets is larger than voids at $z<5$ for $\lambda_{th}=0.2$. Forero-Romero et al. (2009) showed that the volume fraction of sheets at $z=0$ would large than voids for $\lambda_{th}<0.25$ and a lower $\lambda_{th}$ closing to $0.0$ would make the voids becoming too small and isolated.

In terms of the mass fraction, clusters grows gradually with time, while filaments demonstrate a rapid growth. On the contrary, the mass content in voids keeps shrinking towards lower redshifts, becoming more sharply for higher $\lambda_{th}$. The mass fraction in sheets varies slowly at $3<z<5$ and decreases with time since $z \sim 3$. At redshift $z=5$, about $33\%-45\%$ of the mass resides in sheets, while about $18\%-30\%$ and $2\%-3\%$ have collapsed into filaments and clusters respectively. Down to $z=0$, more than half of the mass are found in the filaments. For threshold values between $0.2$ and $0.6$, the filaments become the primary collapsed structures, in replacement of sheets, as measured by the mass fraction at around $z \sim 2-3$. The phenomenon that baryonic matter is less residing in filaments and clusters occurred as early as $z>3$.  It is noticed that, the short of baryon has been largely remedied in clusters at low redshifts, while the relative difference remains around $10-20\%$ between $z=0$ and $z=5$ in filaments. The short of gas is mildly more apparent for higher $\lambda_{th}$.

\begin{table*}
\begin{center}
\begin{tabular*}{16cm}{clcccccccc} 
\hline
\hline
\multirow{2}{*}{Simulation}&  & \multicolumn{4}{c}{Mass fraction} & \multicolumn{4}{c}{Mean density}\\
& & Cluster & Filament & Sheet & Void &  Cluster & Filament  & Sheet & Void \\
\hline
\hline
\multirow{2}{*}{L100}& cdm & 14.8\% & 49.1\%  & 23.0\% & 13.1\% & 50.91 & 5.37  & 0.79 & 0.21  \\
& gas  & 12.8\% & 47.2\%  & 25.3\% & 14.6\% & 51.85 & 5.62  & 0.87 & 0.24 \\
\hline
\multirow{2}{*}{L050}& cdm & 15.3\% & 54.7\%  & 19.8\% & 10.1\% & 72.24 & 5.67  & 0.66 & 0.17 \\
& gas & 13.2\% & 51.6\%  & 22.9\% & 12.2\% & 91.02 & 5.98  & 0.77 & 0.20  \\
\hline
\multirow{2}{*}{L025}& cdm  & 14.8\% & 60.0\% & 17.1\% & 8.1\% & 94.93 & 5.83  & 0.56 & 0.14 \\
& gas & 14.6\% & 52.0\%  & 21.5\% & 11.8\% & 181.7 & 5.89  & 0.73 & 0.19  \\
\hline
\hline
\end{tabular*}
\caption{The mean density and mass fraction of each cosmic web component identified by \textbf{d-web} in three simulations at $z=0$ with $\lambda_{th}=0.6$ and $R_s=R_g$. The term cdm and gas refer to cold dark matter and baryon respectively.}
\end{center}
\vspace{-0.2cm}
\end{table*}

\begin{figure*}[tbp]
\vspace{-0.8cm}
\includegraphics[width=0.50\textwidth]{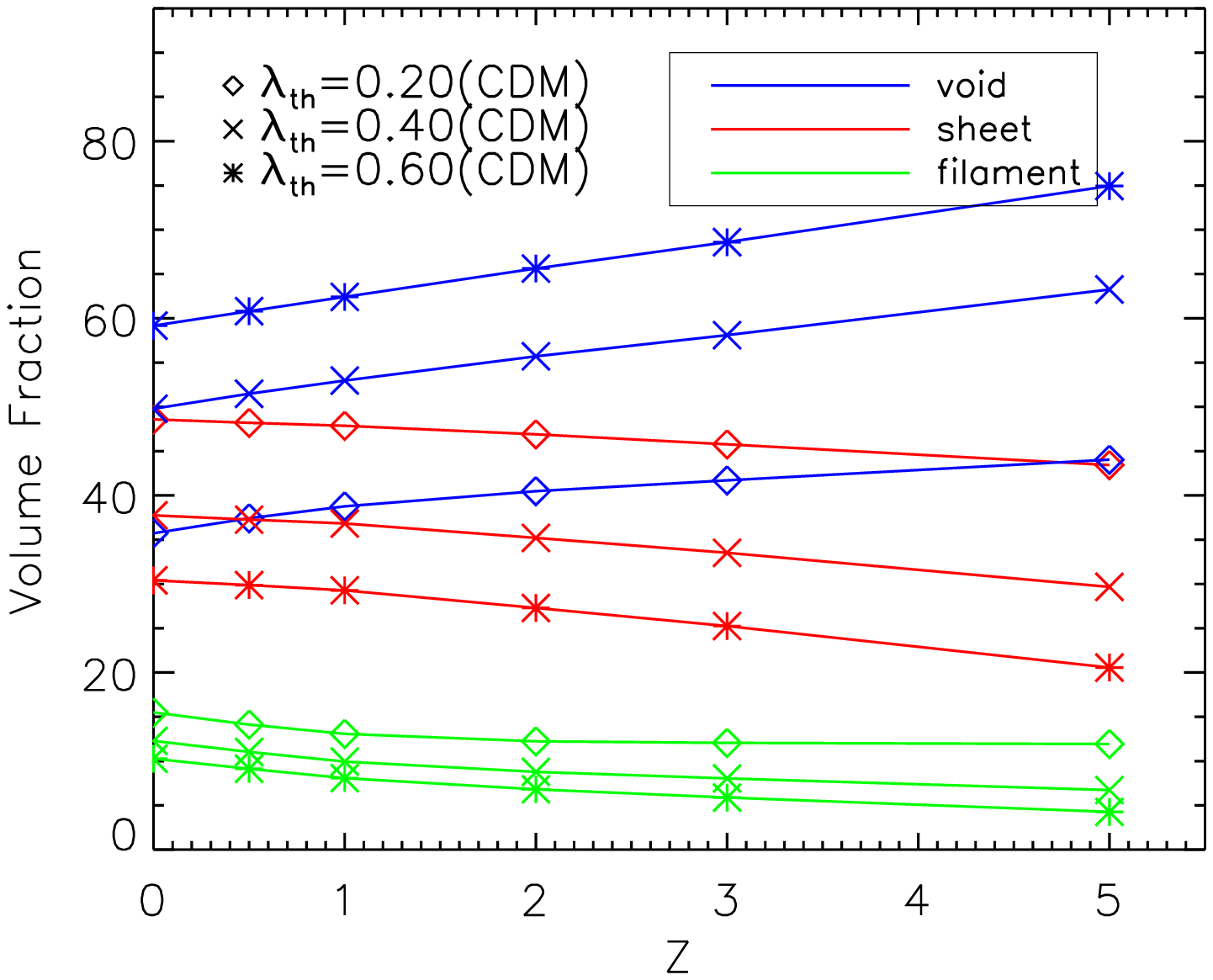}
\includegraphics[width=0.50\textwidth]{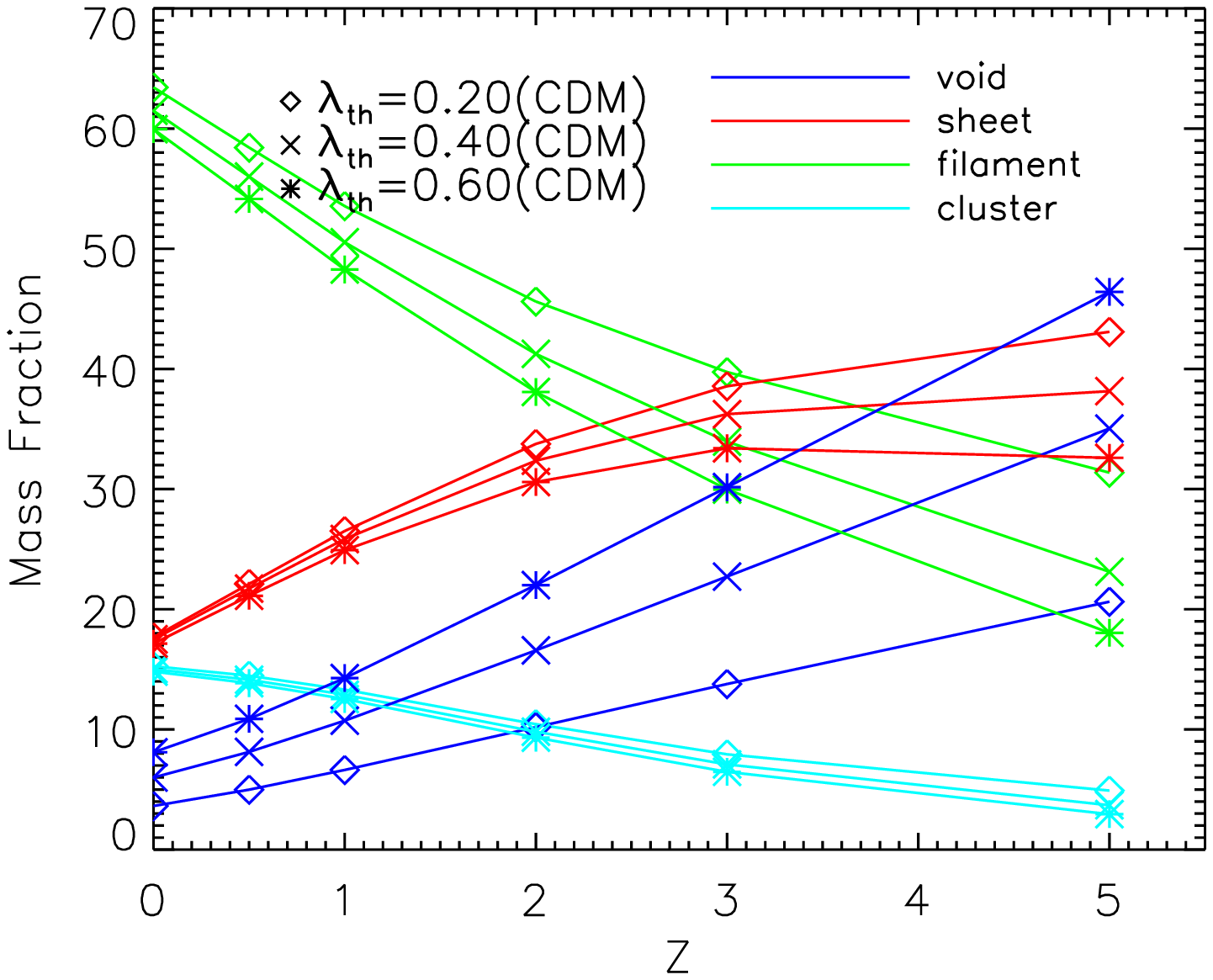}
\vspace{-0.1cm}
\includegraphics[width=0.50\textwidth]{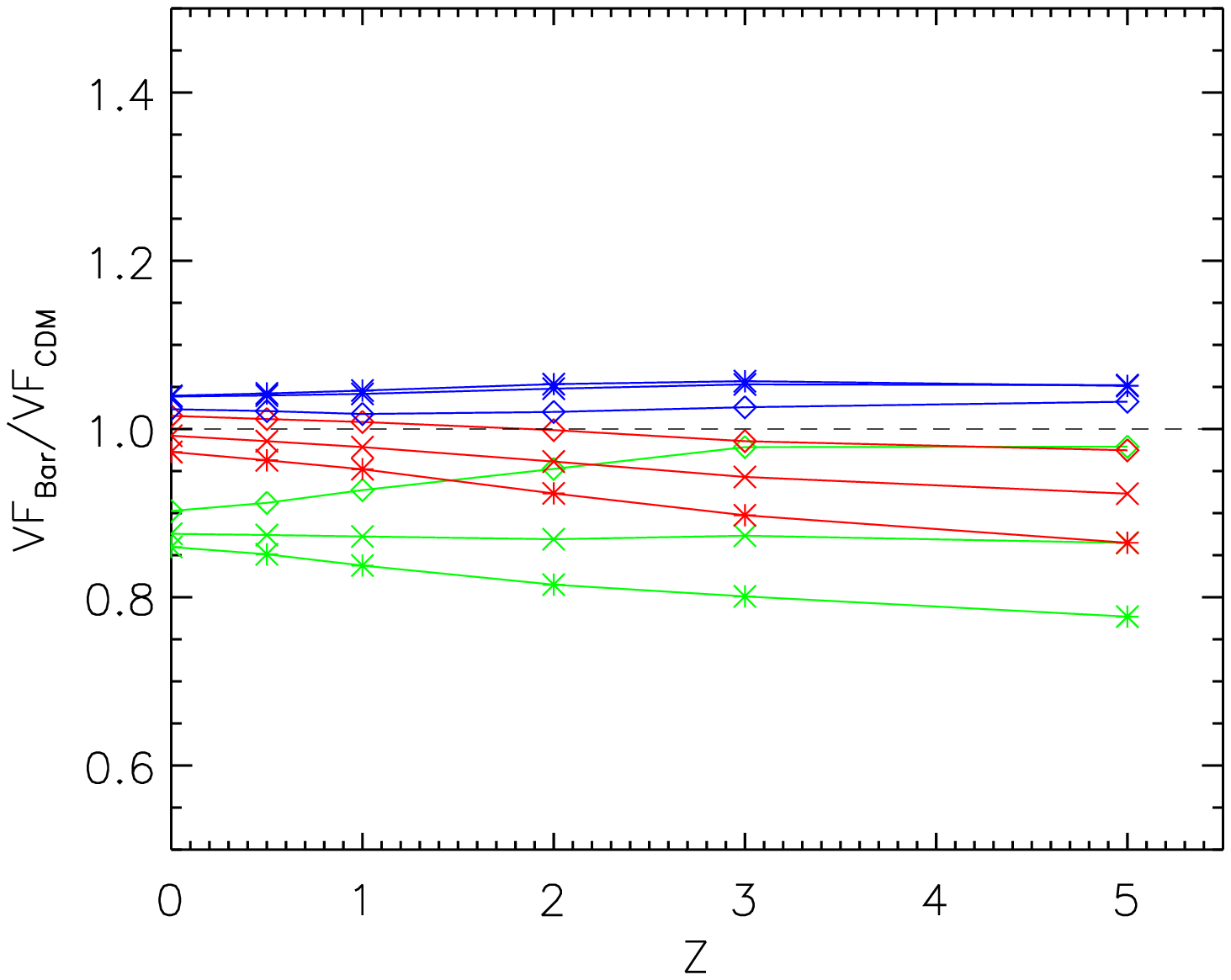}
\includegraphics[width=0.50\textwidth]{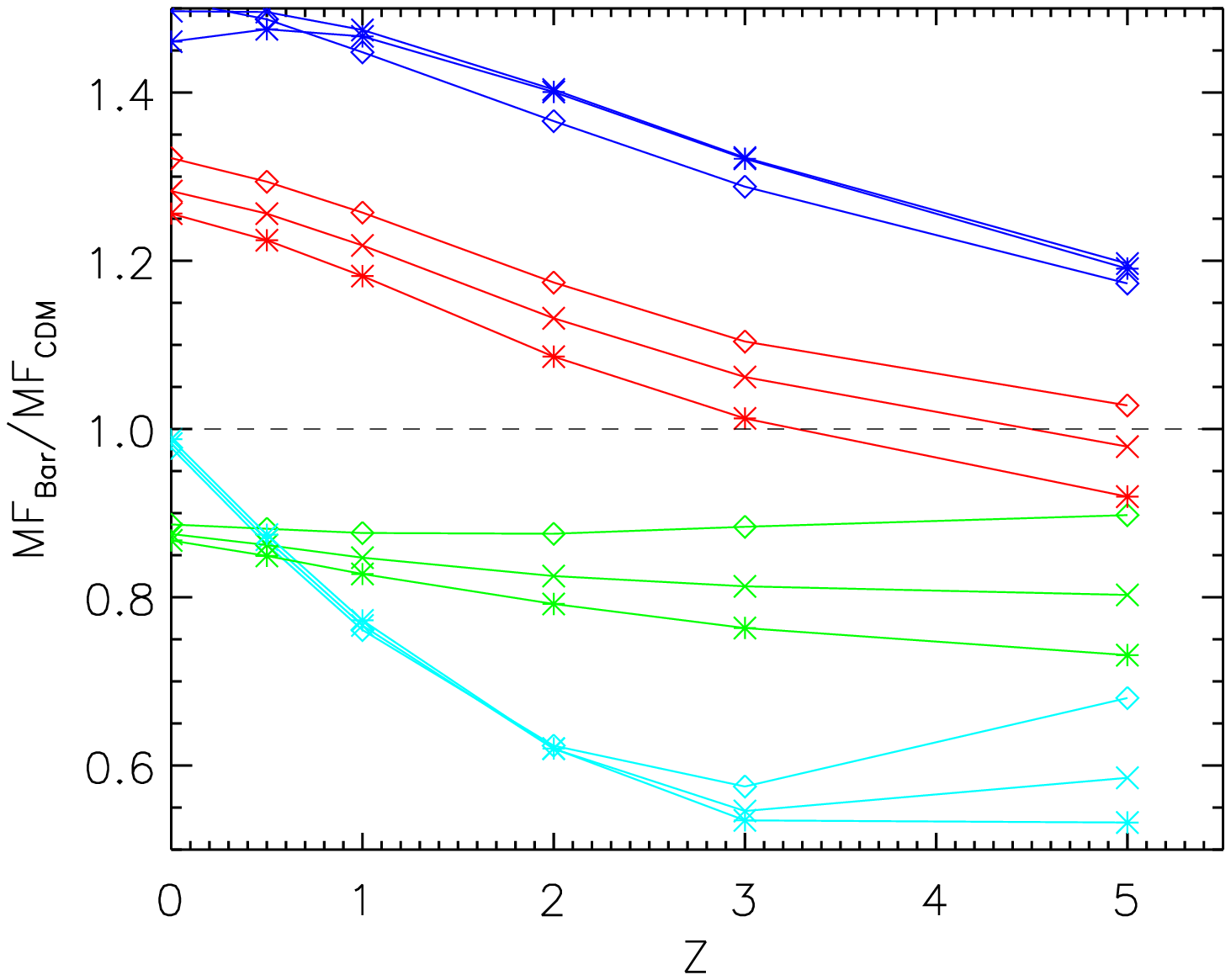}
\caption{The volume and mass fraction of void, sheet, filament and cluster identified by \textbf{d-web} with $\lambda_{th}=0.2, 0.4$, and $0.6$ and $smt01$ in L025 since $z=5.0$. Top row give results of dark matter, while the bottom row show the ratio of baryonic matter to dark matter on volume and mass fraction.}
\label{figure4}
\end{figure*}

\begin{figure*}[tbp]
\vspace{-0.8cm}
\includegraphics[width=0.50\textwidth]{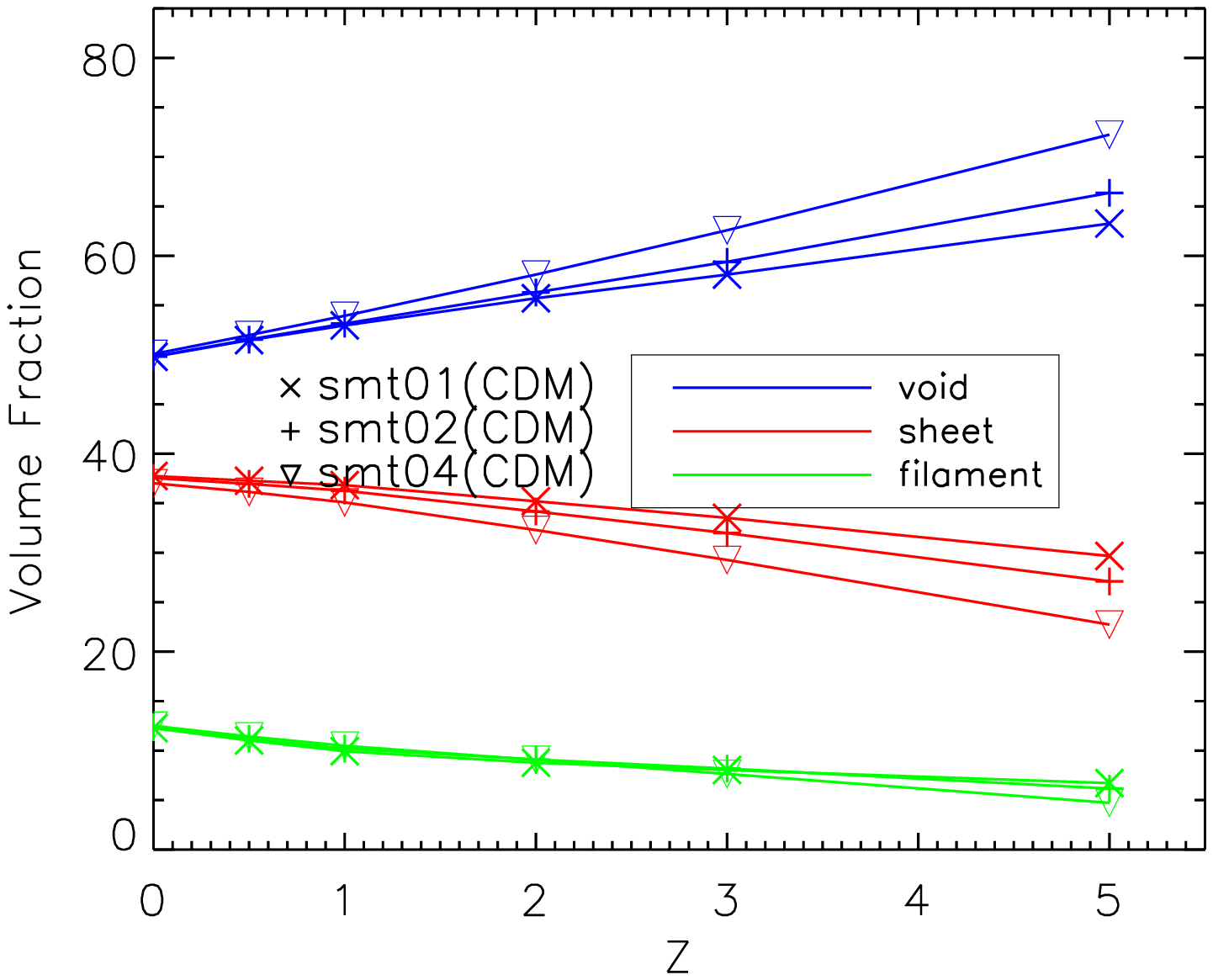}
\includegraphics[width=0.50\textwidth]{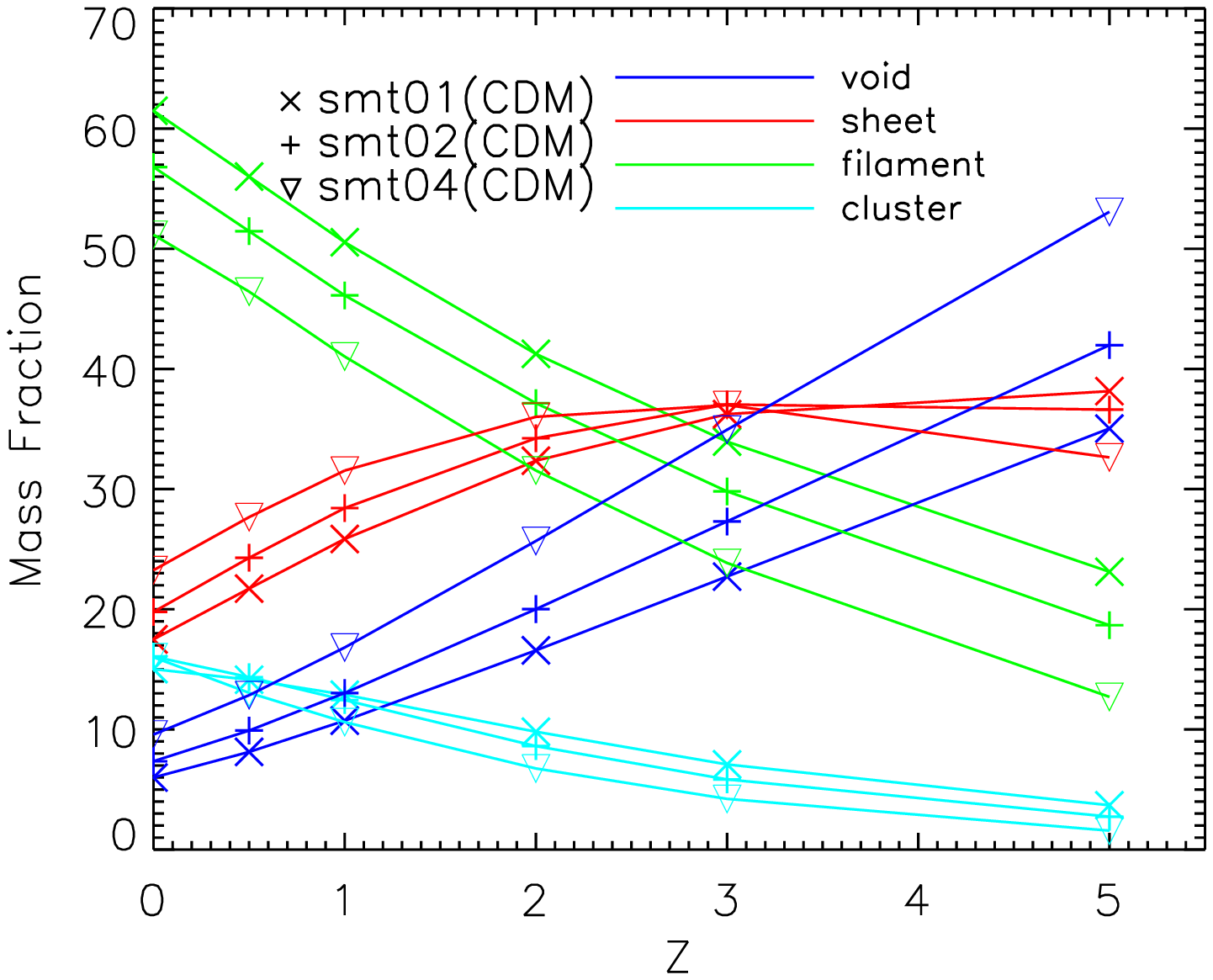}
\vspace{-0.1cm}
\includegraphics[width=0.50\textwidth]{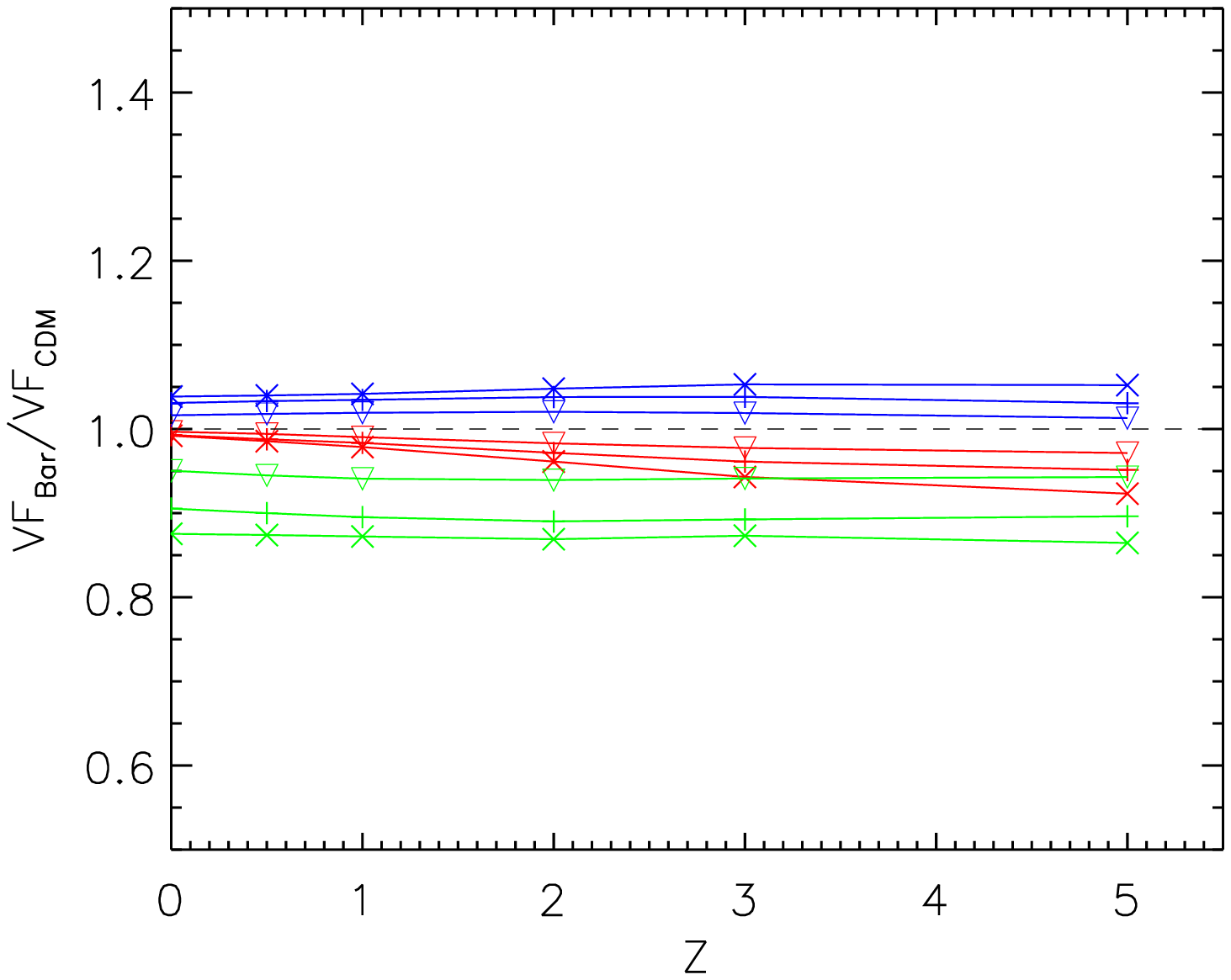}
\includegraphics[width=0.50\textwidth]{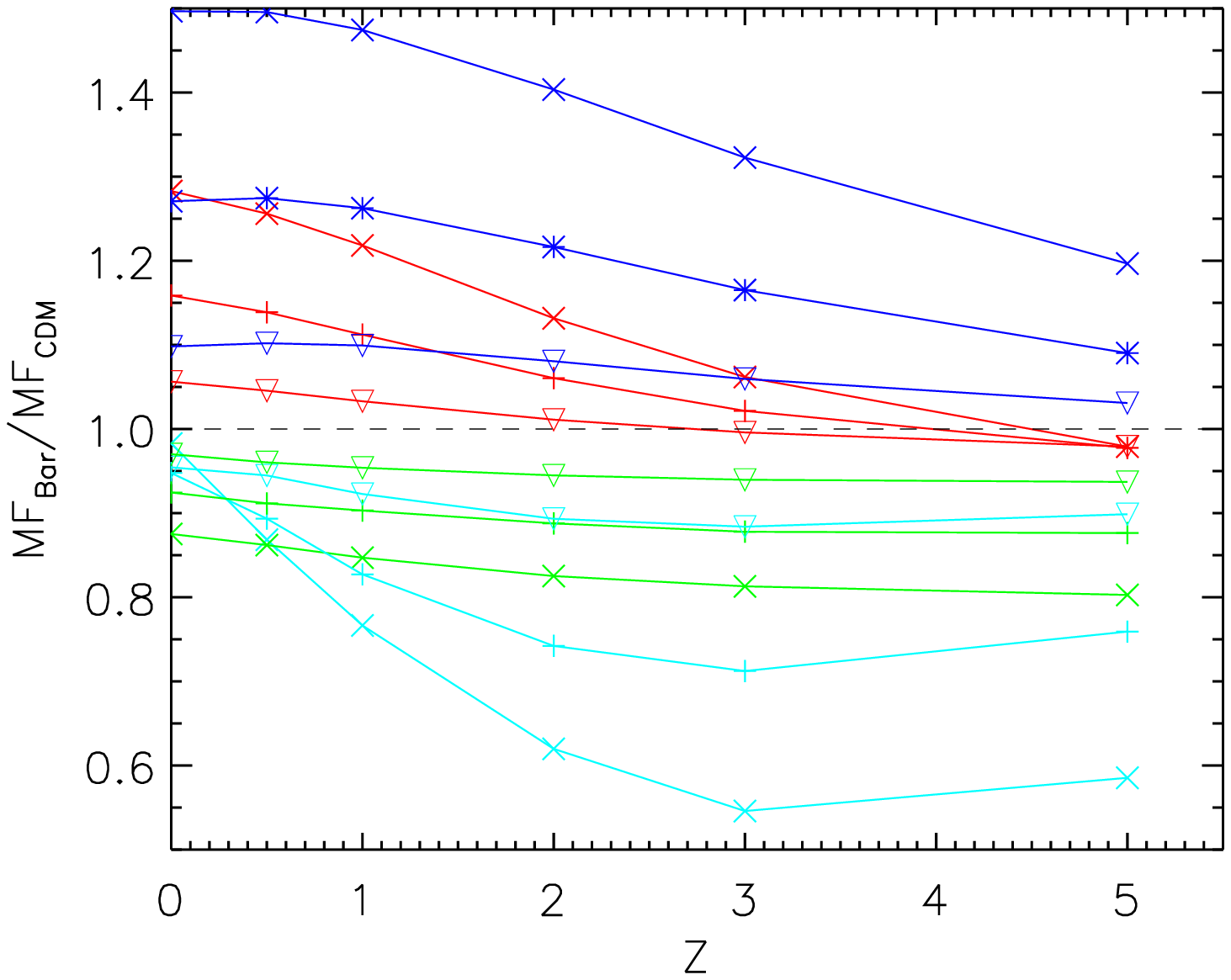}
\caption{The volume and mass fraction of void, sheet, filament and cluster identified by \textbf{d-web} with $\lambda_{th}=0.4$ and smt01, smt02, smt04 in L025 since $z=5.0$. }
\label{figure5}
\end{figure*}

The evolution of volume and mass fraction in L025 with different Gaussian smooth lengths for fixed $\lambda_{th}=0.4$ are demonstrated in Figure 5. 
The effect of different values of $R_s$ is more significant in mass fraction than volume fraction. A larger $R_s$ will lead to notable more mass in voids, and on the contrary, less in filaments. While for sheets, the dependence of mass fraction on $R_s$ makes a turnover before and after $z\sim3$. The redshift that filaments take over sheets as the primary structure decreases moderately for larger smooth length, and is around $z\sim2$ for $R_s = 195 h^{-1} kpc$. The discrepancy between the mass fraction of baryonic and dark matter in filaments and clusters is narrowed by increasing $R_s$. For L025, with the space resolution of $24.4 h^{-1} kpc$, the relative difference is reduced to $\sim 5-10\%$ while taking $R_s = 195 h^{-1} kpc$.

Figure 6 presents the evolution of fractions in all the three simulations with $\lambda_{th}=0.4$. Results in L025 with $R_s=2R_g, 4R_g$ are compared to L050 and L100 with $R_s=R_g$ respectively. Both the volume and mass fractions of dark matter in different simulations are very close, once the smooth length is set equal. Since we have used the same random seeds and phases to produce the initial conditions, it is obviously not out of our expectation. In addition, except for mild differences, the deficiencies of baryonic matter in filaments and clusters can be seen in all simulations. In term of the mass fraction, baryonic matter is less than dark matter by $\sim 5-15\%$ relatively in filaments and clusters at a effective smooth scale of $R_{eff}=0.29 h^{-1} Mpc$ in L100.

\begin{figure*}[tbp]
\vspace{-1.0cm}
\includegraphics[width=0.50\textwidth]{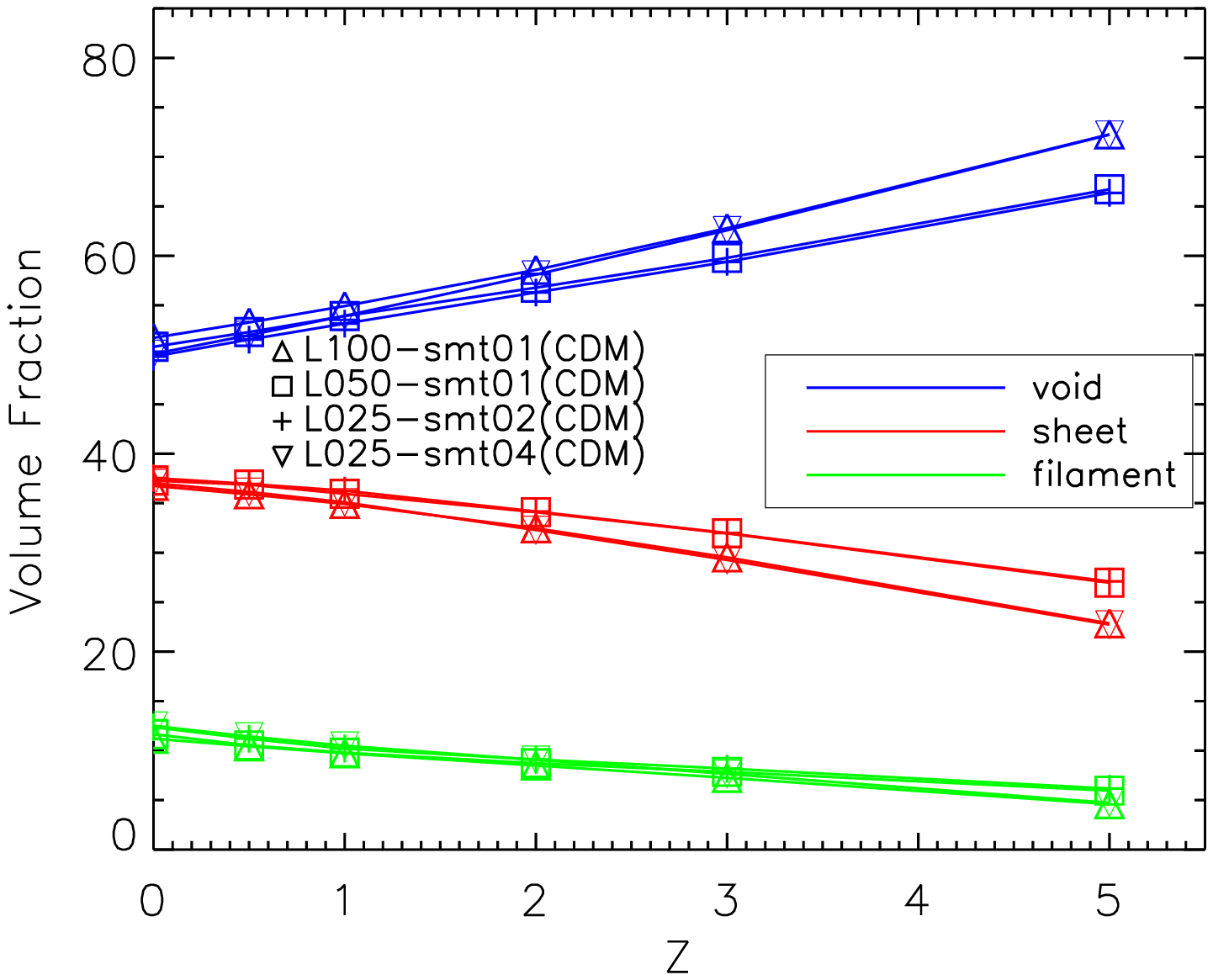}
\includegraphics[width=0.50\textwidth]{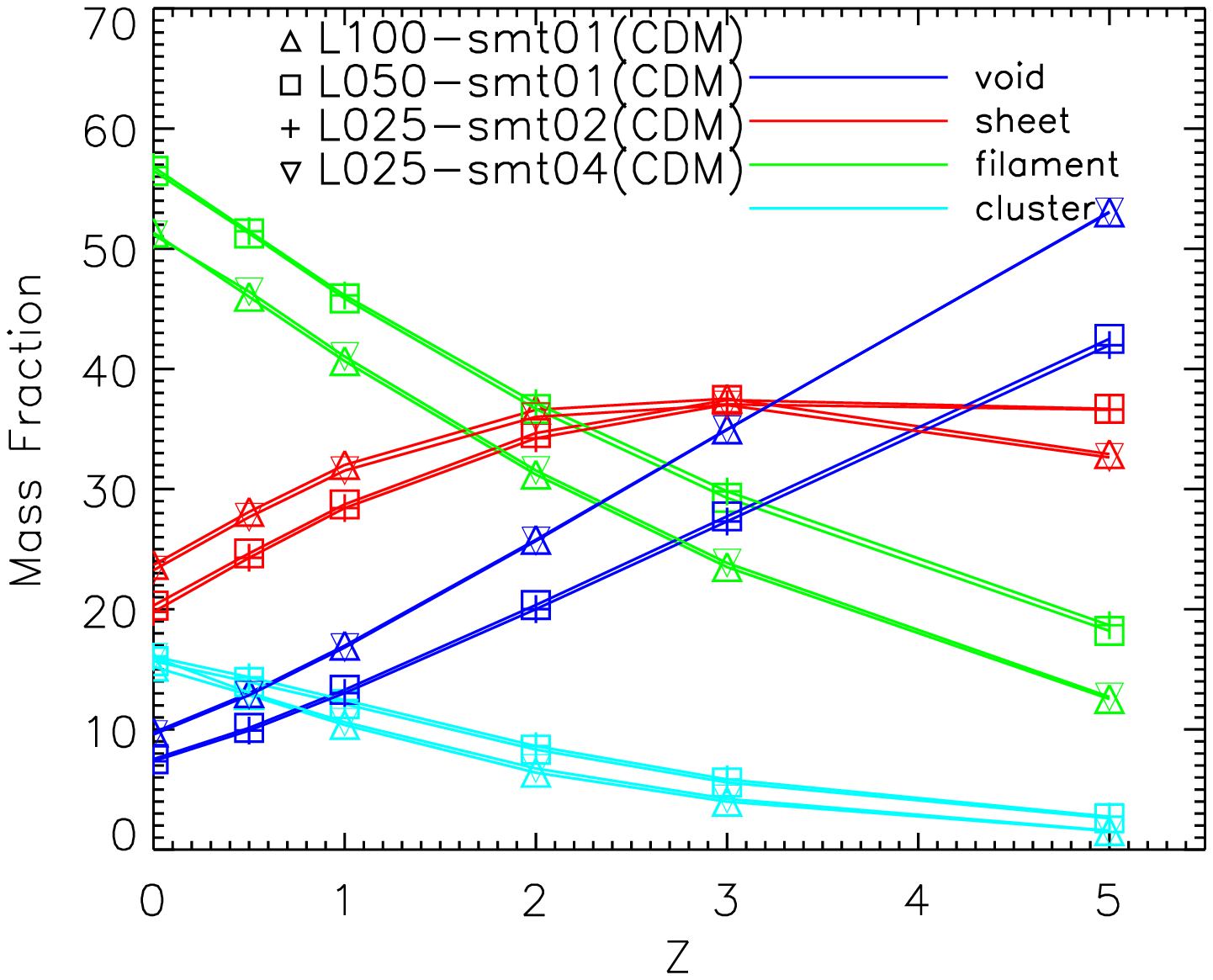}
\vspace{-0.1cm}
\includegraphics[width=0.50\textwidth]{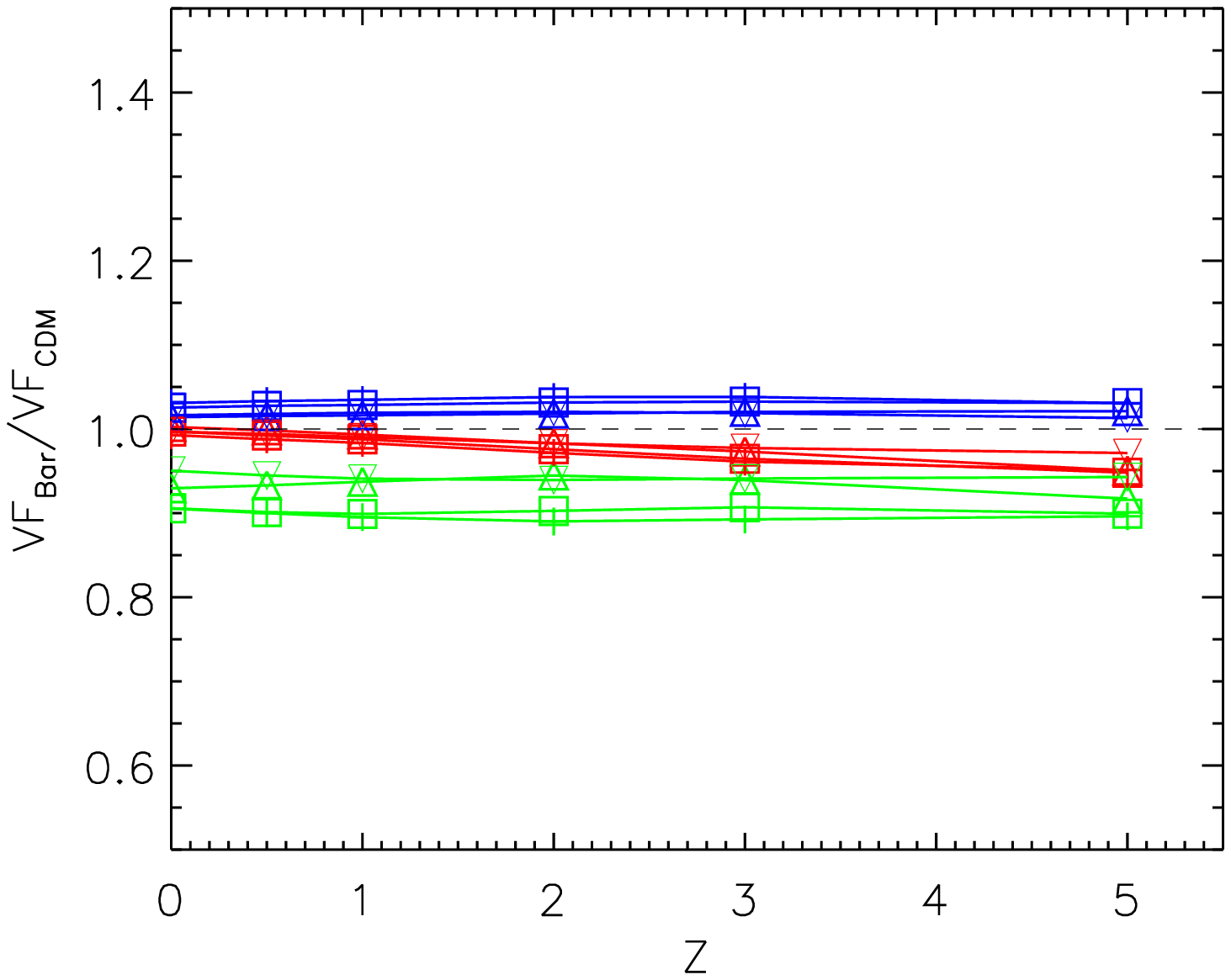}
\includegraphics[width=0.50\textwidth]{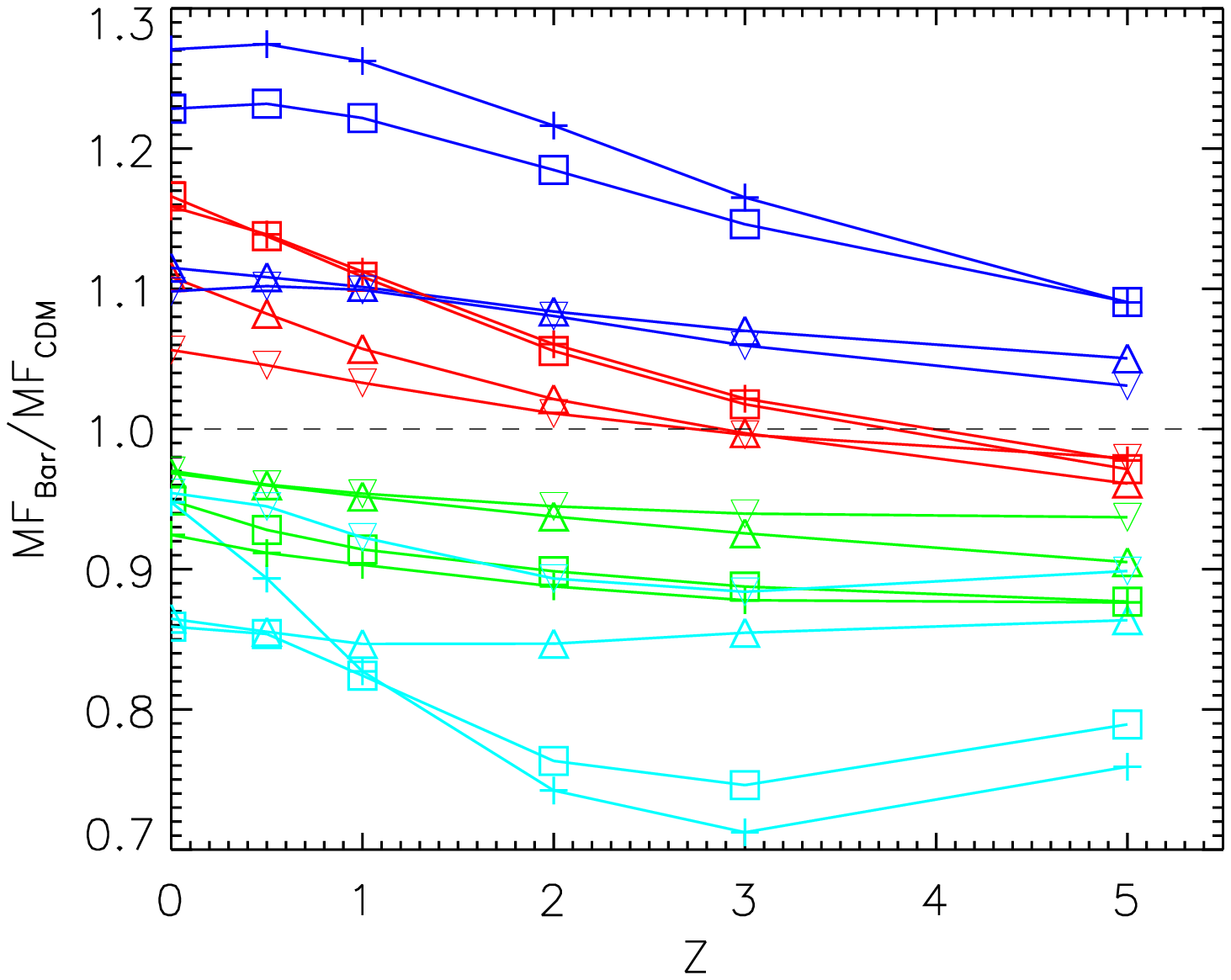}
\caption{The volume and mass fraction of baryonic and dark matter(CDM) in void, sheet, filament and cluster identified by \textbf{d-web} with $\lambda_{th}=0.4$ in three simulations since $z=5.0$.}
\label{figure6}
\end{figure*}

For a short summary, a nice bit of mass have fallen into sheets and proto-sheets, as well as filaments at $z=5$. In terms of mass, sheets has been dominant over filaments since $z>2-3$ for the d-web web classification scheme with $\lambda_{th}=0.2-0.6$. The fast increasing of the mass fraction in filaments with decreasing redshifts is not conflict with the expectation from the structure growth history in the LCDM universe. Actually, filaments are structures experiencing two dimensional collapse. The rapid growth of mass in filaments since $z \sim 3$ should come from the further collapsing of sheets, and accretion from nearby under-dense environment as well. 

\begin{figure*}[tbp]
\vspace{-0.8cm}
\includegraphics[width=0.50\textwidth]{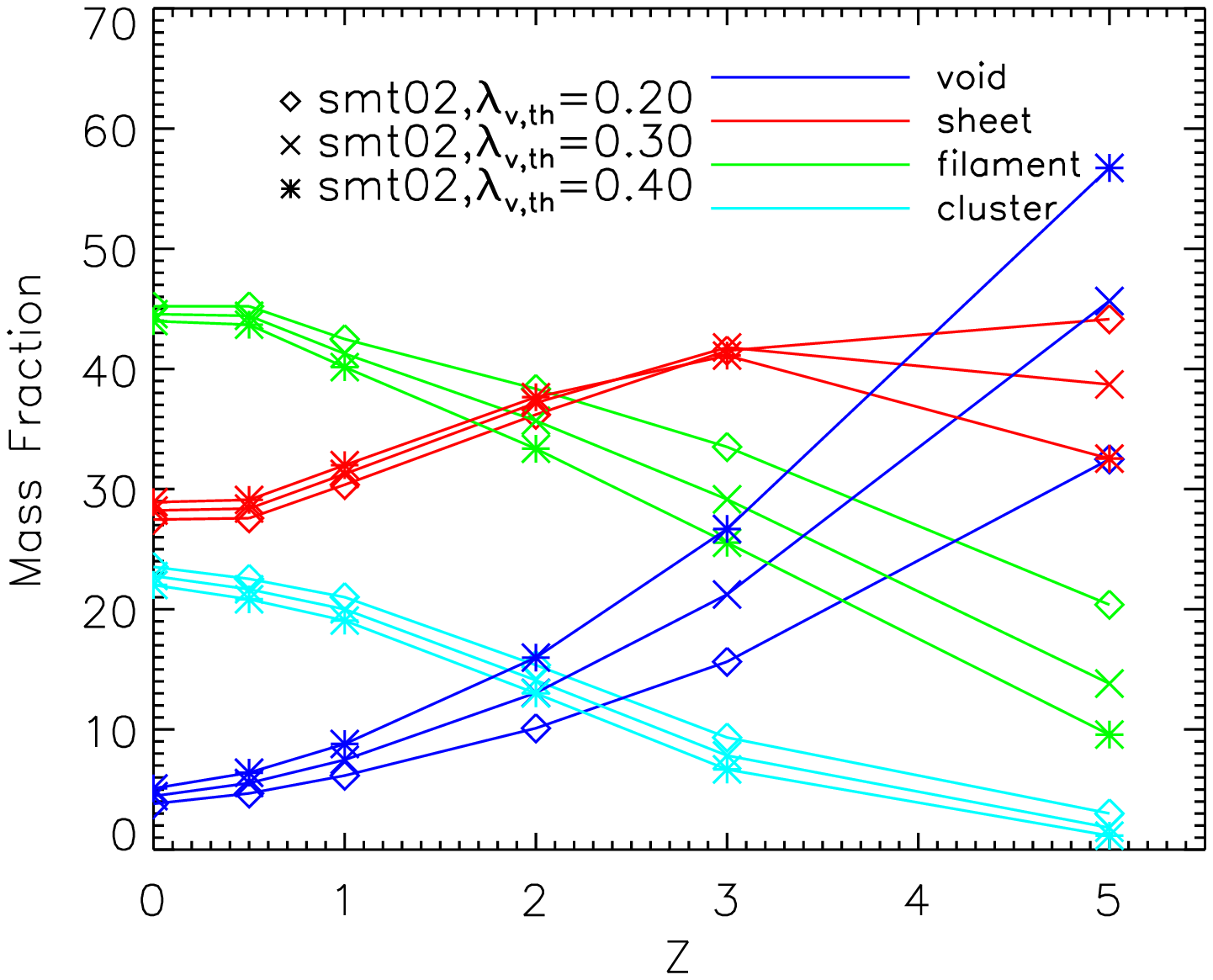}
\includegraphics[width=0.50\textwidth]{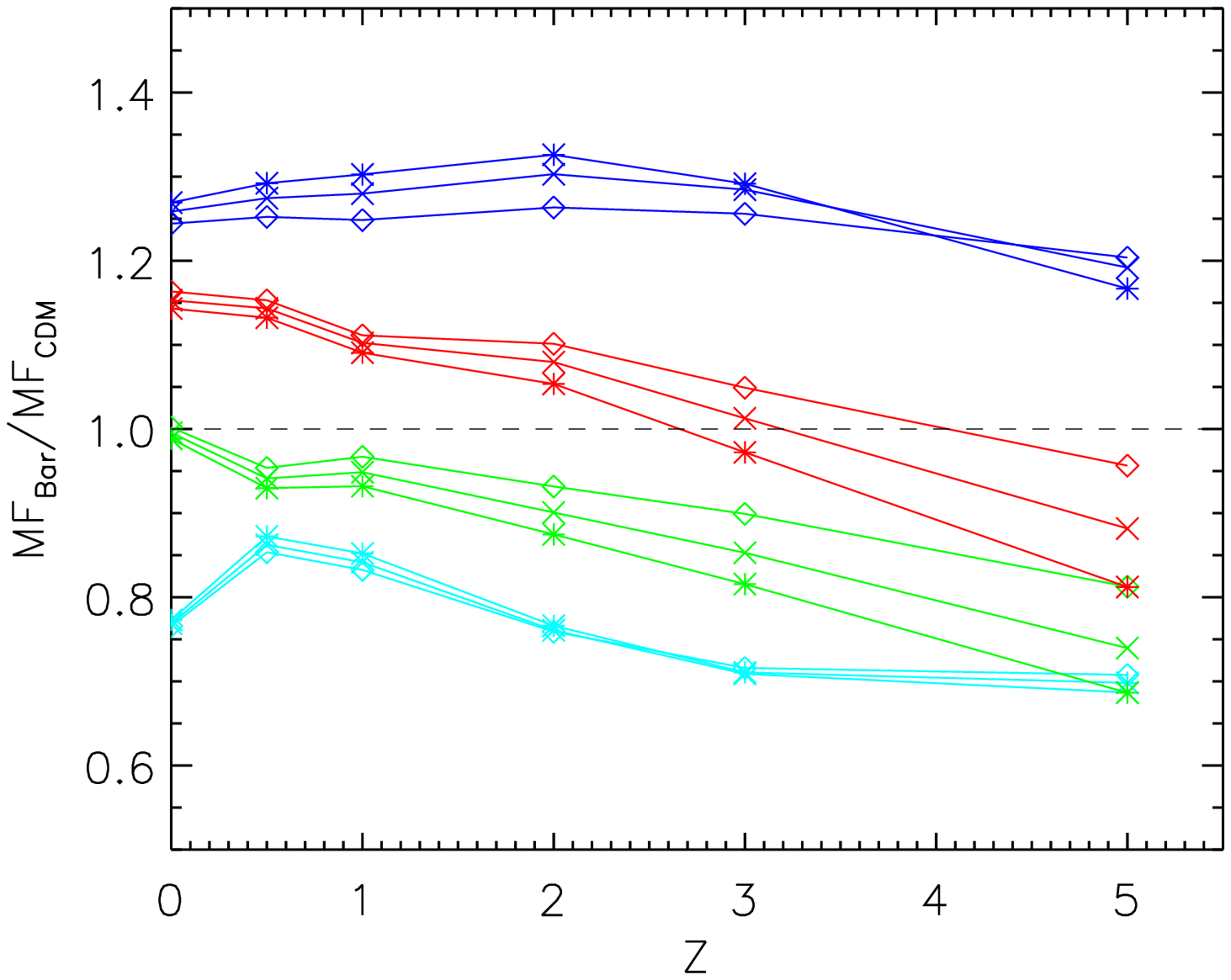}
\caption{Left: The mass fraction of dark matter in void, sheet, filament and cluster identified by \textbf{v-web} in L025 with $\lambda_{v,th}=0.2,0.3,0.4$ since $z=5.0$; Right: The ratio of baryonic matter to dark matter.}
\label{figure7}
\end{figure*}

We further investigate the impact of web classification scheme, by looking into the results of v-web. Figure 7 shows the mass fraction of dark matter in the cosmic web identified by \textbf{v-web} in L025, and the corresponding ratios of baryon to dark matter. $R_s=2R_g$ has been used to identify v-web to reduce the noise in under-sampling regions. $\lambda_{v, th}=0.2-0.4$ are used, smaller than $\lambda_{t,th}$ in d-web, similar as Hoffman et al.(2012). Clearly, the overall evolution is consistent with \textbf{d-web}.  On the other hand, relatively more grid cells are identified as sheets and clusters at low redshifts, rather than voids and filaments in \textbf{d-web}. The mass fraction found in filaments(clusters) is moderately lower(higher) for \textbf{v-web}. Moreover, in relative to \textbf{d-web}, the time since when filaments surpass sheets in mass fraction is postponed to around $z \sim 2$. The deficiency of baryonic matter remains $\sim 5-20\%$, except for in filaments at $z=0$. Our results with two different classification methods are generally consistent with each other.

\begin{figure*}[tbp]
\vspace{-0.8cm}
\includegraphics[width=0.50\textwidth]{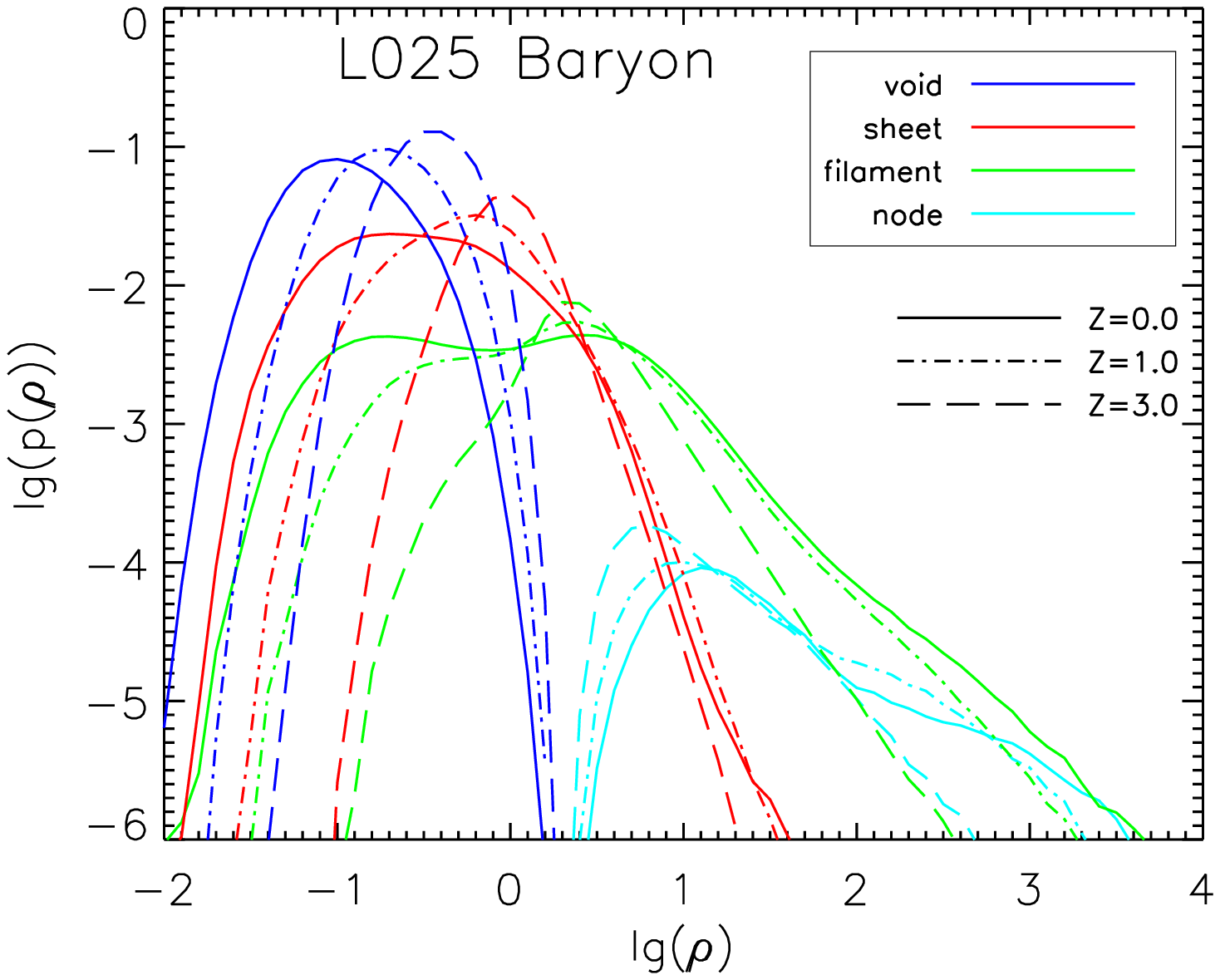}
\includegraphics[width=0.50\textwidth]{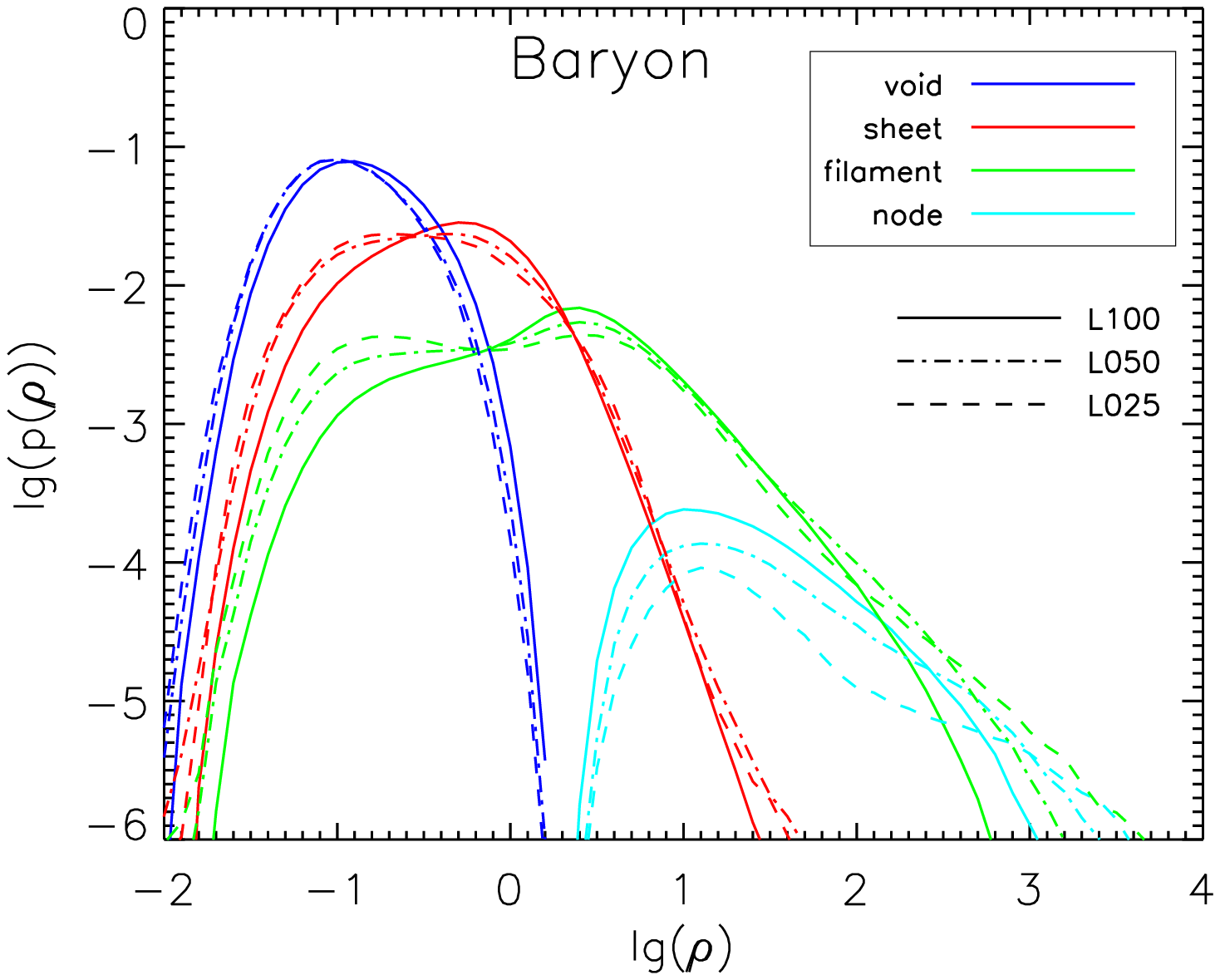}
\includegraphics[width=0.50\textwidth]{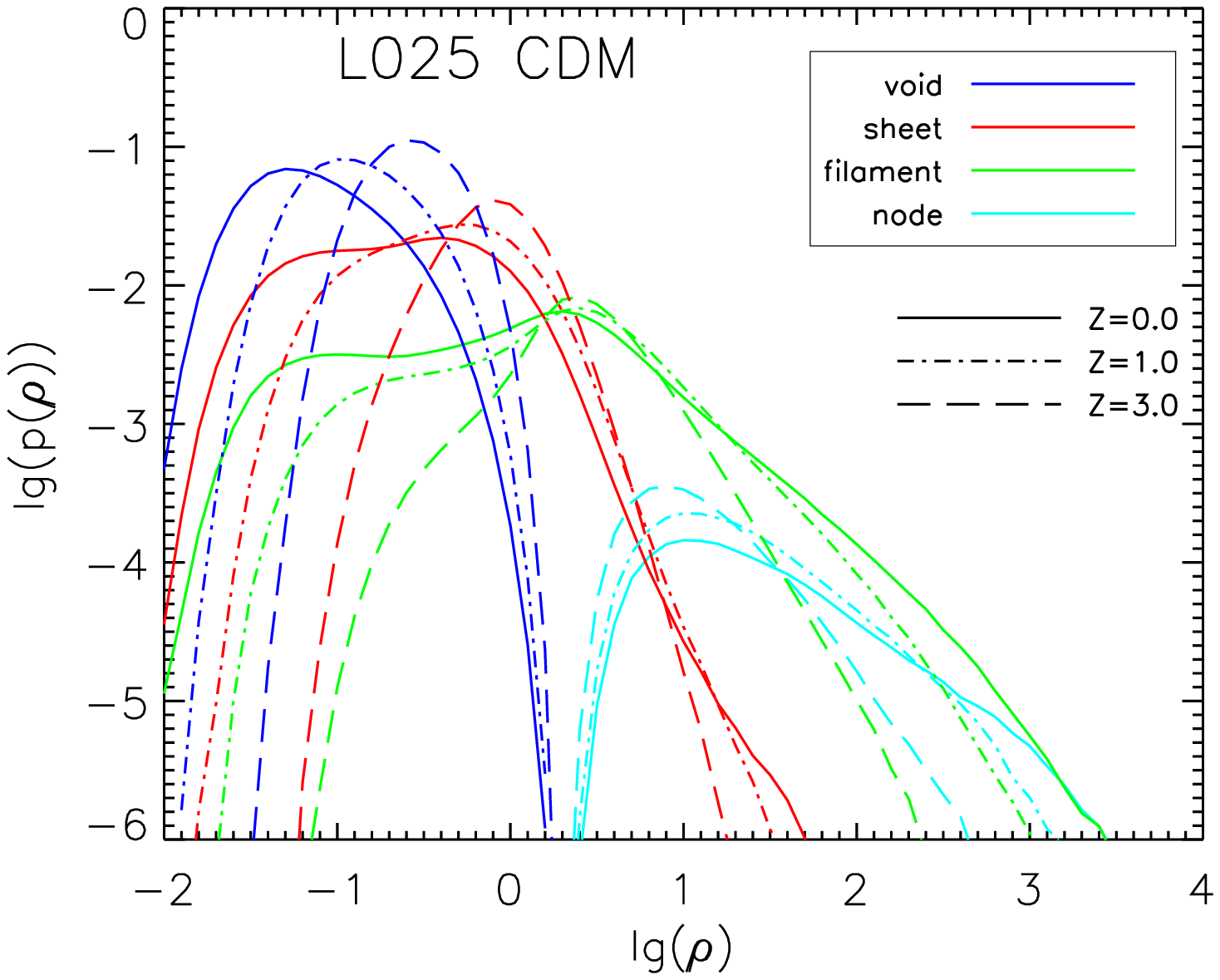}
\includegraphics[width=0.50\textwidth]{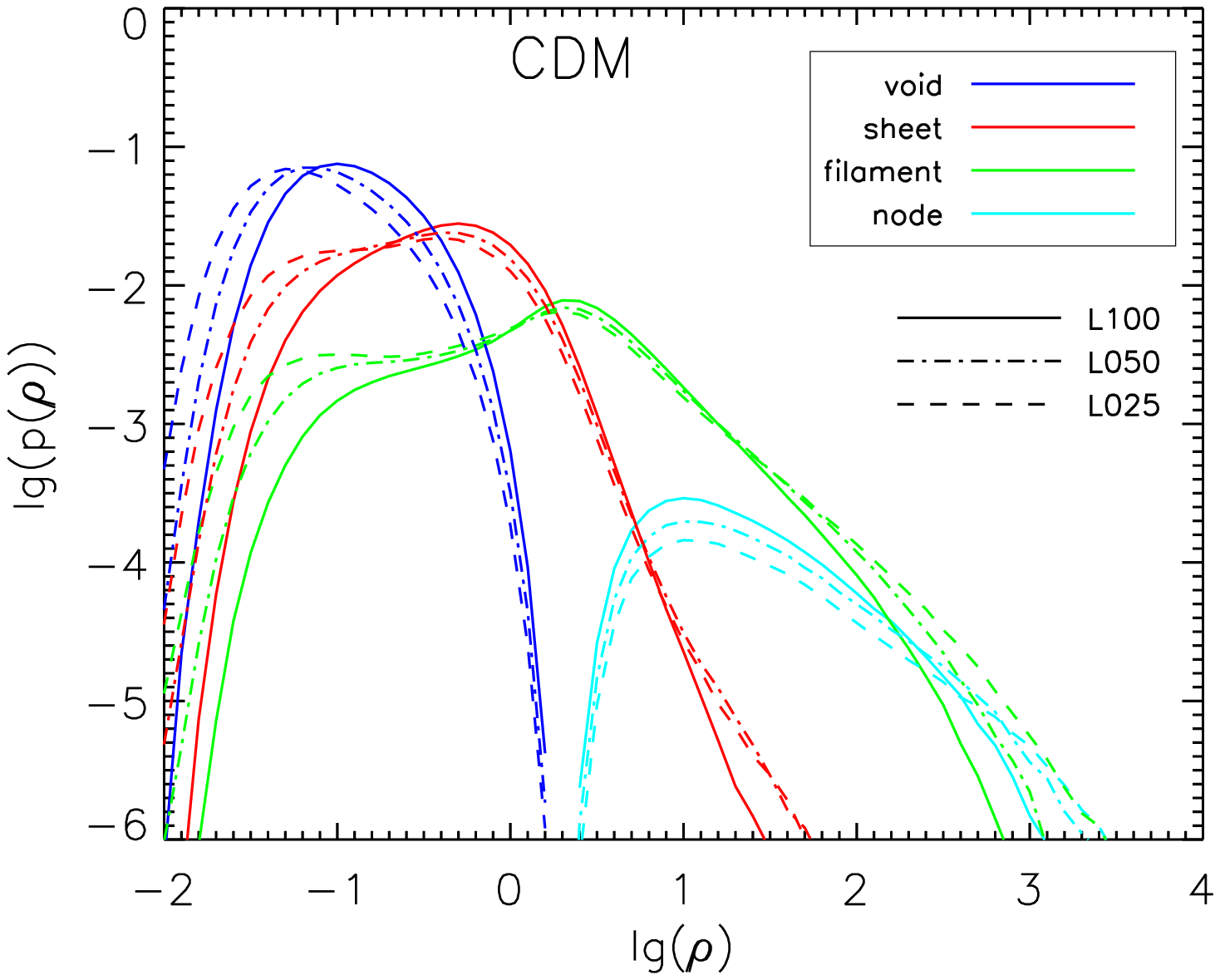}
\caption{The probability distribution of density in each cosmic web environment with $\lambda_{th}=0.6$ in L025 at $z=3.0, 1.0, 0.0$(Left), in three simulations at $z=0.0$(Right).}
\label{figure8}
\end{figure*}

Figure 8 gives the probability distributions of density of baryonic and dark matter in each cosmic web environment. The overtaking of sheets to voids, filaments to sheets, and clusters to filaments by cell number occurs at around $\delta \sim -0.6, 1.0, 200.0$. Each component covers a large density range. Especially, the density range in filaments almost covers the whole density range in the simulation sample. The cluster has the highest peak density, and then filament, sheet and void in decreasing sequence, which is consistent with the theoretical expectation and previous studies. Remarkable changes in time are observed regarding the density distribution function in filaments. In particular, the single peak distribution at high redshifts gradually transits to a double peak distribution at $z=0$, while a long tail at the high density end has been well developed at $z=1$. 

Moreover, a plateau appears in the density probability distribution function in sheets, and the covering density range expands progressively with time. The peak value decreases from  slightly larger than the cosmic mean ($\delta > 0$) at $z=3$ to smaller than it, i.e., $\delta <0$ at $z=0$. The probability distribution in voids remains a single peak pattern all the time, while the width at half peak grows gradually from high redshift to low redshift. The density of cell in voids decreases moderately with time, the peak value  is $\sim 0.30$ at $z=3$, and about $\sim 0.06$ at present epoch. A long tail at higher density has been also developed in clusters at $z=1$, which is the main change of the pdf of clusters. The density range of each cosmic morphology environment in L025 is slightly wider than L050 and L100, while the probability peak is shallower. As we have argued in the earlier paragraph, the higher resolution in L025 is helpful to develop more refined structures and sharper dense contrast in the nonlinear regimes, i.e., resolving higher density peak and lower density floor. 

\subsection{Comparison with C14}

\begin{figure*}[tbp]
\includegraphics[width=0.50\textwidth]{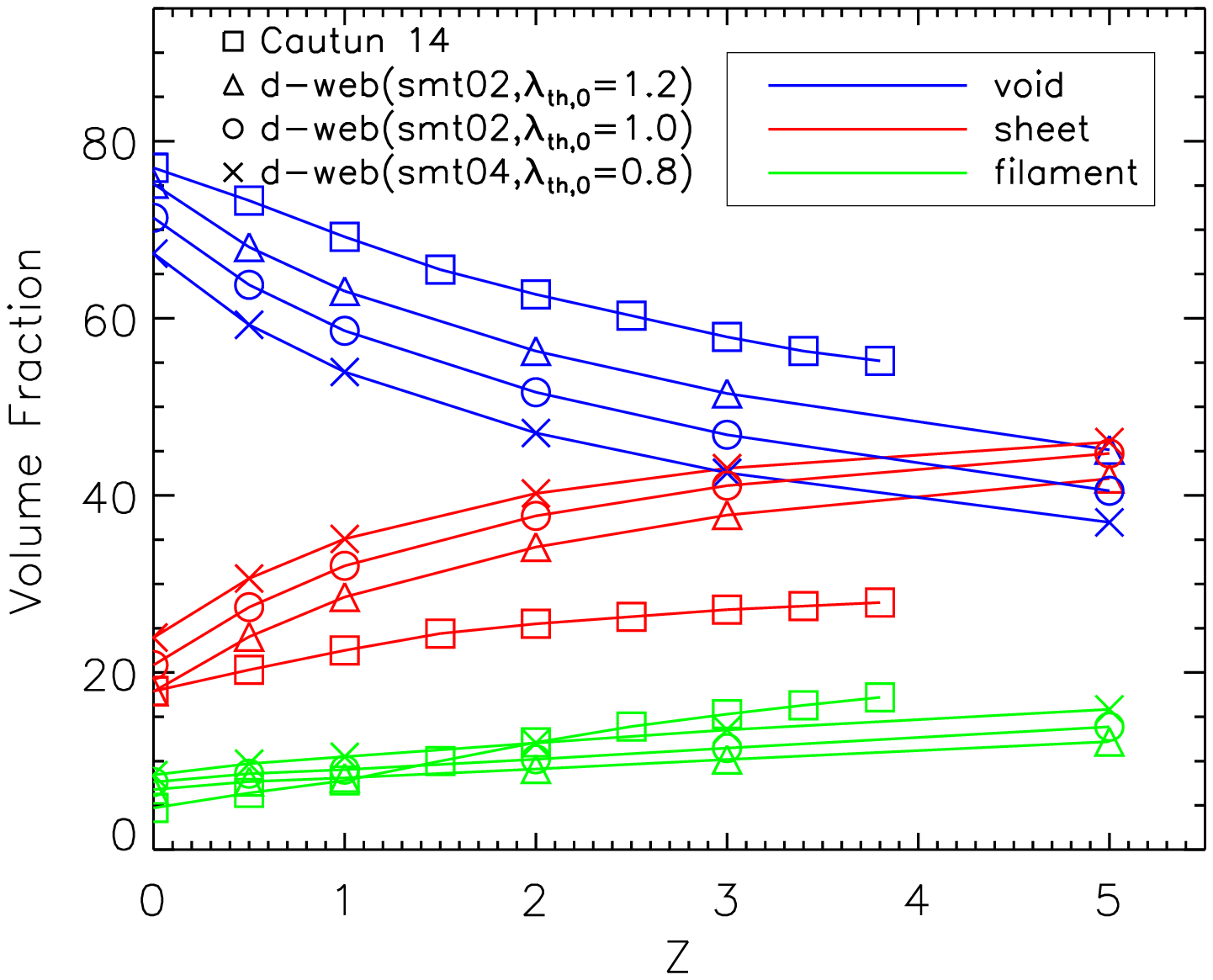}
\includegraphics[width=0.50\textwidth]{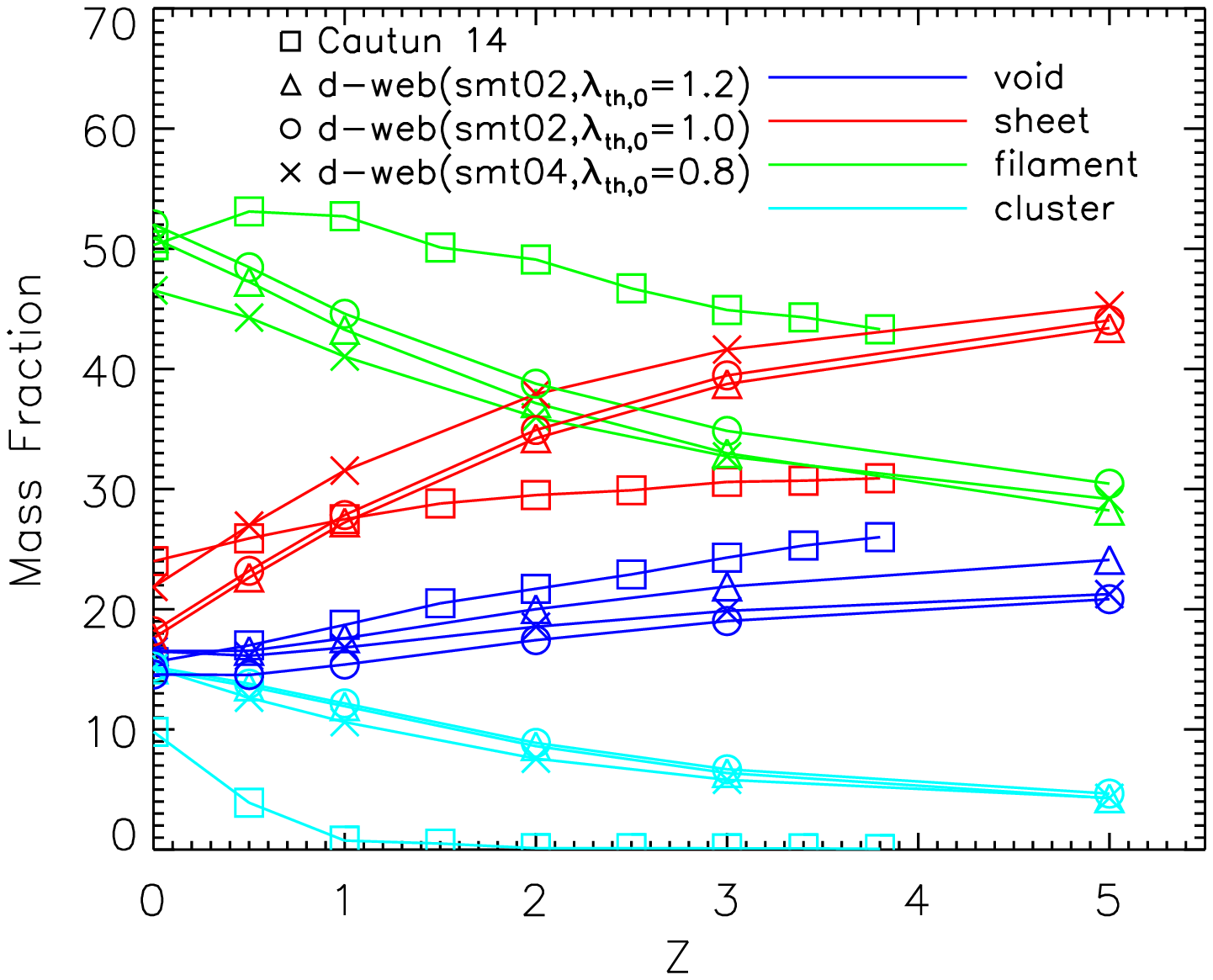}
\includegraphics[width=0.50\textwidth]{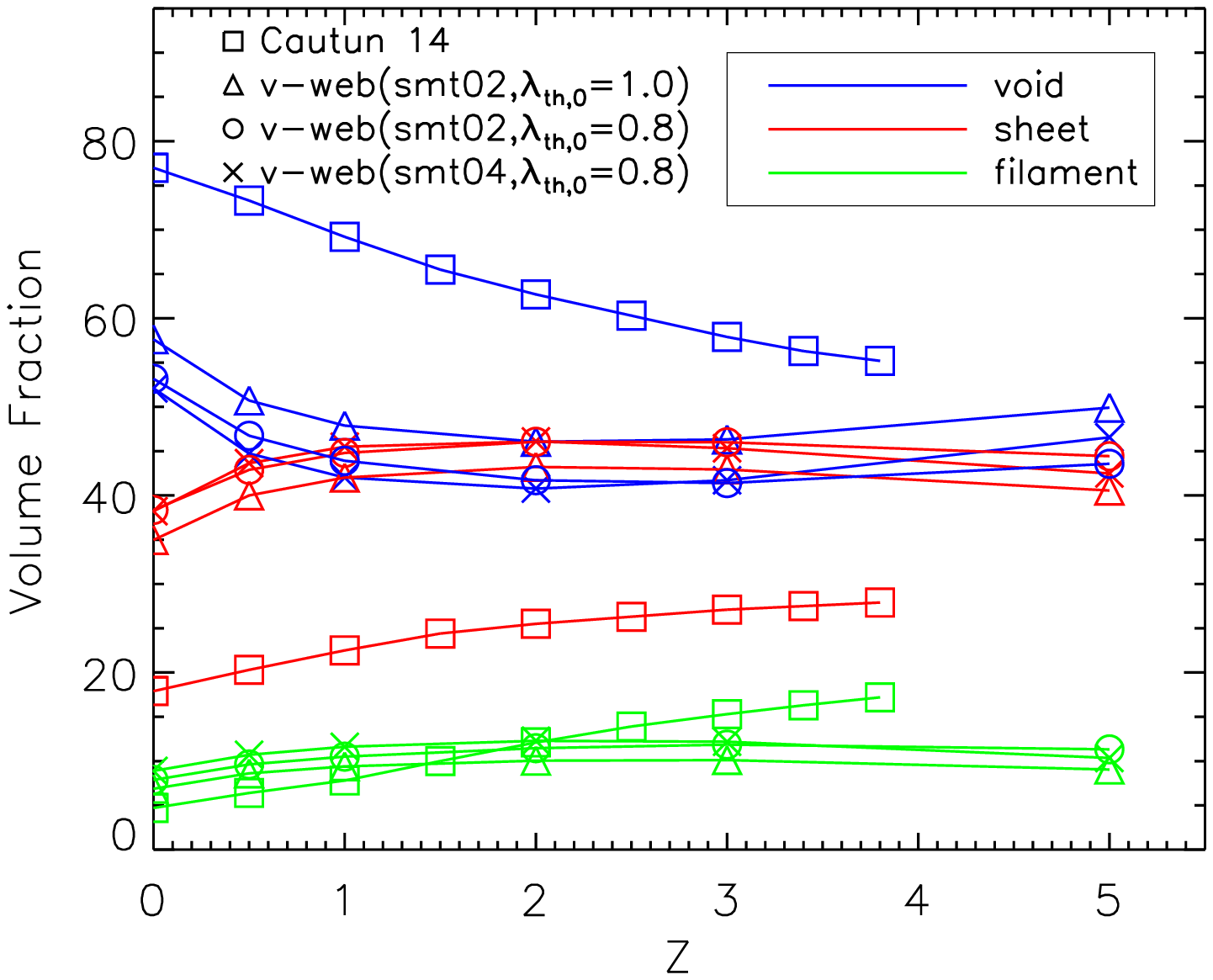}
\includegraphics[width=0.50\textwidth]{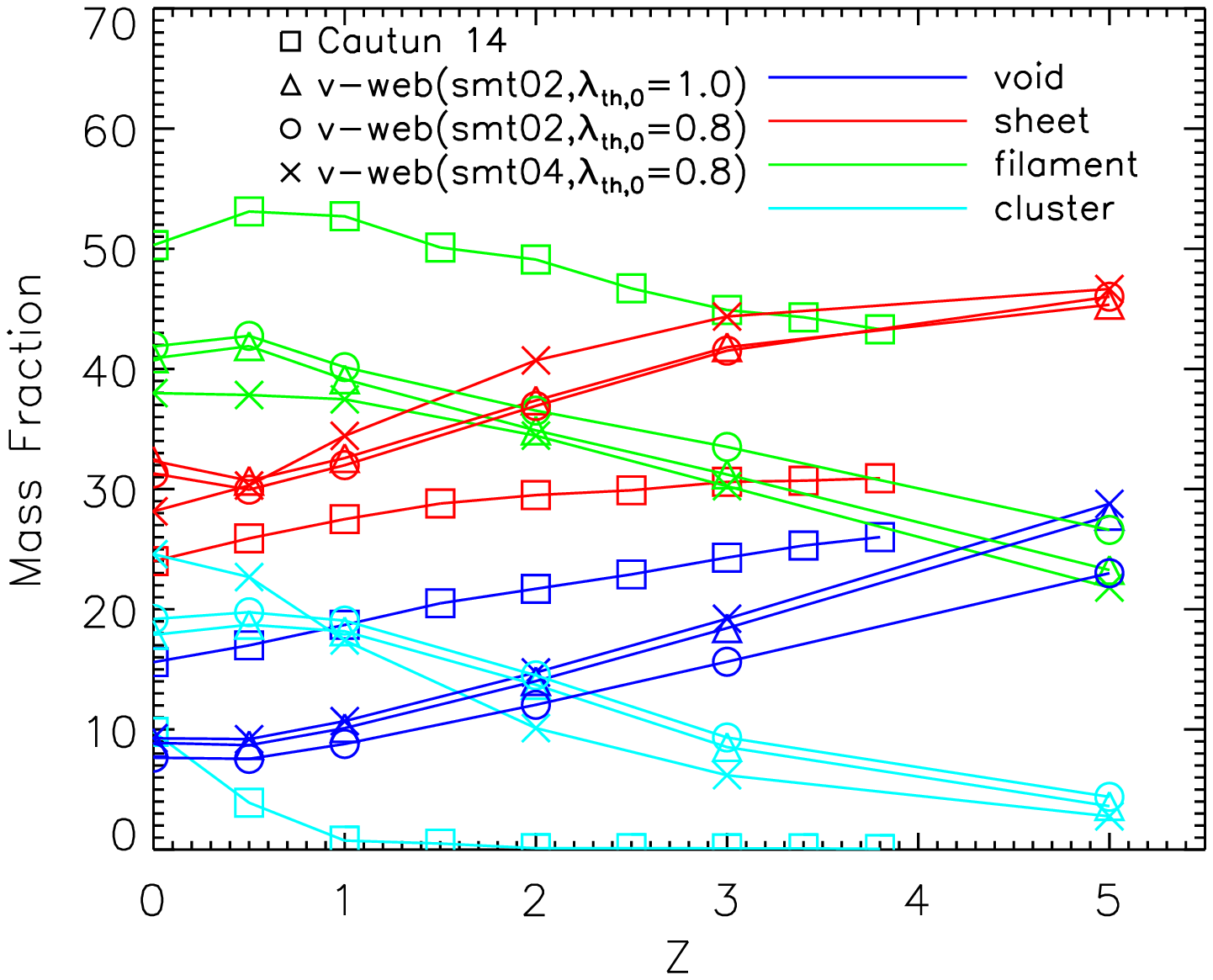}
\caption{The redshift evolution of volume and mass fraction in L025 with time dependent $\lambda_{th}(z)=\lambda_{th,0}/(1.+z)$ for \textbf{d-web}(Top), and \textbf{v-web}(Bottom).}
\label{figure9}
\end{figure*}
The mass distribution of dark matter in the cosmic web at $z=0$ revealed by the \textbf{d-web} and \textbf{v-web} in this work is basically in agreement with the results obtained from dark matter only simulation in C14. The characteristic density in each environment presented here(see Figure 8) is also consistent with C14. However, our results of the volume and mass fractions, and their evolution in voids and filaments at $z>0$ show different behaviors in comparison with C14 in several aspects. Firstly, in contrast to C14, the volume fraction in voids/sheets is decreasing/increasing with time in our simulation samples. This is partly because the \textbf{d-web} (also \textbf{v-web}) method tends to identify regions with under-developed density and velocity fluctuations as voids. Namely, none of the three eigenvalues is larger than $\lambda_{th}$, due to lack of significant collapsing activity. For a constant threshold value, the volume and mass fraction in voids would be higher while going to higher redshift. Secondly, the mass fraction in filaments changes slowly toward $z=3.8$ in C14, which is different from our results significantly, even in L100 that has a comparable box size. A larger/smaller share of mass has been found in voids/filaments in our samples, becoming more apparently as the redshift increasing. The margin between filaments and sheets in terms of mass fraction at $z>0.5$ is decreasing as going to higher redshift in C14, however in a much slower pace comparing to our results. The epoch of sheets surpassing filaments in C14 is expected to be much earlier than $z \sim 4$. In addition, the probability distribution functions in our identified web are more widespread.

\begin{figure*}[tbp]
\vspace{-0.8cm}
\includegraphics[width=0.50\textwidth]{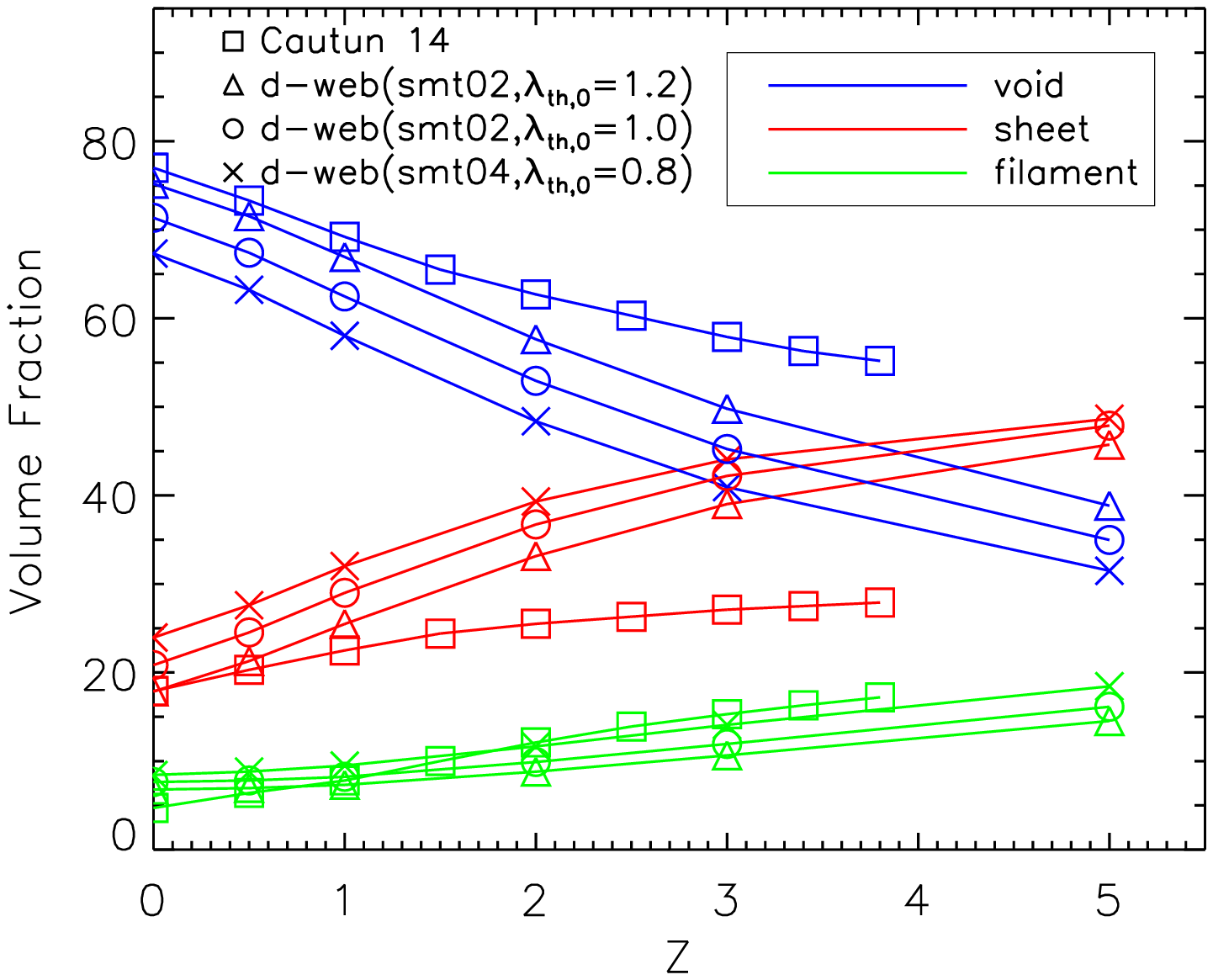}
\includegraphics[width=0.50\textwidth]{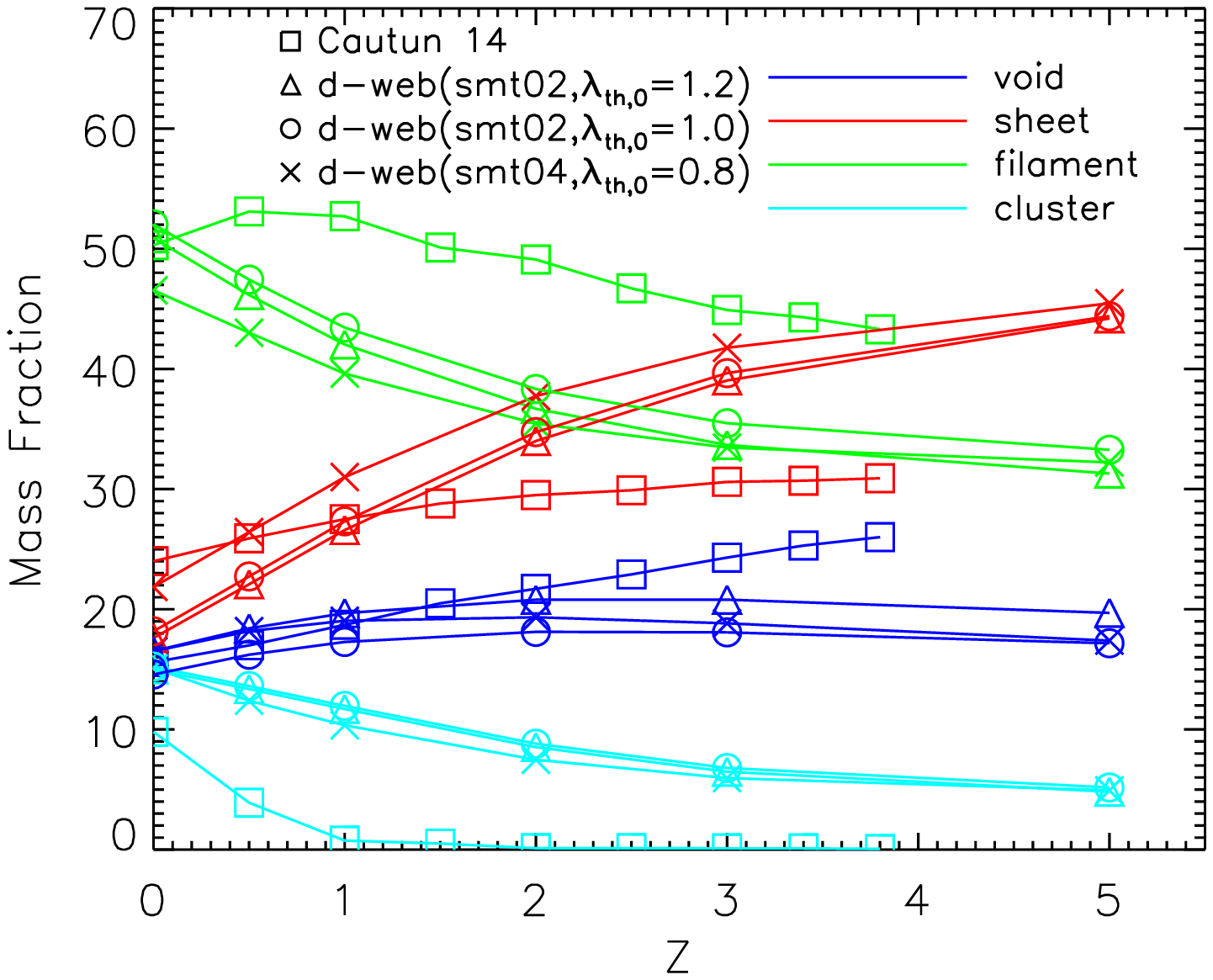}
\caption{The redshift evolution of volume and mass fraction in L025 with time dependent $\lambda_{th}(z)=\lambda_{th,0}*H_0/H(z)$ for \textbf{d-web}.}
\label{figure10}
\end{figure*}

The key factor leading to these differences is the cosmic web identification method. The NEXUS+ scheme used in C14 is a multi-scale morphological analysis tool implemented with filtering of the density field in logarithmic space, and is different from the \textbf{d-web} and \textbf{v-web} with single-scale filter. Cautun et al.(2013) shown that the NEXUS+ scheme was successful in capturing filaments and walls with a wide range of sizes. C14 further showed that the mass fraction is insensitive to the identification tracer, i.e., either density field or tidal tensor, as well as velocity divergence etc, in their NEXUS scheme.  However, it has been justified only at z=0, and its validity is not clear while going to high redshifts.  On the other hand, it needs more insightful investigations to set a time dependent $\lambda_{th}$ that is physically meaningful for the \textbf{d-web} and \textbf{v-web} schemes. Though the \textbf{d-web} with a fixed $\lambda_{th}$ could identify both filaments and walls with different sizes at low and high redshifts in our simulation samples, it would be interesting to implement a time-varying $\lambda_{th}$ in our scheme and justify whether it is able to reduce the difference between our statistics and C14.

Figure 9 shows the evolution of volume and mass fraction of dark matter in L025 with $\lambda_{th}(z)=\lambda_{th}(0)/(1+z)$. Generally, the \textbf{d-web} provides a better match than \textbf{v-web}. Taking a fixed threshold value of $\lambda_{th}(0)=1.2$ with $R_s=2R_g$ in the \textbf{d-web} scheme, the volume fractions of all the structures, and the mass fraction of filaments and voids at $z=0$ are found to be almost identical to C14. The mass fractions in sheets and clusters, however, still shows differences about $5\%$ while comparing with C14. At $z>0$, once a time dependent $\lambda_{th}(z)$ is applied, the consistency of overall evolution trend in both mass or volume fractions between \textbf{d-web} and C14 can be somewhat improved, whereas there still exist distinct differences in absolute value of volume and mass fractions. The \textbf{d-web} scheme predicts relatively higher mass fractions in sheets and clusters, but lower fractions in filaments at high redshits, in contrast to C14.

The remaining discrepancy can hardly be reduced by adjusting either $\lambda_{th,0}$ or smooth length. The numerical results with an alternative monotonic $\lambda_{th}(z)$, $\lambda_{th}(0)*H_0/H(z)$, are given in Figure 10 and showing mild changes. The differences in volume and mass fractions between \textbf{d-web} and C14 are apparent, especially at high redshifts. It should result from the different kernels we have used in identification of cosmic web components. In addition, the \textbf{d-web} scheme with time varying $\lambda_{th}$ predicts also a turnover redshift of $z\sim 2-3$, at which the universe goes through a transition from sheet-dominated to filaments dominated, in agreement with results presented in the last sub-section where fixed $\lambda_{th}$ has been used.

\section{The Evolution of Velocity Field in the Cosmic Web}

The velocity fields of baryonic and cold dark matter record the anisotropic collapse of the structures and the mass transportation among different components in the cosmic web. The flow of cosmic matter is almost curl-free in the linear regime, during which initial vorticity would be diluted by the cosmic expansion. The anisotropic collapse that drives the growth of clustering structures will lead to the appearance of multi-flow regions and hence vorticity in the quasi-linear regimes. The curl velocity would be amplified by further anisotropic collapsing and  become comparable to the irrotational mode in highly nonlinear regimes. The evolution of the two modes of motions, i.e., irrational and curl, are expected to indicate the variation of density, volume and mass content presented in the last section. In order to reveal this connection, we investigate the visual and statistical properties of the flows in this section. 

\subsection{The Divergence and Vorticity Fields}

\begin{figure*}[tbp]
\vspace{-1.5cm}
\gridline{
\hspace{-0.6cm}
\includegraphics[width=0.60\textwidth]{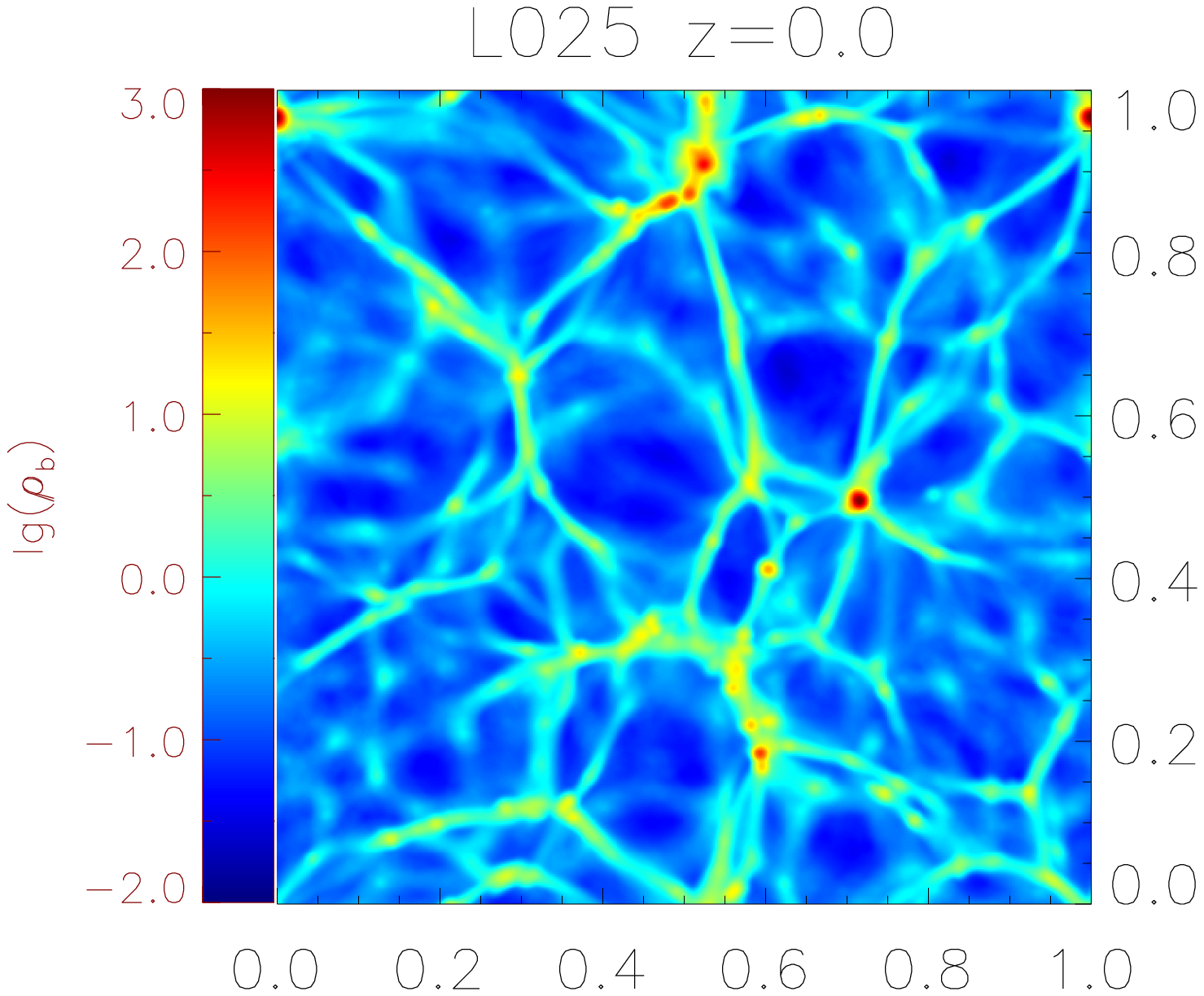}
\hspace{-2.4cm}
\includegraphics[width=0.60\textwidth]{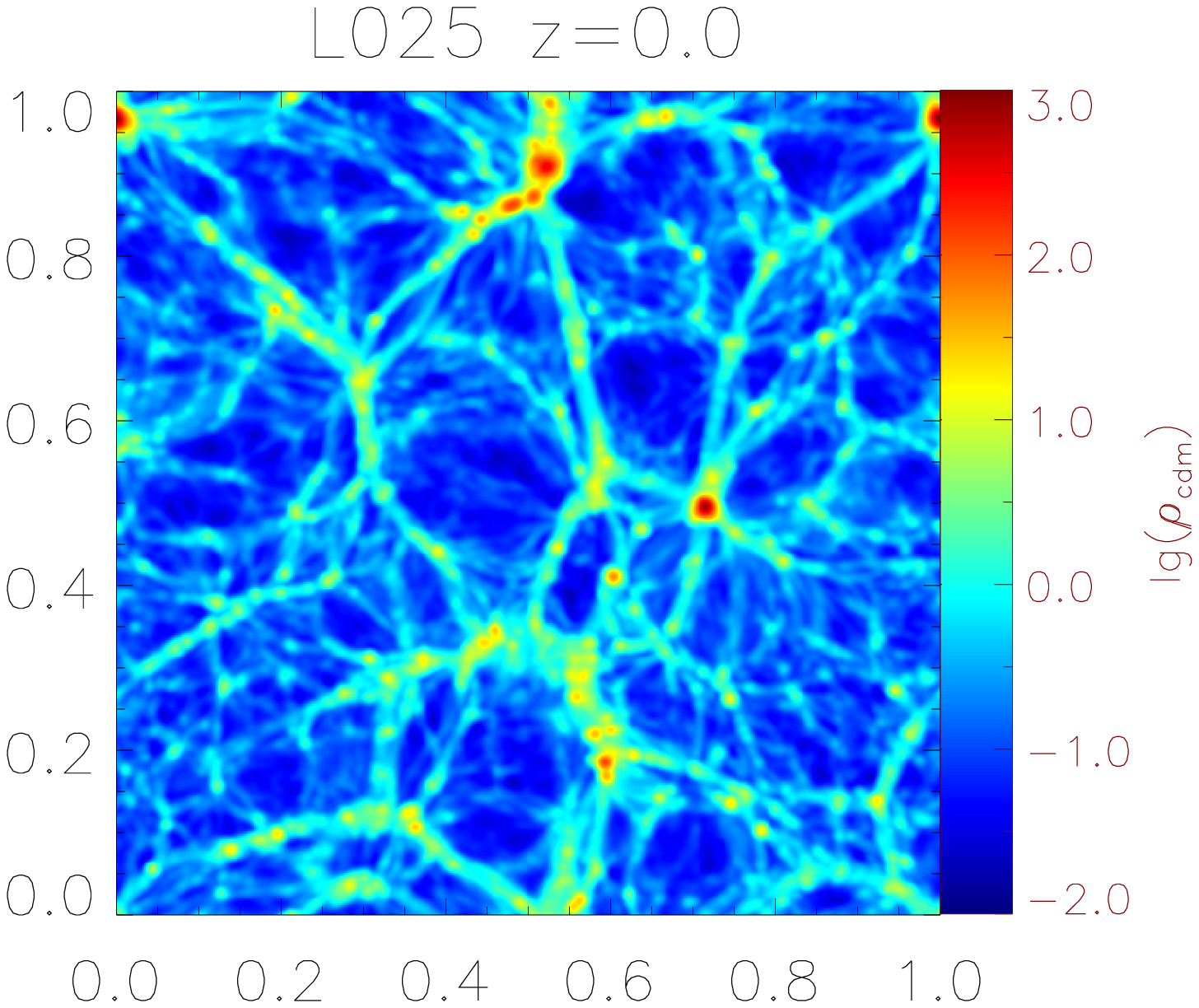}
}
\vspace{-1.0cm}
\gridline{
\hspace{-0.6cm}
\includegraphics[width=0.60\textwidth]{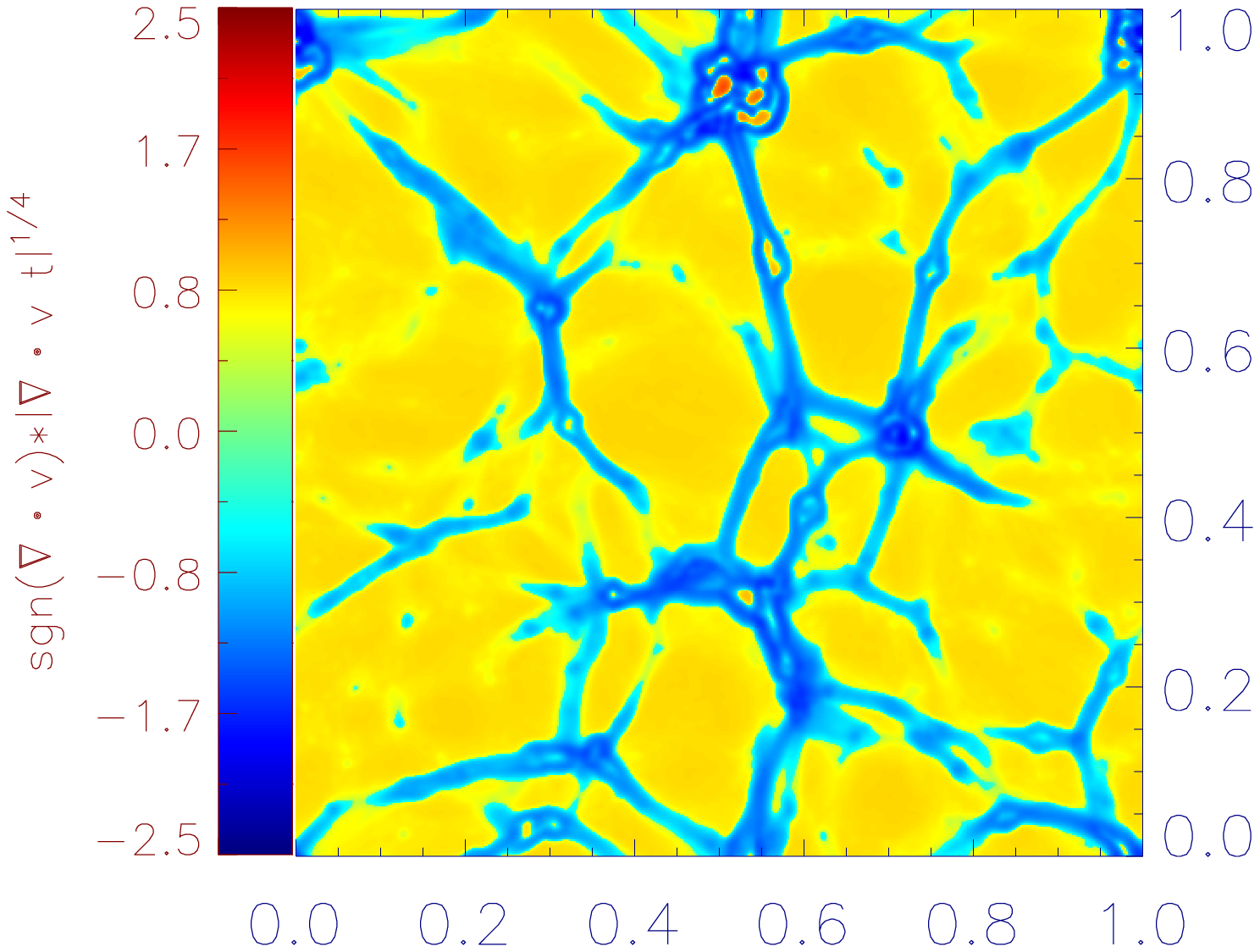}
\hspace{-2.4cm}
\includegraphics[width=0.60\textwidth]{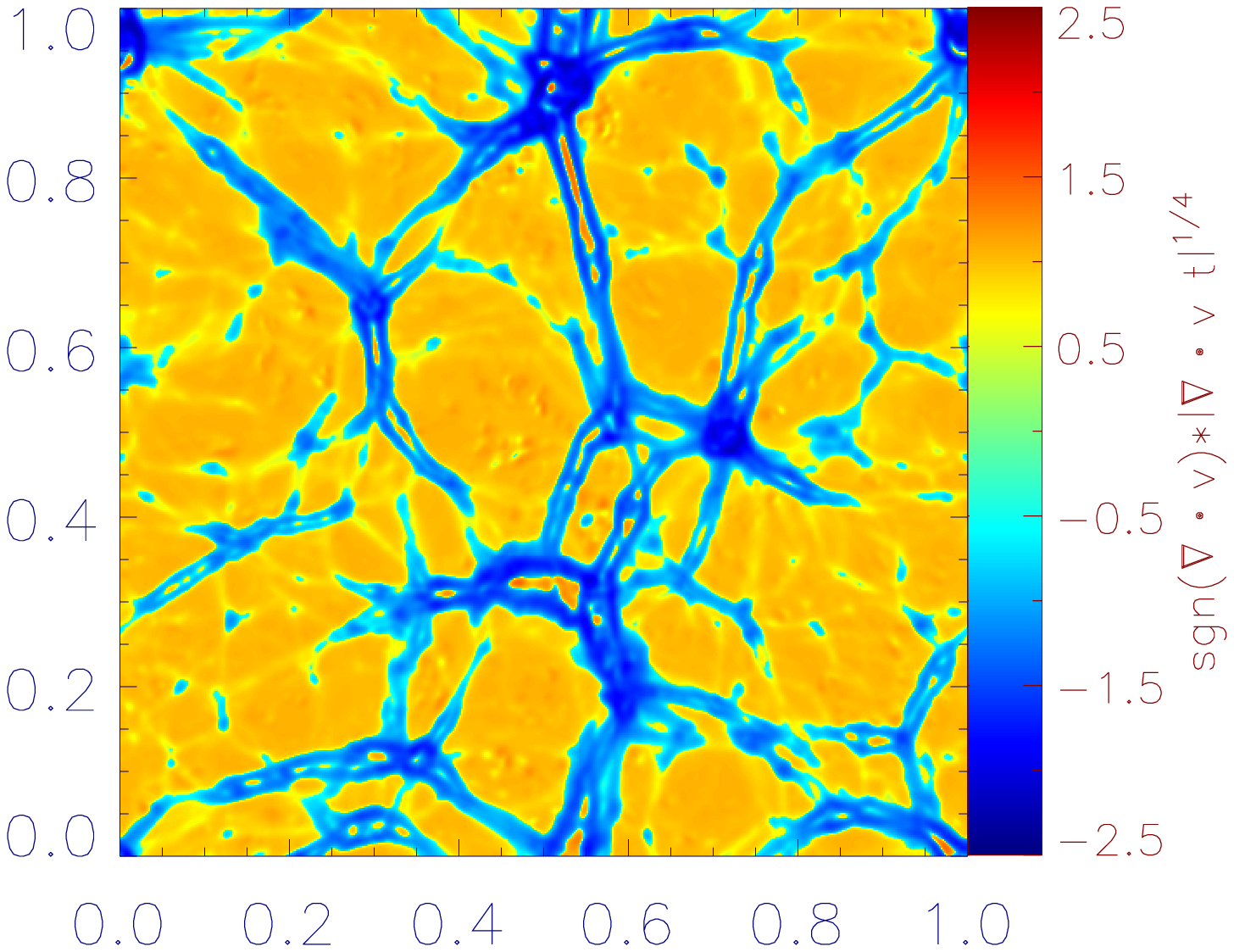}
}
\vspace{-1.0cm}
\gridline{
\hspace{-0.6cm}
\includegraphics[width=0.60\textwidth]{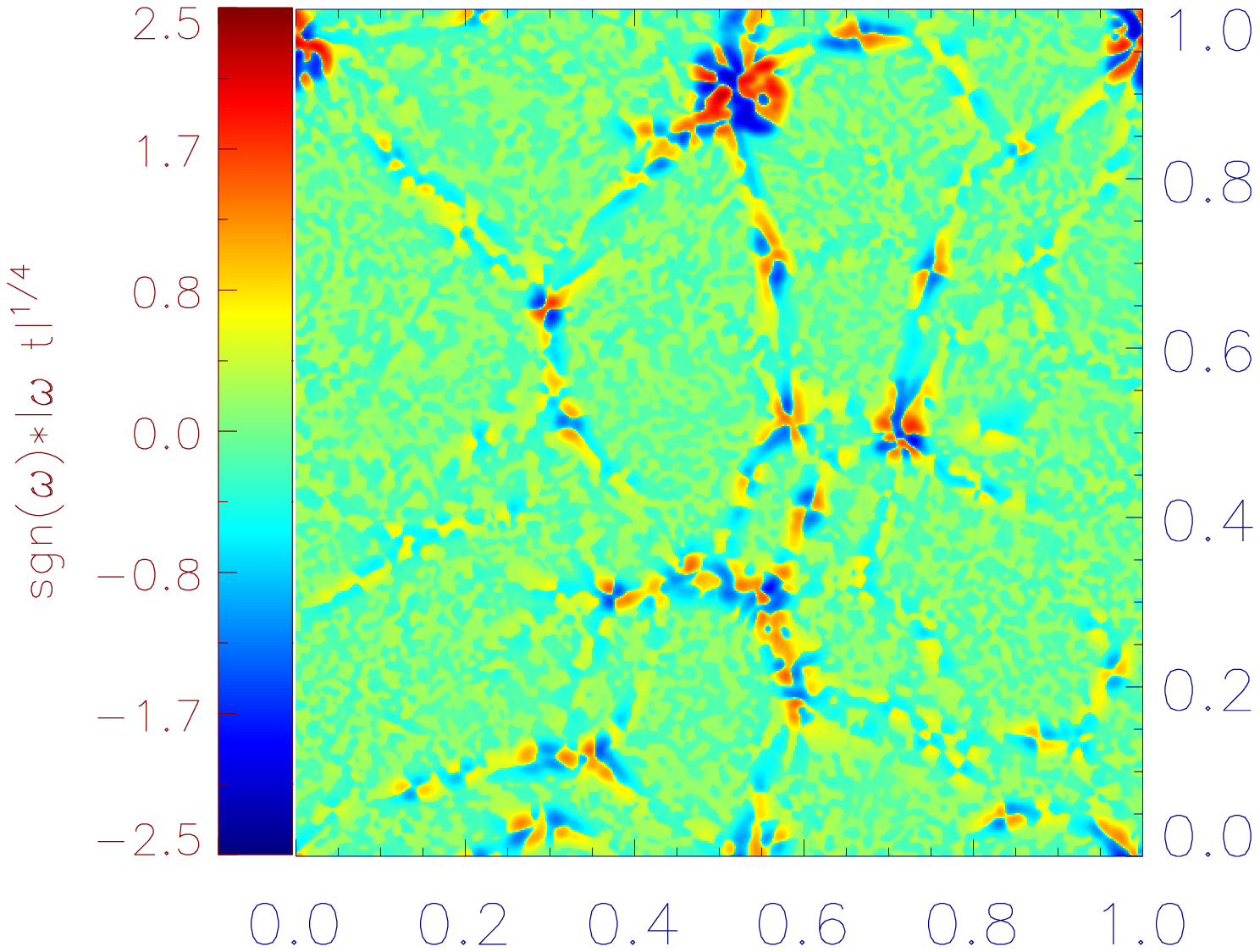}
\hspace{-2.4cm}
\includegraphics[width=0.60\textwidth]{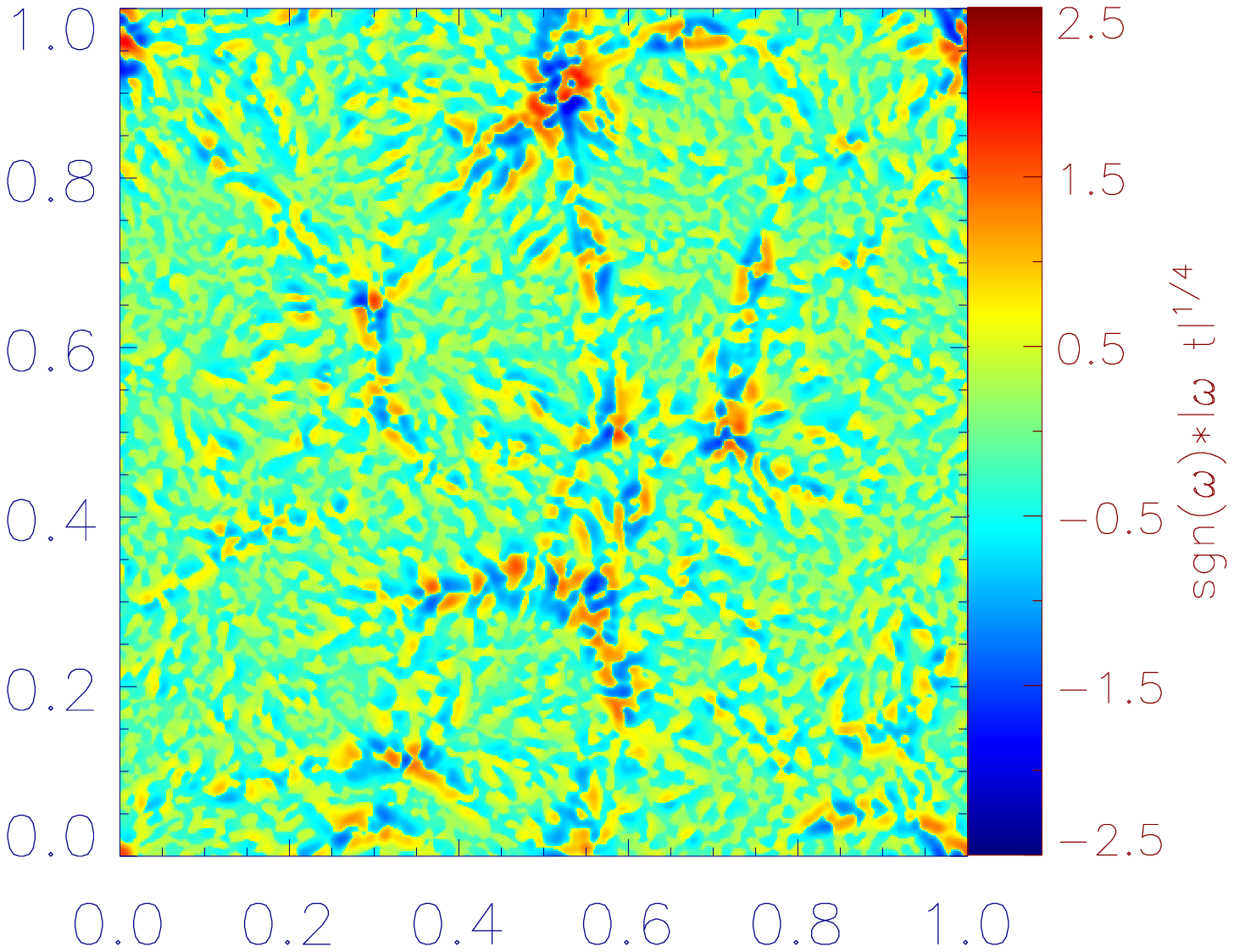}
}
\caption{The density(Top row), divergence(Middle row), and projected vorticity(Bottom row) of baryonic(Left column) and dark(Right column) matter in a slice in L025. Divergence and projected vorticity are rescaled to enhance the contrast. Red/blue color in the rescaled vorticity is towards/away from the observer.}
\label{figure11}
\end{figure*}

\begin{figure*}[tbp]
\vspace{-0.5cm}
\gridline{
\hspace{-0.7cm}
\includegraphics[width=0.40\textwidth]{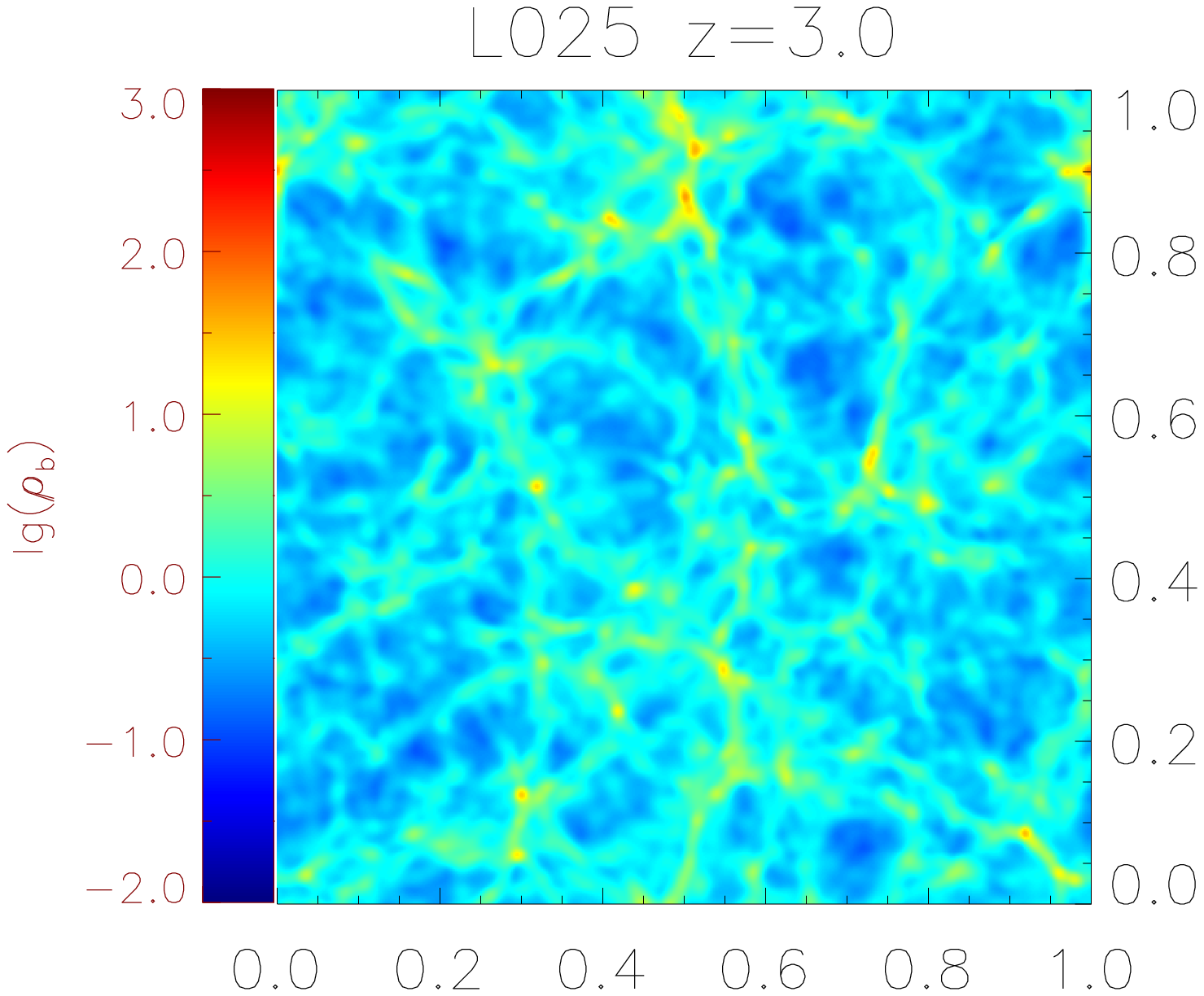}
\hspace{-0.7cm}
\includegraphics[width=0.40\textwidth]{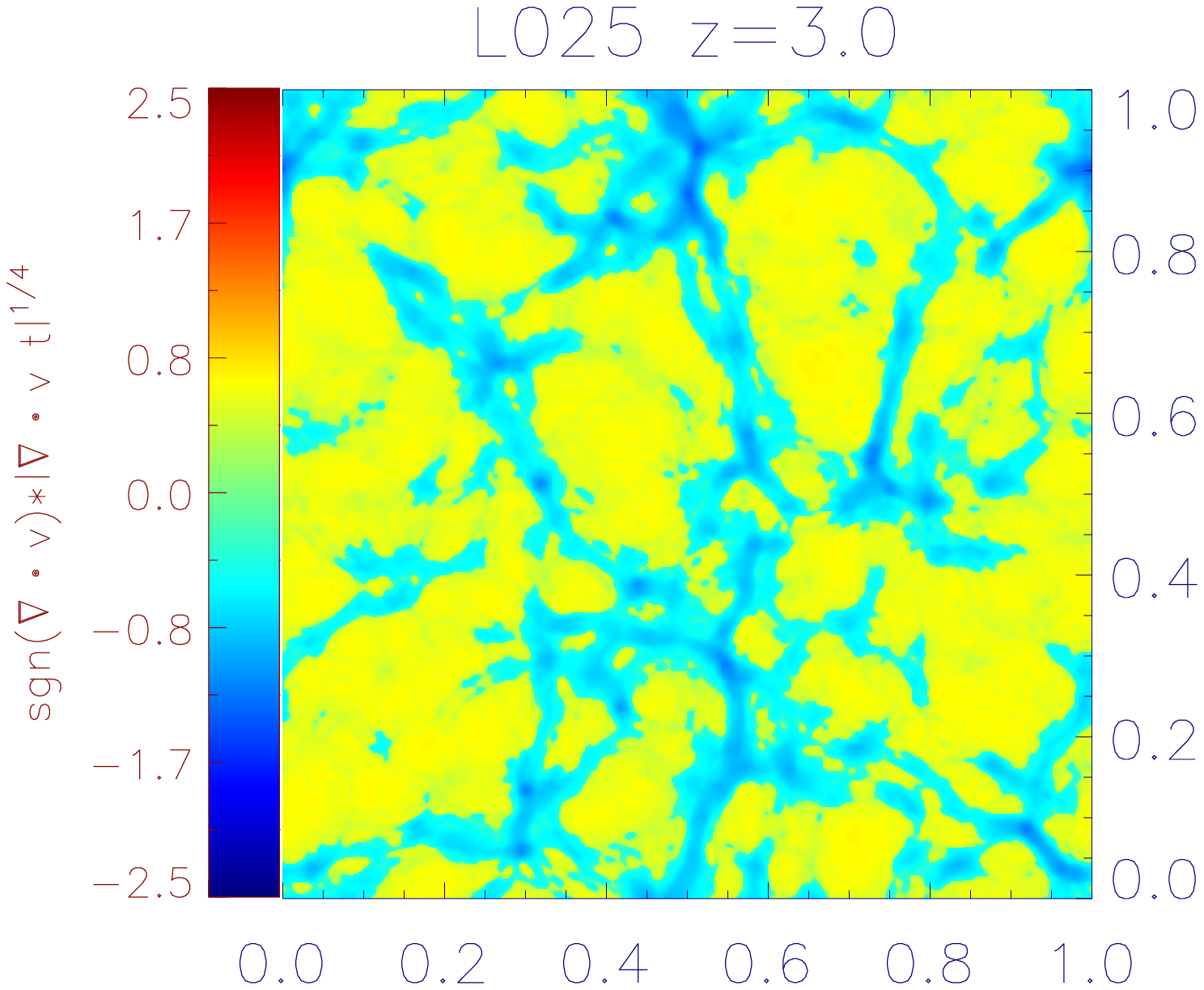}
\hspace{-1.0cm}
\includegraphics[width=0.40\textwidth]{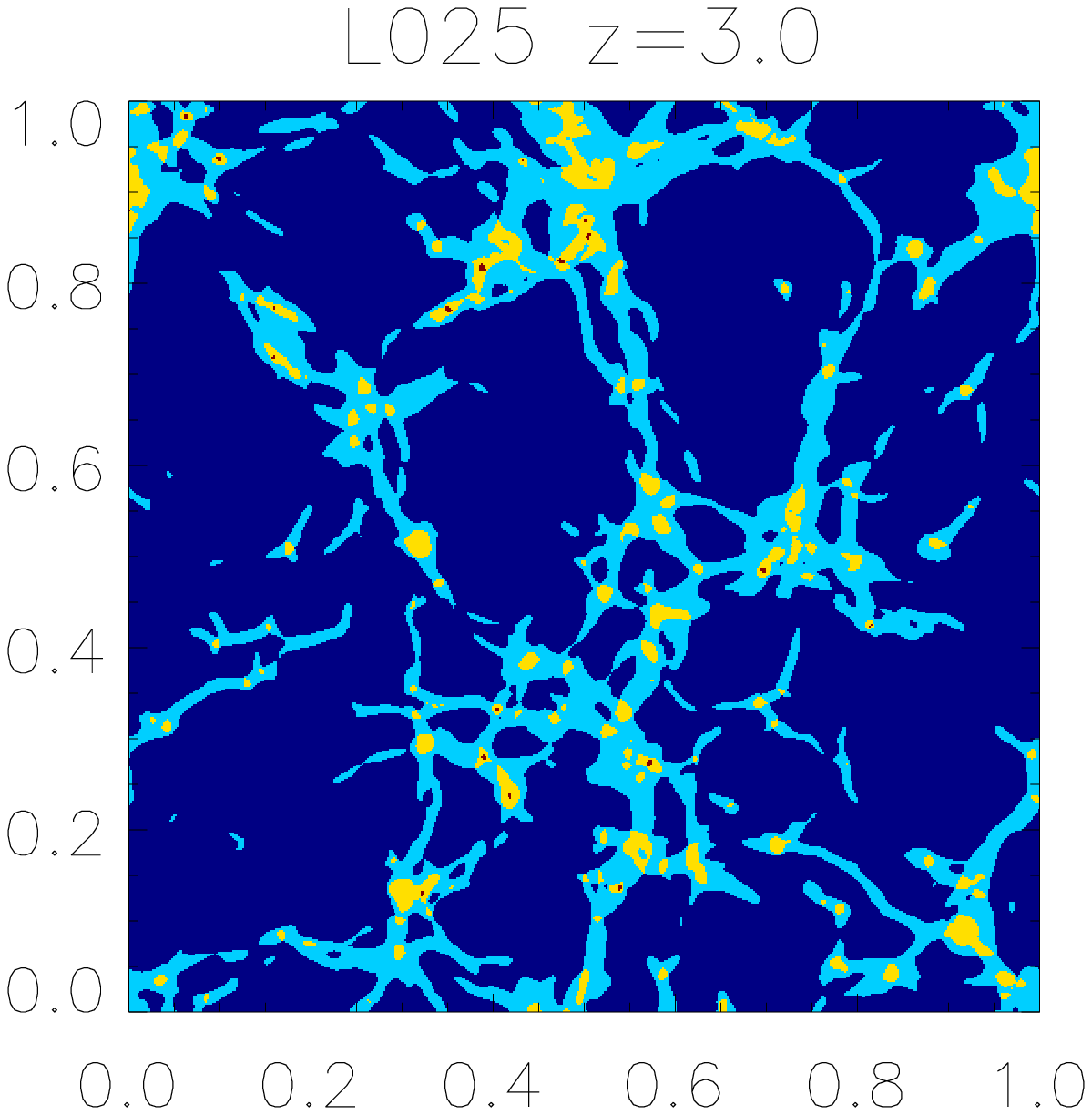}
}
\vspace{-0.2cm}
\gridline{
\hspace{-0.7cm}
\includegraphics[width=0.40\textwidth]{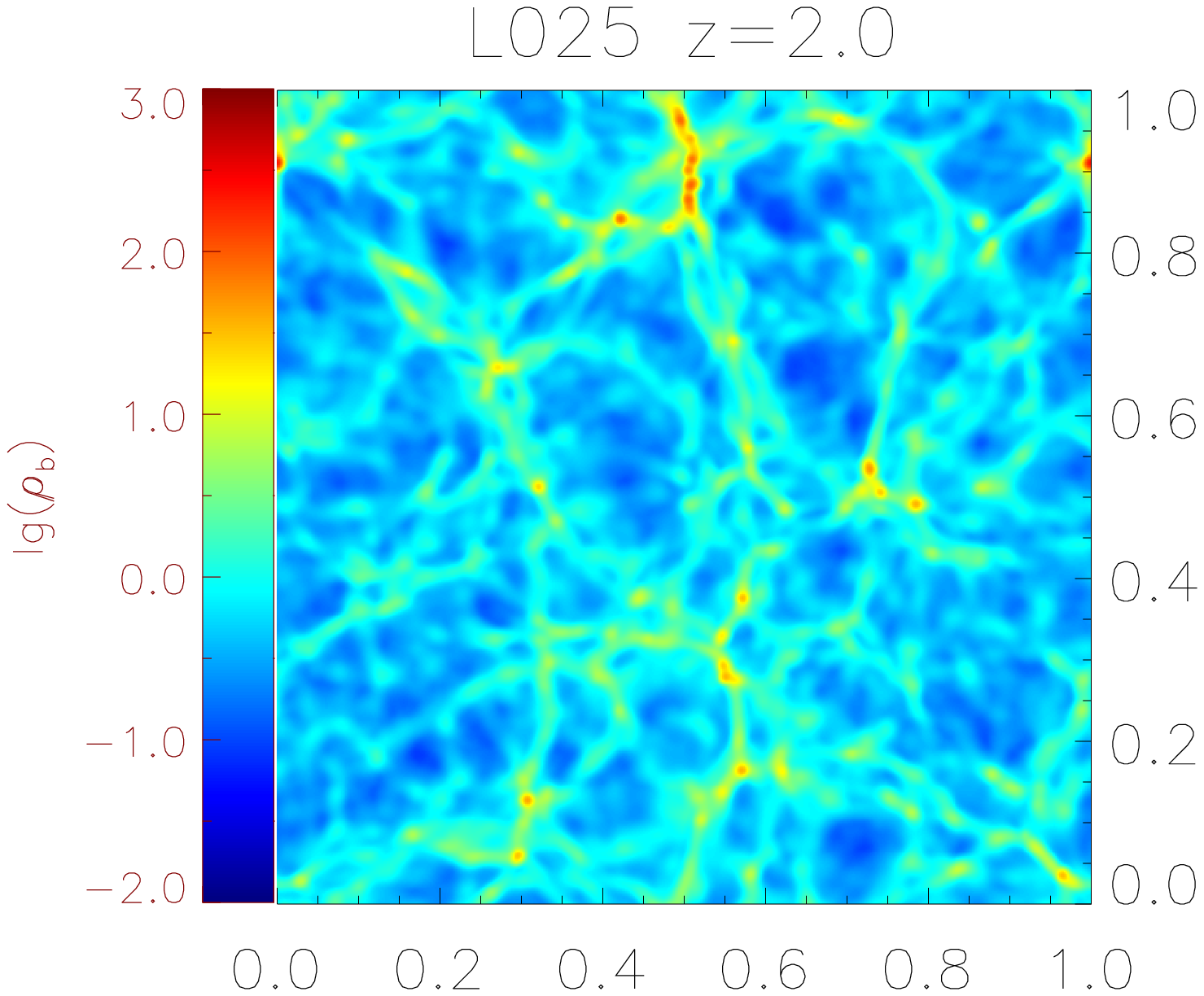}
\hspace{-0.7cm}
\includegraphics[width=0.40\textwidth]{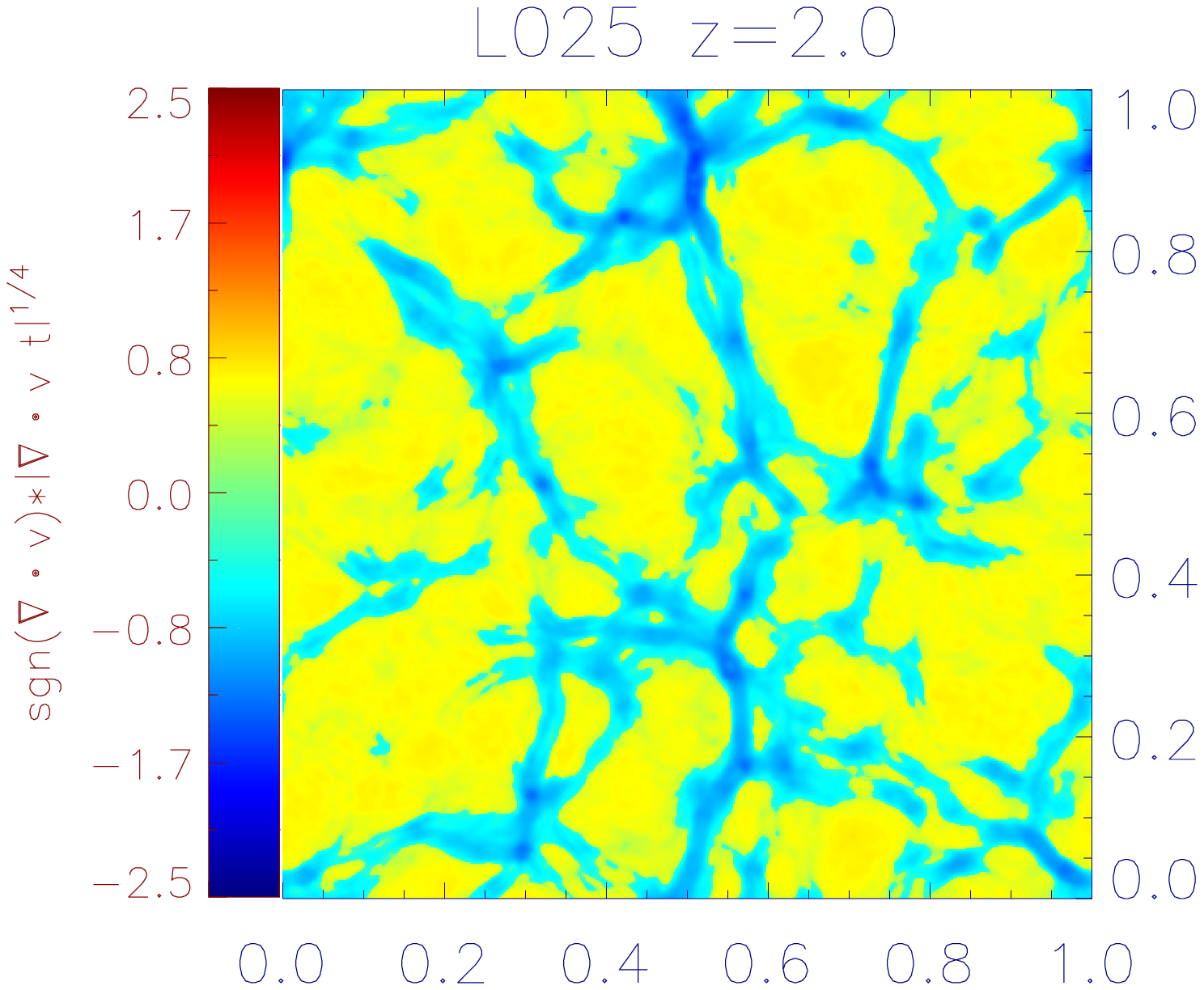}
\hspace{-1.0cm}
\includegraphics[width=0.40\textwidth]{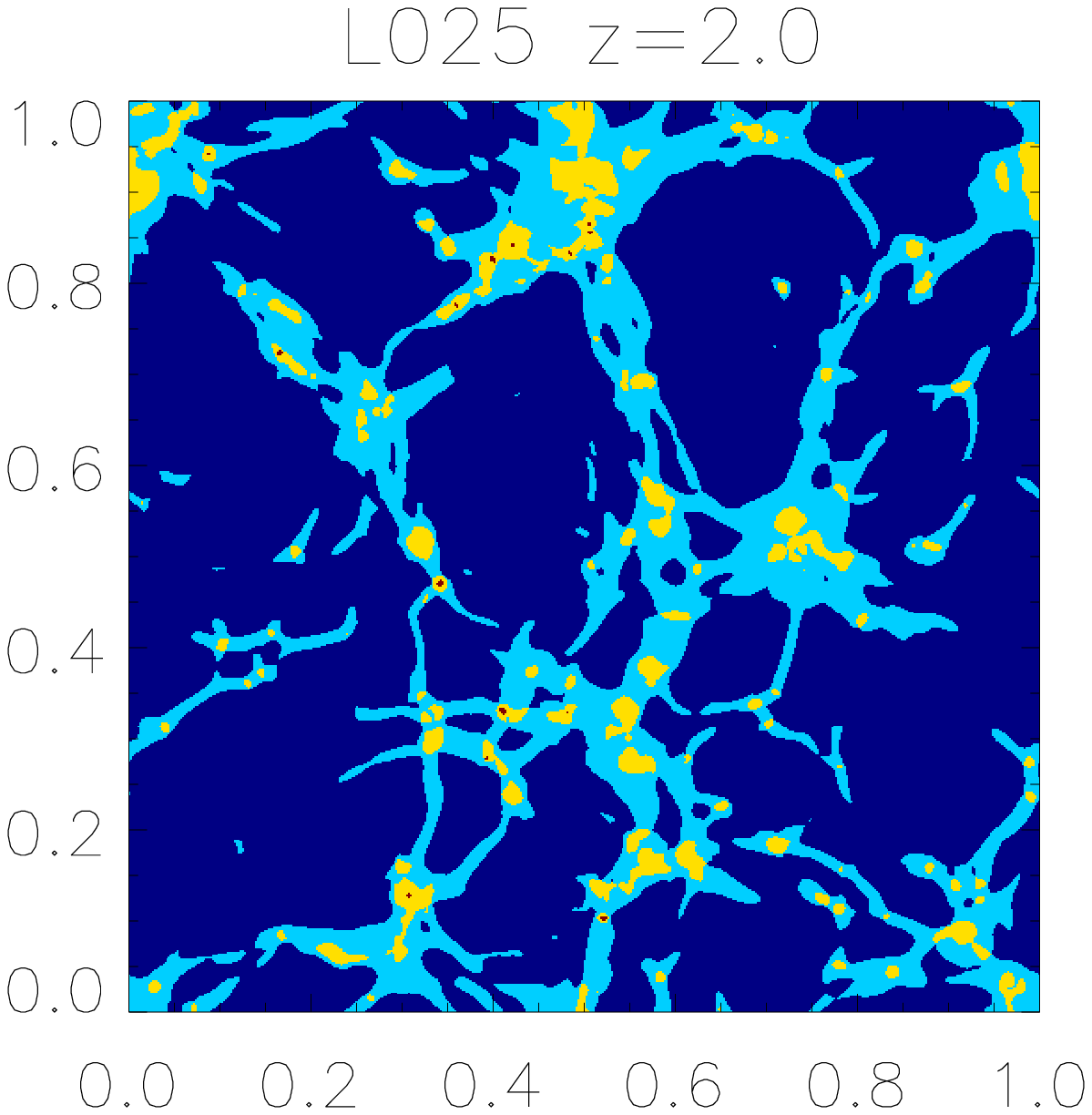}
}
\vspace{-0.1cm}
\gridline{
\hspace{-0.7cm}
\includegraphics[width=0.40\textwidth]{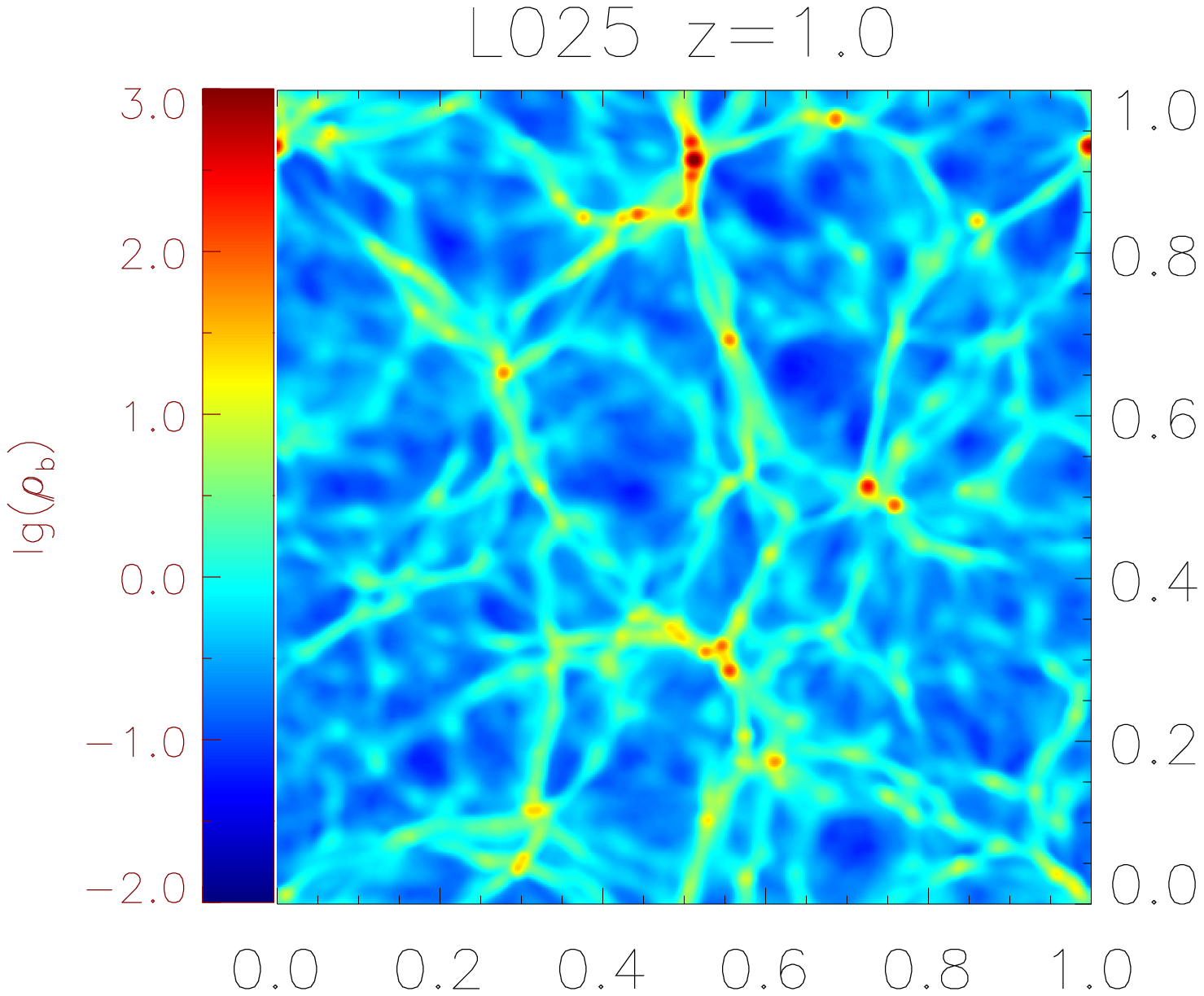}
\hspace{-0.7cm}
\includegraphics[width=0.40\textwidth]{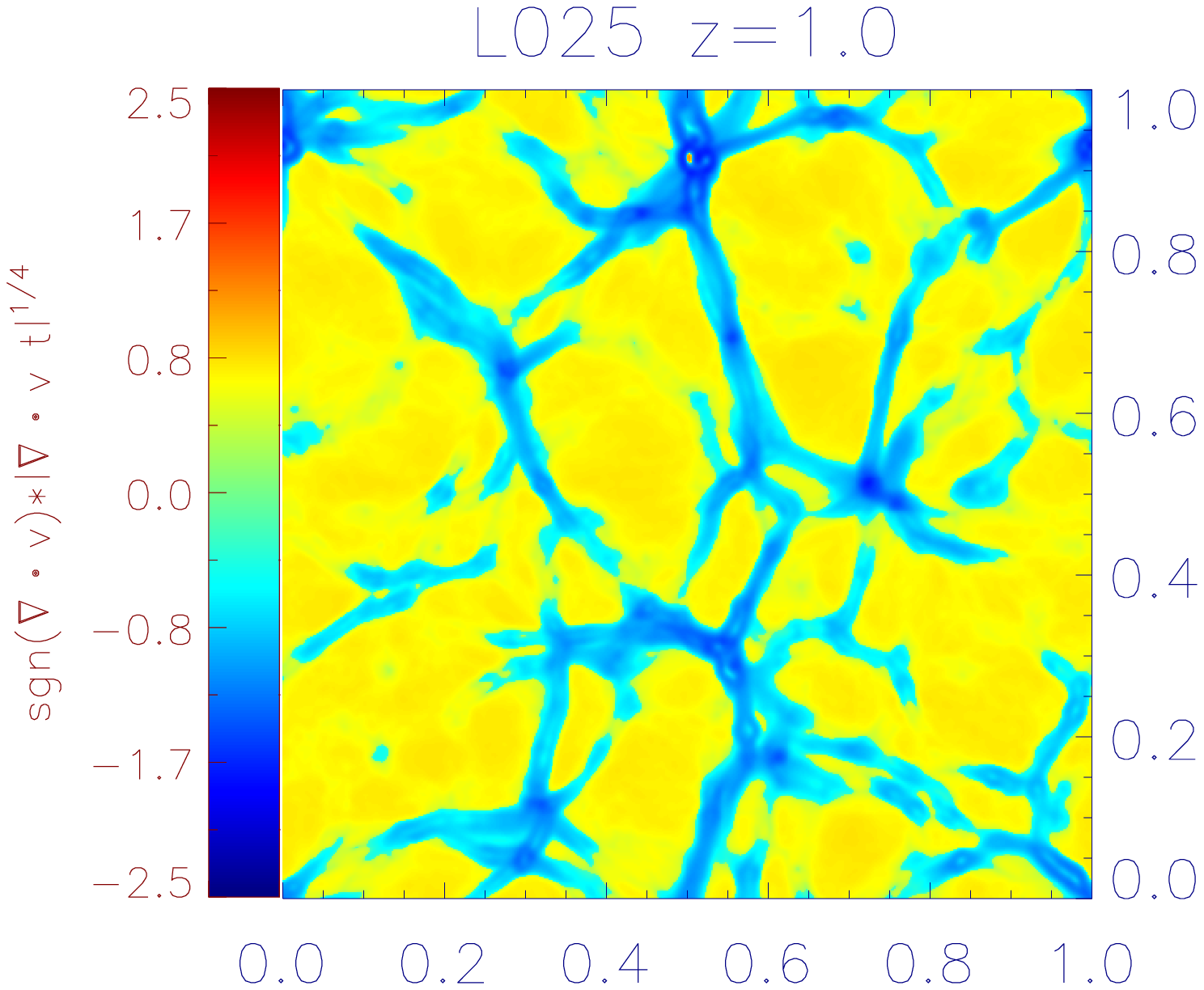}
\hspace{-1.0cm}
\includegraphics[width=0.40\textwidth]{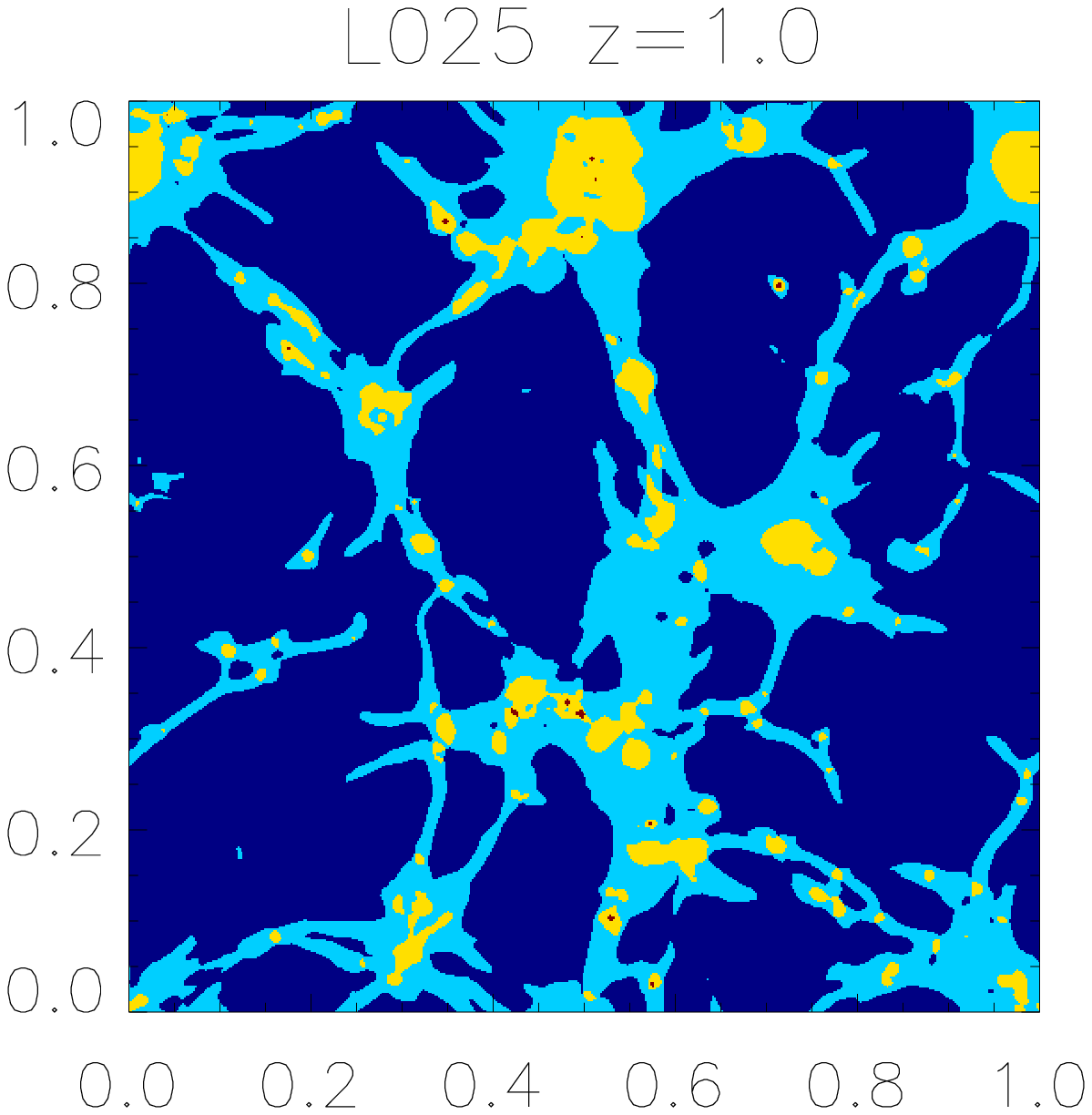}
}
\hspace{0.2cm}
\caption{The evolution of the density(Left column), rescaled divergence(Middle column) and web environment(Right column, dark blue, light blue, yellow and red color stand for voids, sheets, filaments and clusters respectively) of baryonic matter in the same slice of Figure 11 from $z=3.0$ to $z=1.0$.}
\label{figure12}
\end{figure*}

Before studying the properties of velocity in the cosmic web, we firstly take a look at the divergence and vorticity field of baryonic and dark matter. Figure 11 displays the spatial distributions of divergence and projected vorticity \textbf{${\omega}$} with $R_s=2R_g$ in a slice at $z=0$, both have been rescaled to enhance the contrast. Specifically, $sgn(\omega) (|\omega| t)^{1/4}$, and $sgn(\nabla \cdot \vec{v}) (|\nabla \cdot \vec{v}| \ t)^{1/4}$ are plotted, where t is the cosmic time. The structure in the baryonic matter looks smoother than dark matter. For the divergence, it makes a transition from positive to negative usually at the boundaries between voids and over-dense structures, which enclose its minima (negative) in mildly over-dense regions. It is just what we expect from the simple physics, the collapsing process will slow down and even cease in the central region of collapsed structures, where the relaxing takes place. The vorticity is well developed within multi-streaming regions. Evident signs of quadripolar multi-streaming regions, which was firstly demonstrated by Laigle et al.(2015), can be found in the baryonic matter. Furthermore, more complex multi-streams can be located at the cross section of prominent structures. The pattern in dark matter is somewhat vague in the over dense region and shows relative more small scale structures. 

In order to manifest the growth of divergence in the evolutionary background of cosmic web, we show the distribution of density, divergence and web classification of baryonic matter in the same slice from $z=3$ to $z=1.0$ in Figure 12. Voids, sheets, filaments and clusters are represented by dark blue, light blue, yellow and red color respectively. The collapsed structures are mainly sheets and filaments in this slice. Numerous sheets with varies sizes have formed by $z=3$. At the same epoch, the associated divergence had been developed substantially. Many of the sheets at early time have disturbed boundaries as the three dimensional rendering has illustrated in Figure 3. Afterwards, small-size sheets would either consolidate with neighbours of similar scale or be absorbed by their large-size kin along with time. The right column of Figure 12 shows that the spacial distribution of sheets and voids on large scale has been set up by $z=3$ and experienced mildly change at later time, indicating that the outline of cosmic web may have formed at $z>3$ and confirming our interpretation of Figure 3. The divergence is gradually enhanced around the prominent sheets, while the configuration on large scale is slightly changed. We note that at $z=1.0$, part of the cells identified as sheets has positive divergence, about  $(\nabla \cdot \vec{v}) t < 0.1$. These cells are lying adjacent to prominent sheets and filaments, where the potential tidal field should have a complex structure. Cells with one eigenvalue large than $\lambda_{th}$ and two negative eigenvalue might be classified as sheets by \textbf{d-web}, i.e., collapsing in one dimension while expanding in the other two dimensions. The region belongs to filaments is much less than sheets at $z=3$, in good agreement with our volume and mass fraction statistics presented in the last section. Afterwards filaments keep emerging and growing, usually in the cross region of multi sheets. In comparison with Figure 11, the divergence is mildly evolved between $z=1$ and $z=0$ in sheets. Significant changes only take place in filaments having large cross section. 

\begin{figure}[tbp]
\vspace{-1.4cm}
\includegraphics[width=0.48\textwidth]{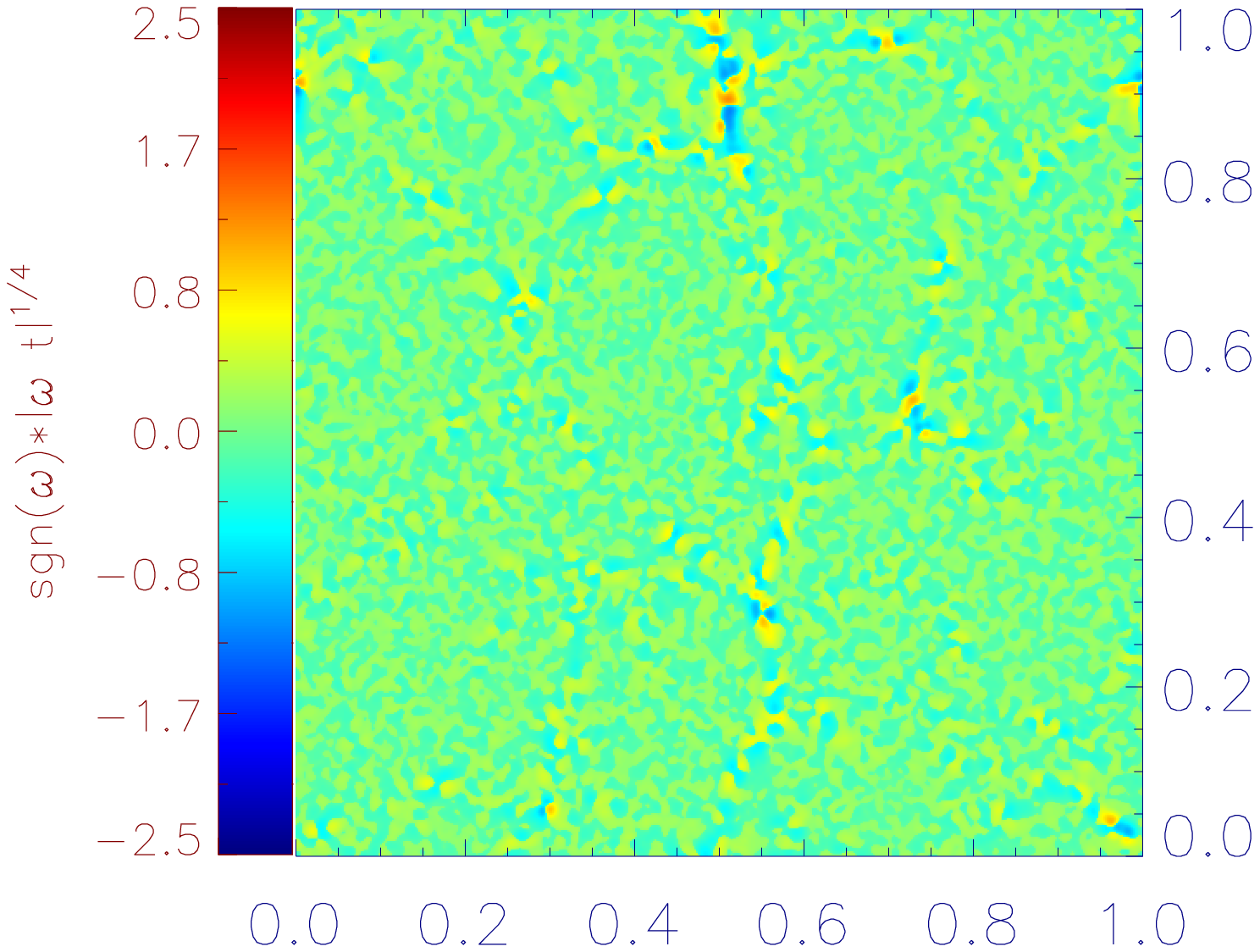}
\hspace{-0.5cm}
\includegraphics[width=0.48\textwidth]{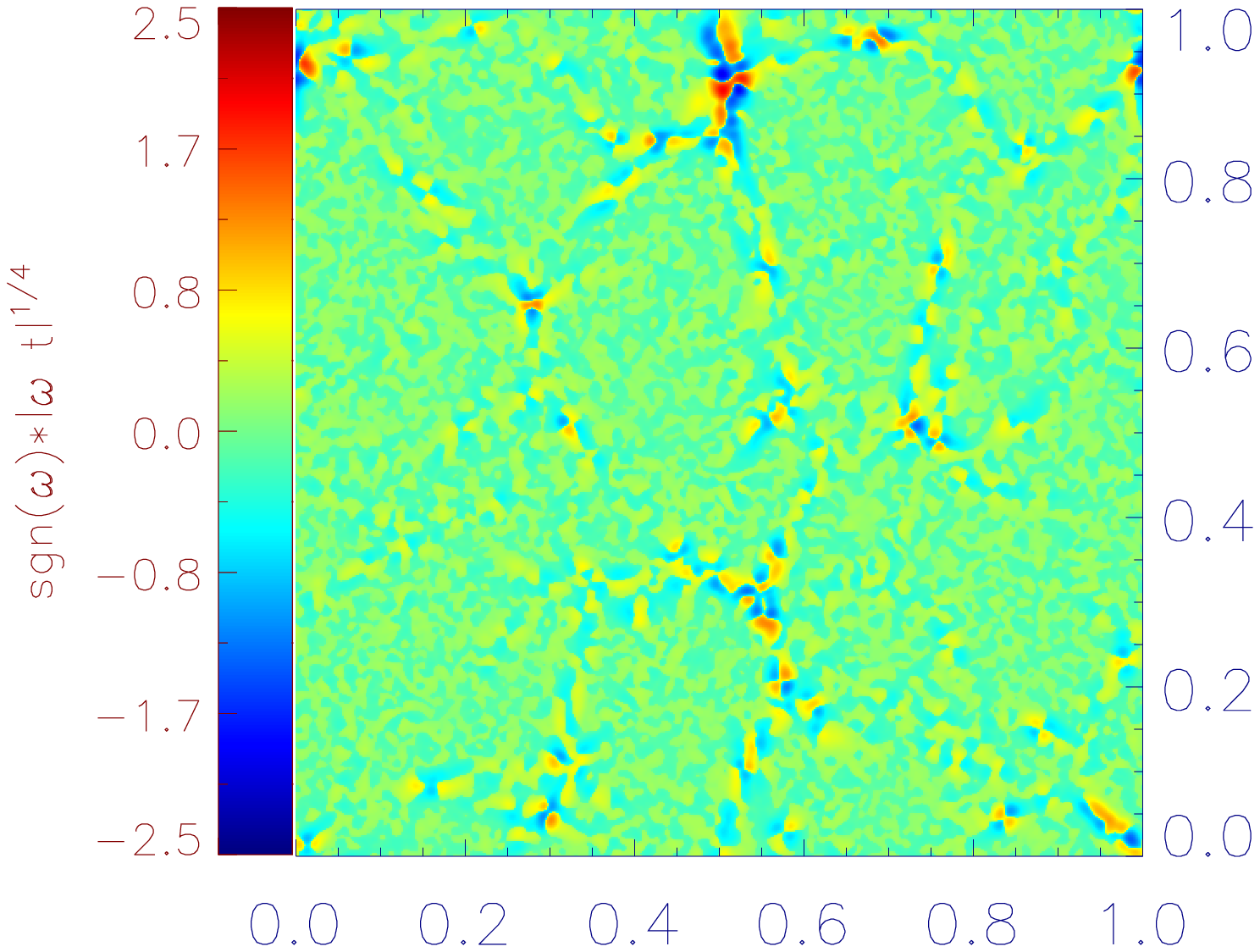}
\hspace{-0.5cm}
\includegraphics[width=0.48\textwidth]{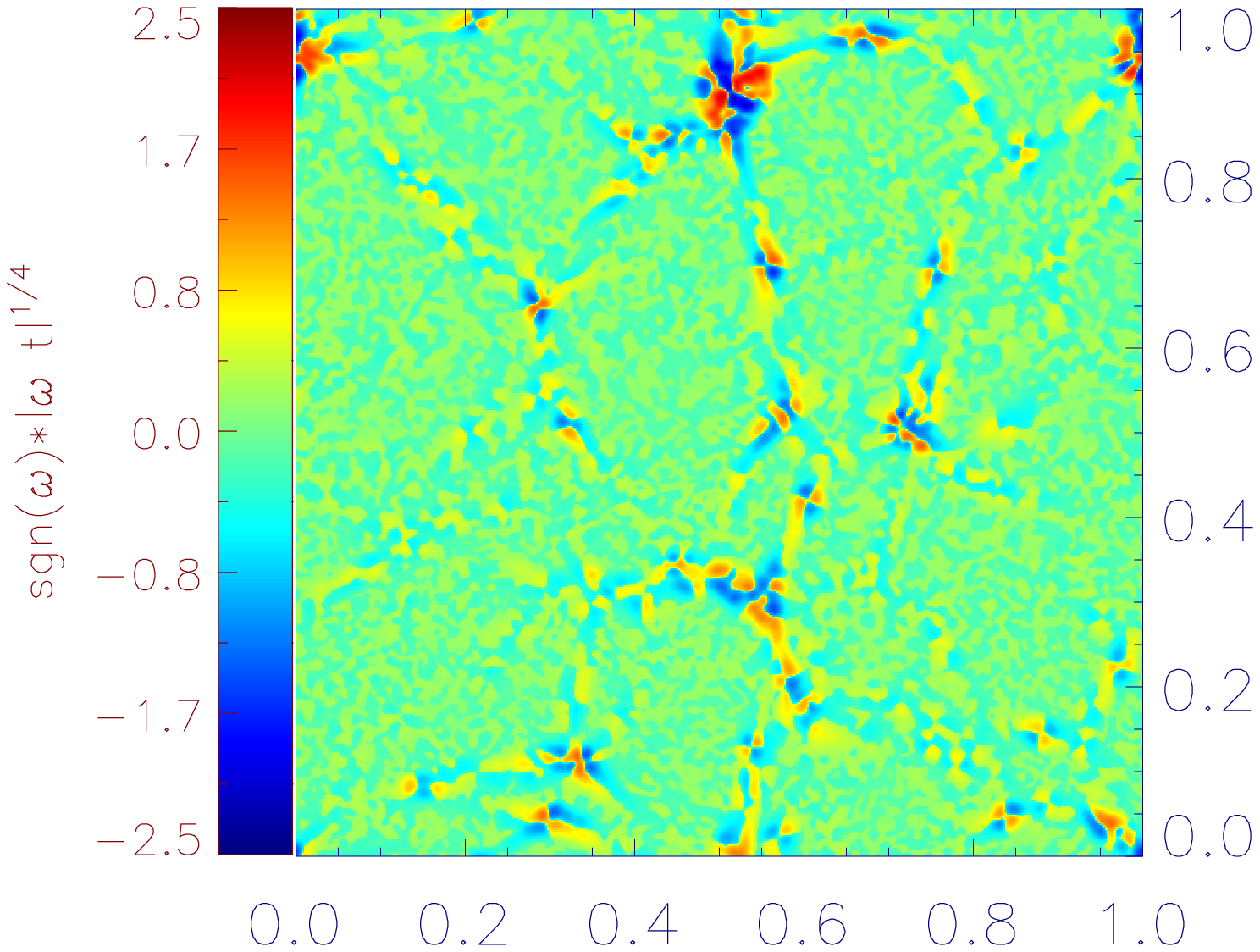}
\hspace{-0.5cm}
\includegraphics[width=0.48\textwidth]{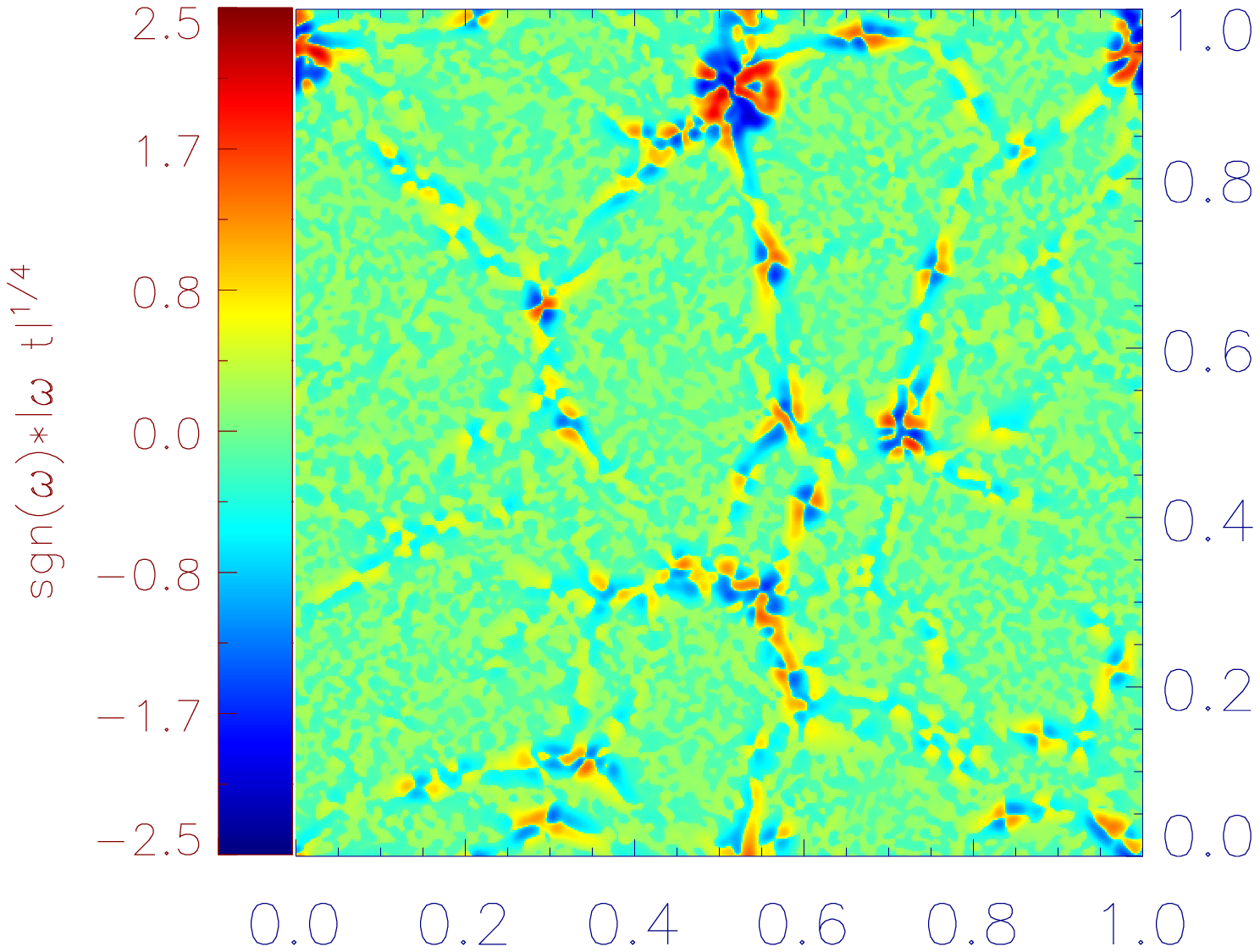}
\vspace{-0.5cm}
\caption{The evolution of rescaled vorticity of baryonic matter in the same slice as Figure11 at $z=2.0, 1.0, 0.5, 0.2$ from top to bottom.}
\label{figure13}
\end{figure}

\begin{figure*}[tbp]
\hspace{0.75cm}
\includegraphics[width=0.9\textwidth]{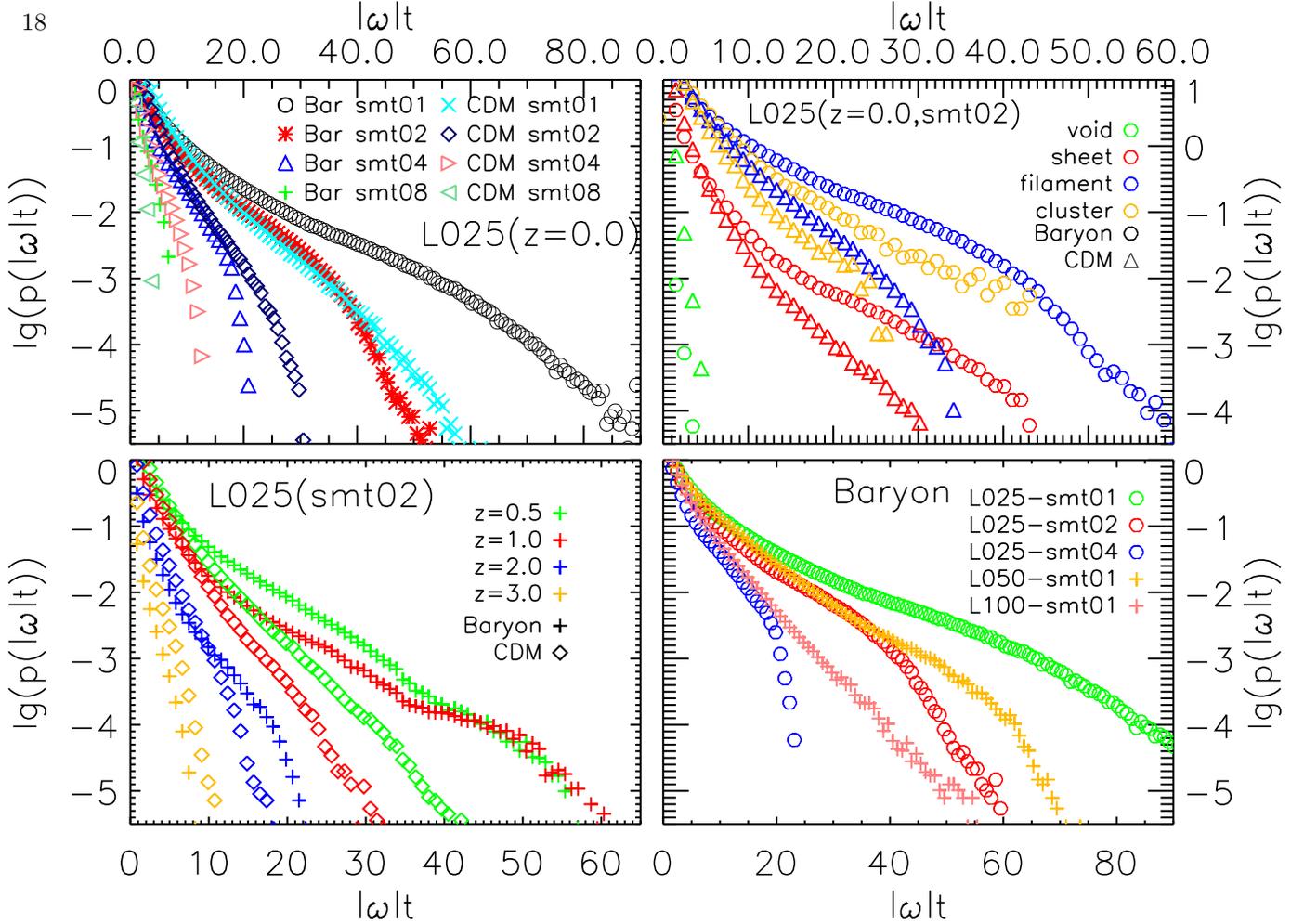}
\vspace{0.7cm}
\caption{The probability distribution of vorticity, rescaled by the cosmic time, of baryonic and dark matter. Top left: in L025 with different smooth length; Top right: in different cosmic environment, red, green, blue, gold color represents void, sheet, filaments and  clusters respectively; Bottom left: in L025 since $z=3.0$; Bottom right: in different simulations at $z=0$.}
\label{figure14}
\end{figure*}

Figure 13 demonstrates the rapid growth of vorticity of gas in the same slice since $z=2.0$. Weak curl motions would develop in the sandwich layer of sheets that is well confined in the normal direction and is the place where the first shell crossing happens. Quadripolar multi-streaming regions tend to arise at the intersections of multi-sheets, i.e, the seeds of filaments. Moderate vorticity have been developed in a few filamentary regions at $z=2.0$. The vorticity is boosted significantly when matter from multi-flows keeps collapsing and swirling around filaments. The baroclinicity produced in strong curved shocks surrounding filaments should be the driving factor for gas(Zhu, Feng, Xia et al. 2013). The sign of quadripolar multi-flow in isolated filaments became much evident at $z=1.0$ and evolved moderately afterwards, in sync with the growth of filaments. Even more complex pattern of vorticity are found in the cross section of prominent filaments. The curl motions in the progenitor of the prominent filaments at position $(x=0.50, y=0.90)$ was  already more complex than a quadripolar pattern at $z=1.0$. Matter should have been accreted into filaments along multi sheets, as well as directly from nearby voids. In a word, the rapid growth of vorticity is in sync with the dramatic rise of filaments at $z<2-3$.

In Figure 14, we plot the probability distributions function(pdf) of vorticity rescaled by the cosmic time, i.e., $|\omega|t$ measured in L025 at $z=0$ with different smooth lengths. The vorticity of both baryonic and dark matter are found to decrease with increasing smooth lengths, in agreement with Libeskind et al.(2014). Vortical motions are effectively boosted in the nonlinear regime, but a large smooth length would reduce the nonlinearity. The vorticity of baryonic matter exhibits stronger heavy-tailed distribution pattern than dark matter, suggesting that vorticity of baryonic matter can be developed more effectively in the highly nonlinear regimes. The pdf of vorticity in baryonic matter at $z=0$ is likely to be described by a Log-normal distribution(Zhu et al. 2010). We also compare the distribution in each cosmic environment in the top right plot of Figure 14.  Consisting with ZF15, the vorticity are triggered and pumped up in sheets, and especially in filaments, and then experienced dissipation in clusters for both baryonic and dark matter. The vorticity of dark matter in the voids is relatively higher than baryonic matter, as displayed in Figure 11, actually, it might be due to the under-sampling issue of dark matter particles in voids. Moreover, the growth of the vorticity since $z=3$ can be clearly seen in the bottom left plot of Figure 14. During the epoch between $z=2$ and $z=1$, the vortical component of velocity field experiences the largest gains. Meanwhile, the difference between vorticities of baryonic and dark matter was widened. The bottom right plot in Figure 14 compares the pdf of vorticity of baryonic matter at $z=0$ in three simulations. Obviously, as the cosmic vorticity is a typical non-linear feature emerging from gravitational collapsing, higher spatial resolution is capable of resolving higher nonlinearity, and hence the global vorticity of L025-smt01 is the highest. On the other hand, for a same smooth length, a larger simulation box could host more collapsed structures, and have relatively more cells with very large vorticities, i.e., a longer tail in the pdf. For instance, L050-smt01 have more cells with $|\omega|t>40$ than L025-smt02.

\subsection{Mean Divergence and Curl Velocity}
\begin{figure}[tbp]
\vspace{-1.0cm}
\includegraphics[width=0.42\textwidth]{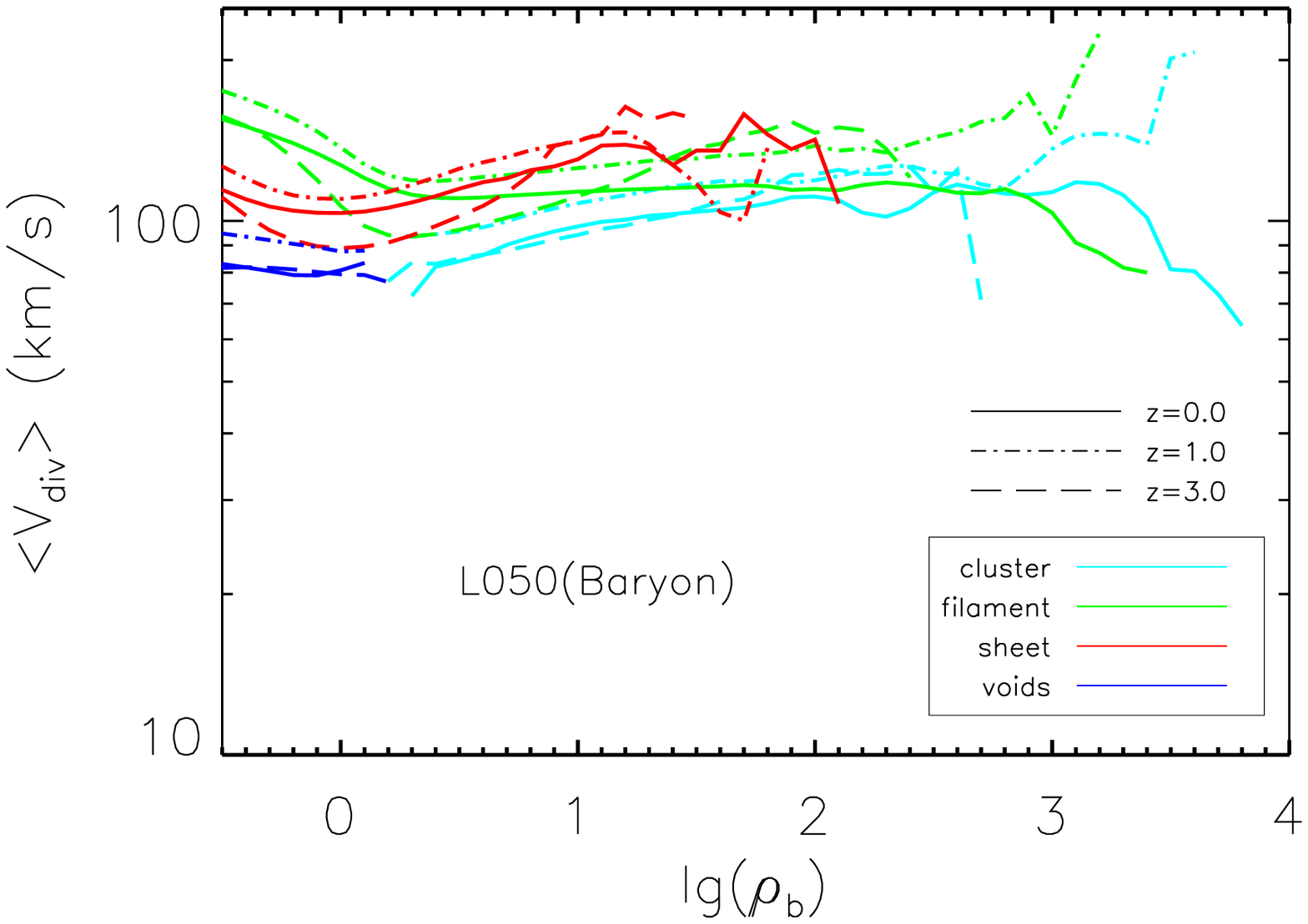}
\includegraphics[width=0.42\textwidth]{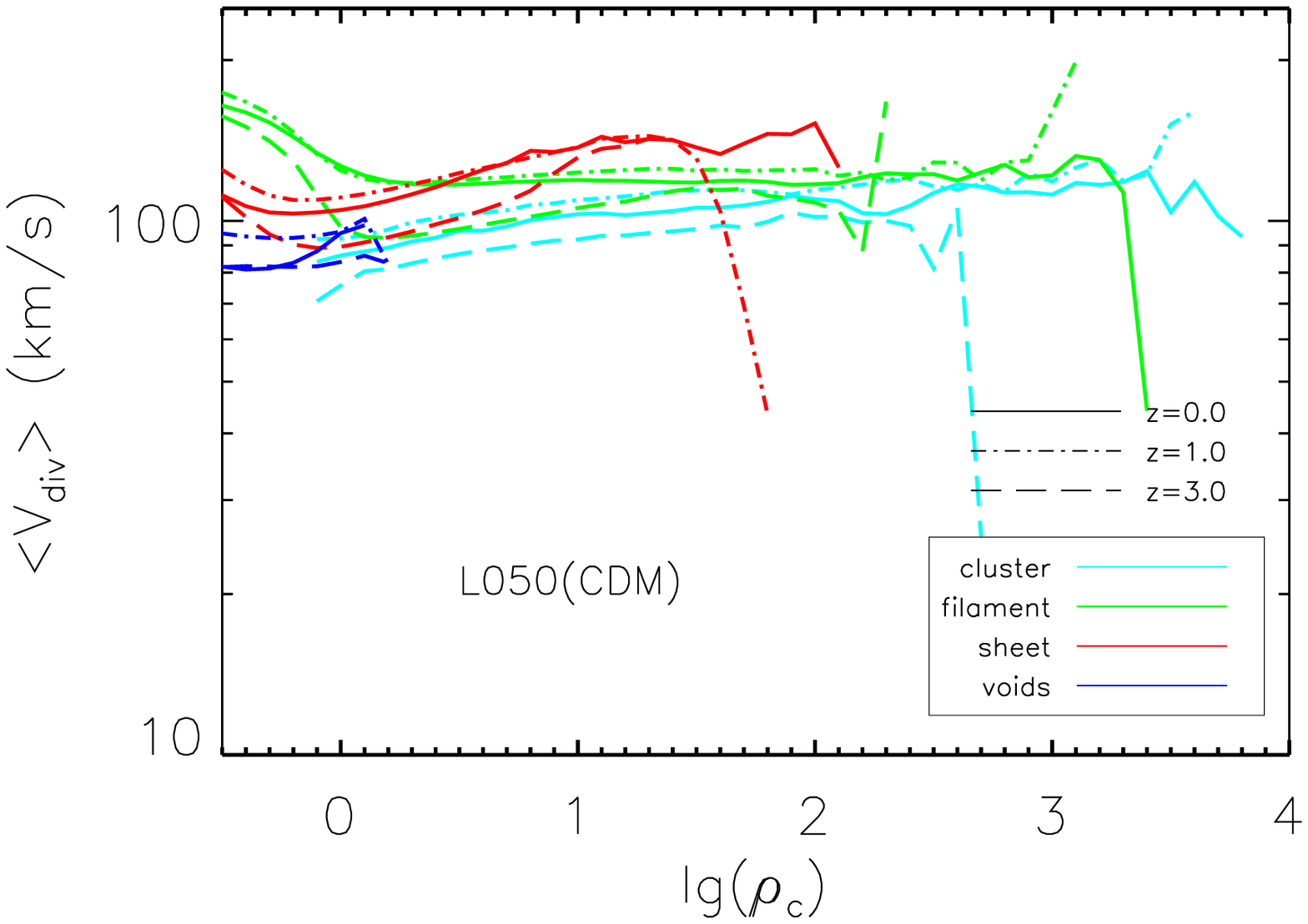}
\includegraphics[width=0.42\textwidth]{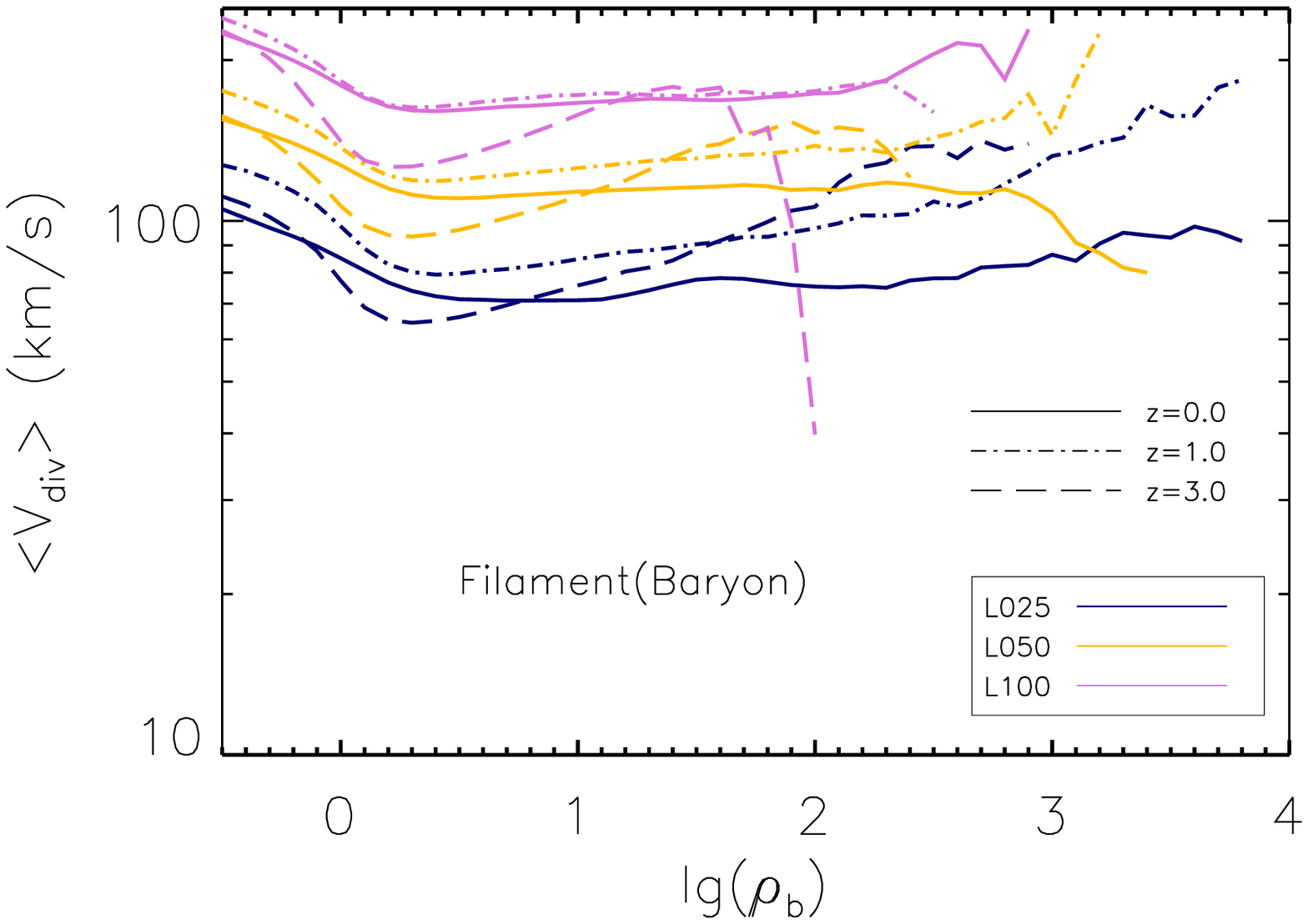}
\includegraphics[width=0.42\textwidth]{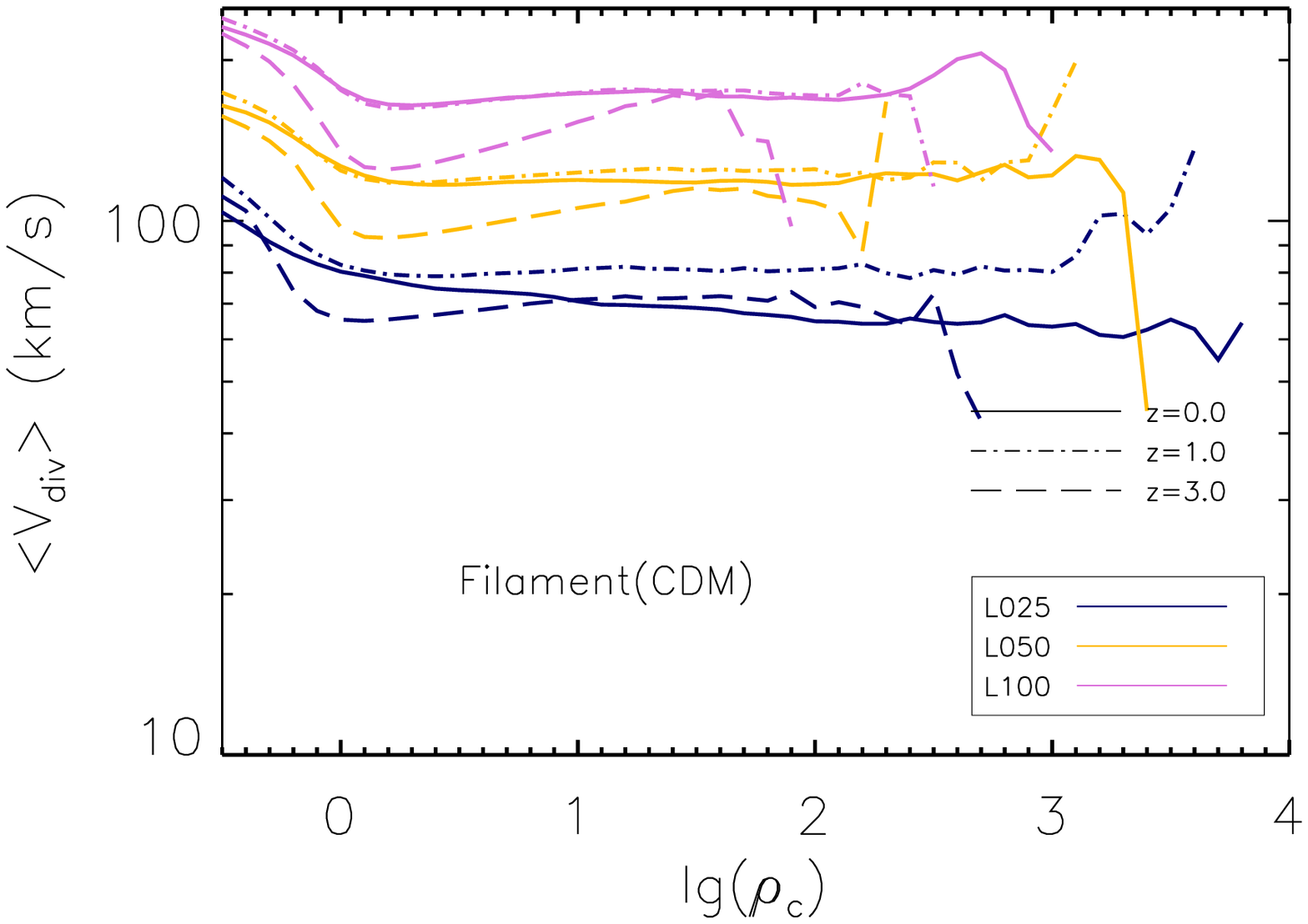}
\vspace{0.2cm}
\caption{The mean divergence velocity in each cosmic web environment in L050 at $z=3.0, 1.0, 0.0$(Top two plots), and in the filaments of all the three simulations at $z=0.0$(Bottom two plots).}
\label{figure15}
\end{figure}

\begin{figure*}[tbp]
\includegraphics[width=0.50\textwidth]{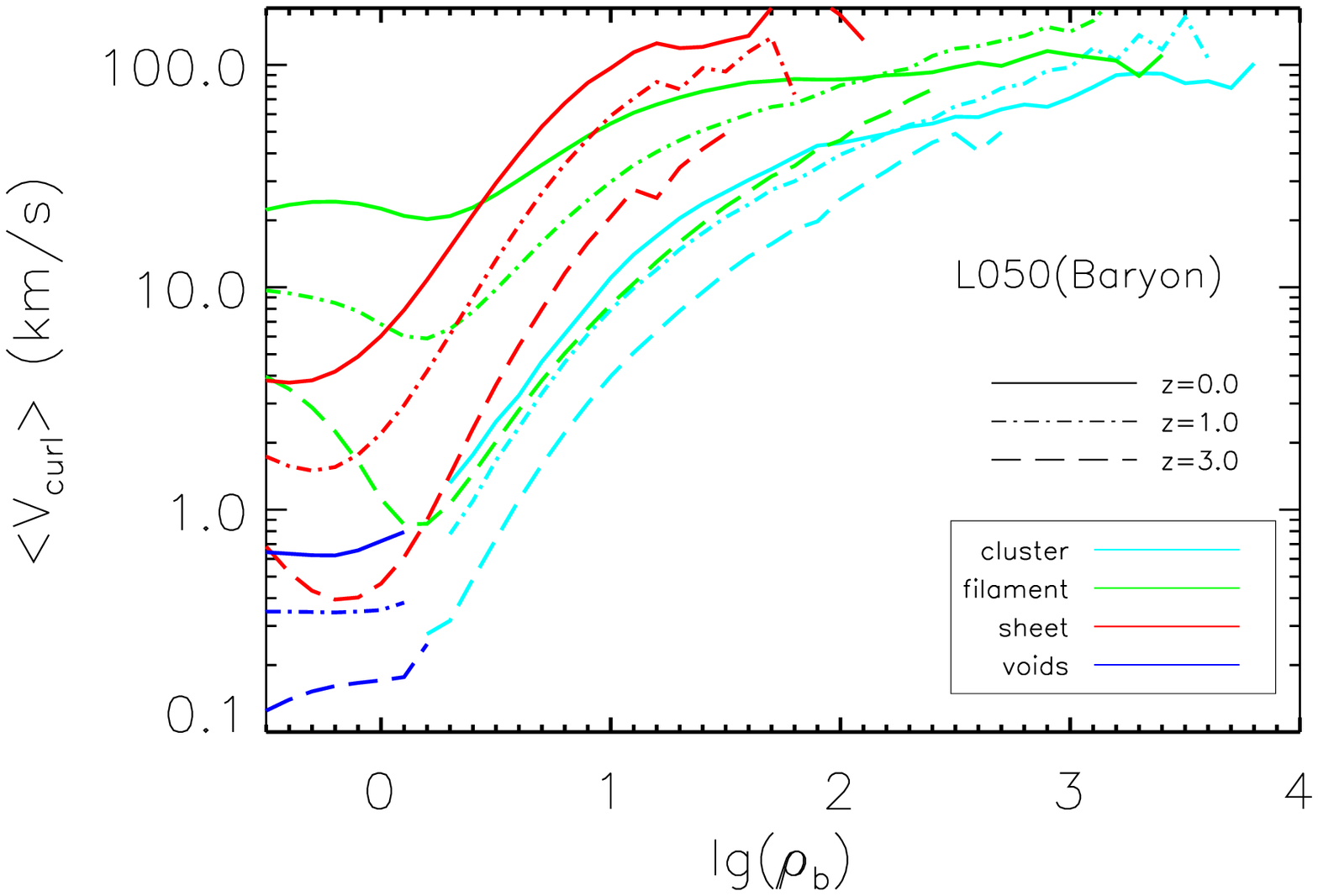}
\includegraphics[width=0.50\textwidth]{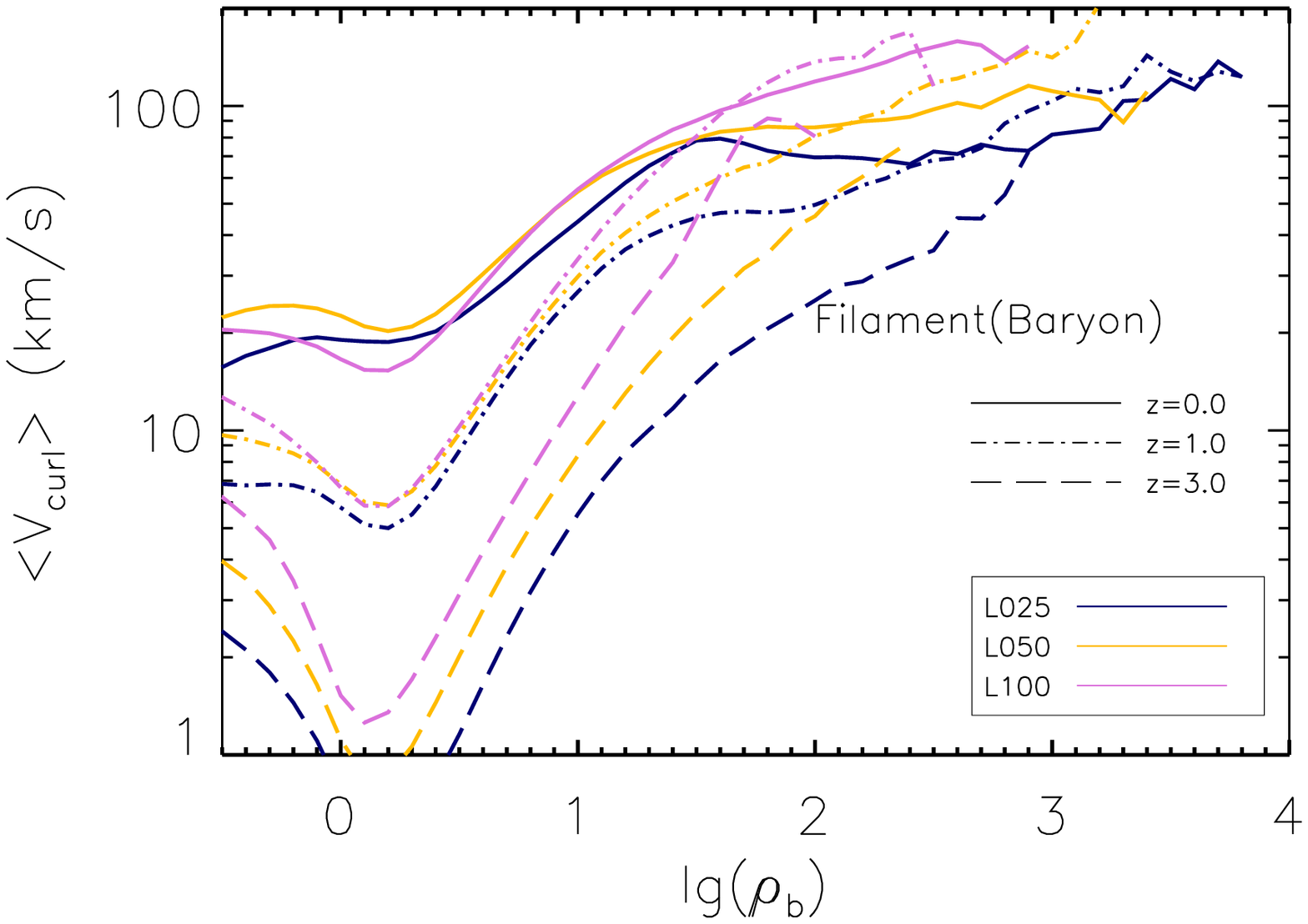}
\includegraphics[width=0.50\textwidth]{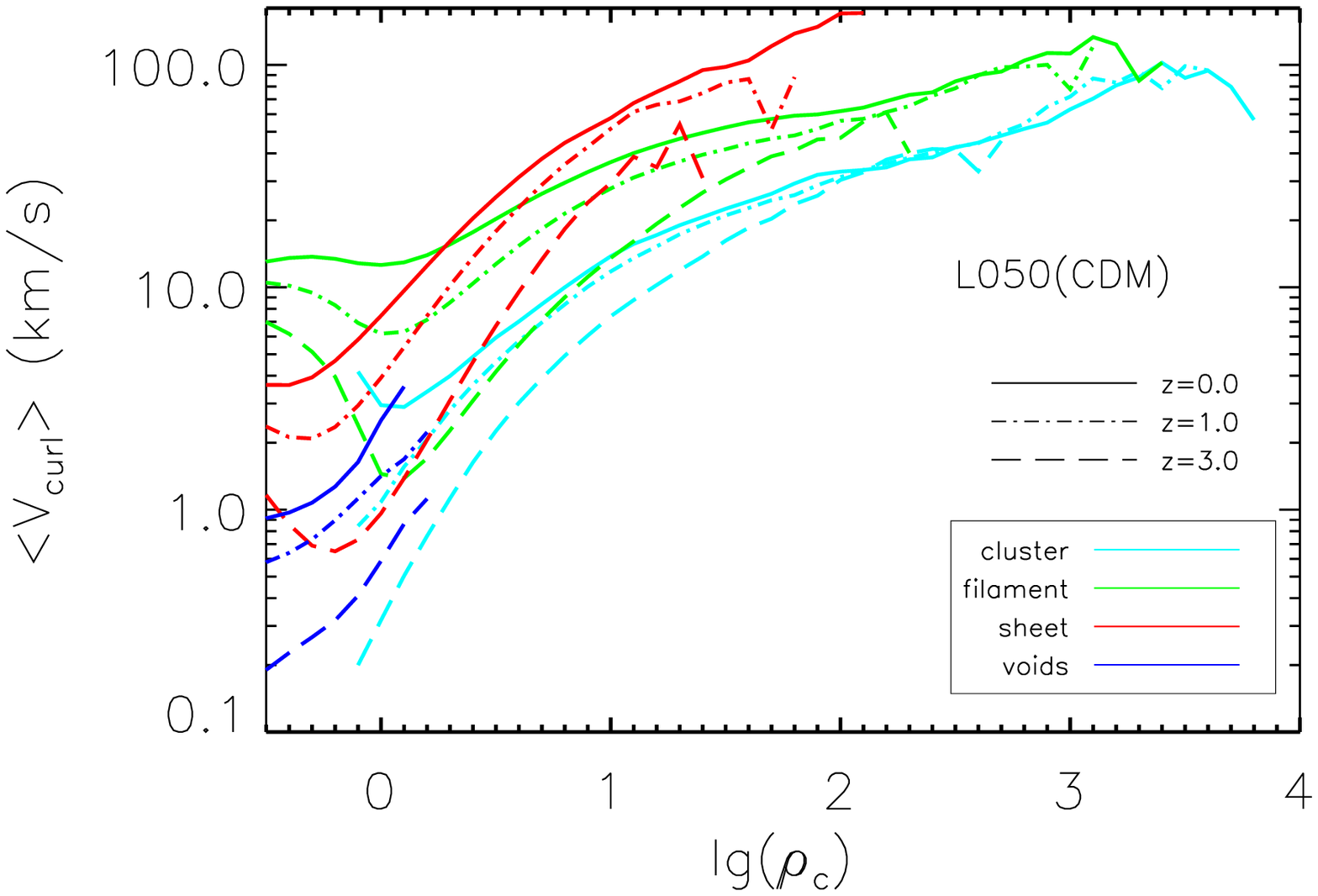}
\includegraphics[width=0.50\textwidth]{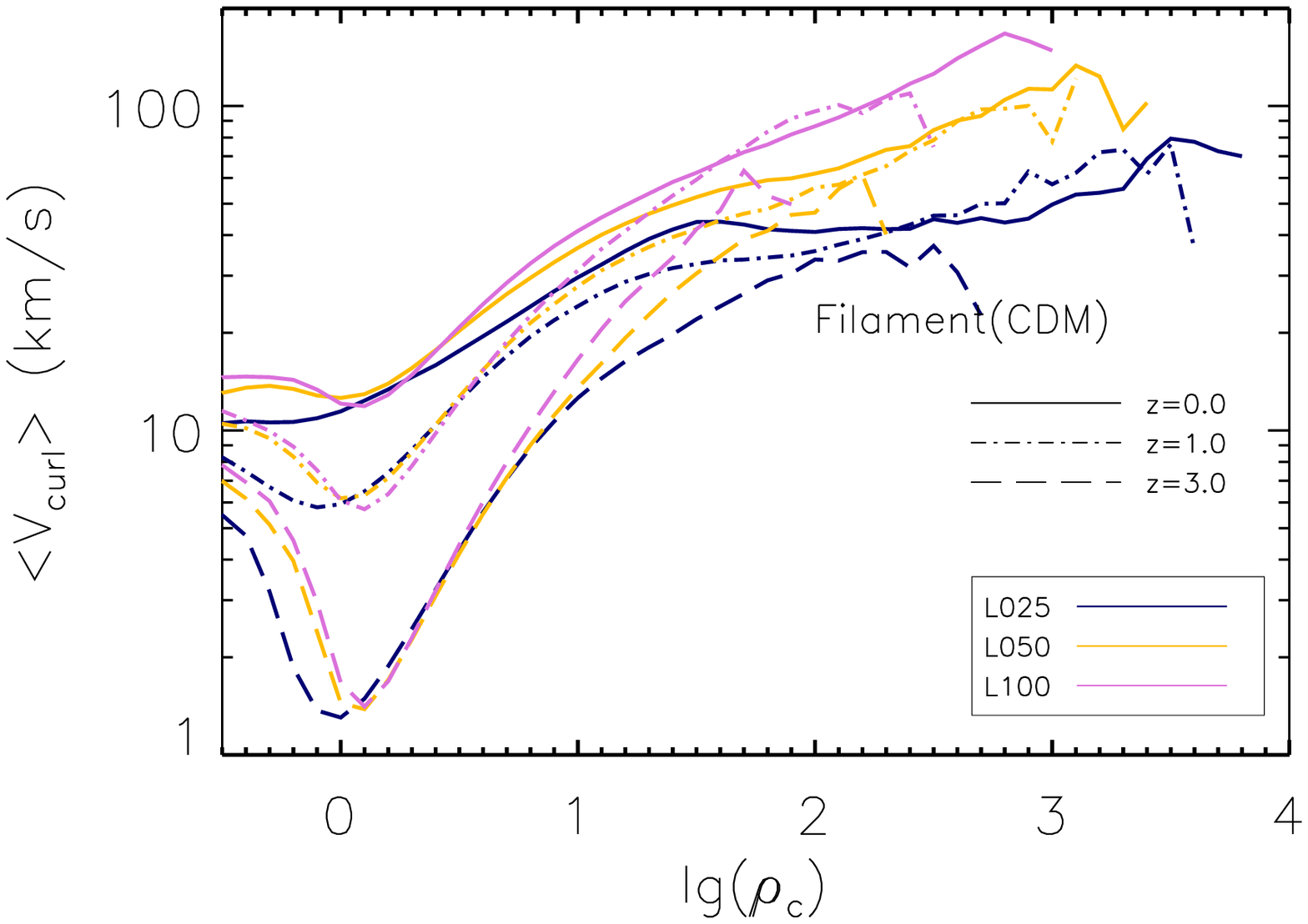}
\vspace{0.5cm}
\caption{The mean curl velocity in each cosmic web environment in L050 at $z=3.0, 1.0, 0.0$(Left), in the filaments of the three simulations at $z=0.0$(Right). The top and bottom row present results in baryonic and cold dark matter respectively.}
\label{figure16}
\end{figure*}

The value of velocity can directly characterize the flows corresponding to divergence and vorticity. The mean divergence velocity $\vec{v}_{div}$ as a function of density in each web component in L050 since $z=3.0$ are shown Figure 15. The magnitude of divergence velocity in each structures has been set up by $z=3$ and changed mildly since then, mainly between $z=3$ and $z=1$. It is not out of our expectation, as the outline of sheets and voids on large scale is found to be already in place by $z=3.0$ in previous paragraphs. In addition, the divergence velocity of matter evolves slowly with respect to the density in each environment till $\delta \sim 100$, during the formation of cosmic web. At high over-density regimes,  the small number of grid cells leads to significant dispersion. The level of divergence motions in sheets and filaments is marginally higher than in voids and clusters, in consistent with the divergence distribution given in Figure 12. For a particular box size, e.g., L050, the magnitude is nearly the same in four morphology environment, around $100 km/s$. Notable difference between baryonic matter and dark matter can only be found in high density region, $\delta > 30$, where the dispersion due to limited number of cells become evident. The distribution of $\vec{v}_{div}$ in the filaments of all the three simulations are also compared in Figure 15. The largest mean divergence velocity is found in L100, can be about two times of the value in L025, because more massive structures can form in larger simulation box and hence result in higher bulk velocity. 

Similarly, Figure 16 gives the distribution and evolution of mean curl velocity, $v_{curl}$, in the comic web. The curl motions are effectively developed when the matter flowing into the central layer of sheets, and filaments, and then experience dissipation when flowing into clusters, which basically confirms our previous result in ZF15 and the slice view in Figure 12. The resampling and smooth processes in this work may have partly blurred the velocity fields at the boundaries between sheets and filaments for $\delta >1$, result in a fast growing of curl motions in sheets with respect to density, compared to ZF15. On the other hand, Figure 8 shows that most of the cells with $1<\delta < \sim 100$ belong to filaments. Hence, the curl velocity in the cosmic web are mainly contributed by filaments, which is conform with the visually evolution presented in Figure 13.

The magnitude of $v_{curl}$ have increased rapidly for $\delta \leq 20-30$ since $z=3$. At $z=0$, the curl velocity is $\sim 10\%$ of the divergence velocity at $\delta \sim 0$, then increases to $\sim 30\% $ at $\delta \sim 3-5$, and eventually become comparable to the divergence velocity at $\delta \sim 20-30$. This evolution pattern is basically consistent with Libeskind et al.(2014), which shows that the vorticity of dark matter grows rapidly as $\omega \propto \rho$ for $\rho<10$, and turn to $\omega \propto \rho^{1/3}$ for $\rho >10$. In filaments, the curl velocities are pumped up to $\sim 100 km/s$ for $\delta \gtrsim 30$ at $z=0$. The curl velocity of baryonic matter is moderately higher than the dark matter in the sheets and filaments within over density range $1-30$. The effect of box size on the curl velocity is the same with divergence velocity.  A valley at $\delta \sim 0$ is presented in the sheets and filaments for both mode of motion, which is more evident for $v_{curl}$ at high redshifts. The valley is observed in all the three simulations. A possible explanation is that a fraction of matter with $\delta \sim 0$ in the sheets and filaments have not experience shell crossing or curved shocks for dark and baryonic matter respectively, the fraction decrease with time because the continuous anisotropic gravitational collapsing. 

\subsection{Velocity Power Spectrum}

\begin{figure*}[tbp]
\includegraphics[width=0.50\textwidth]{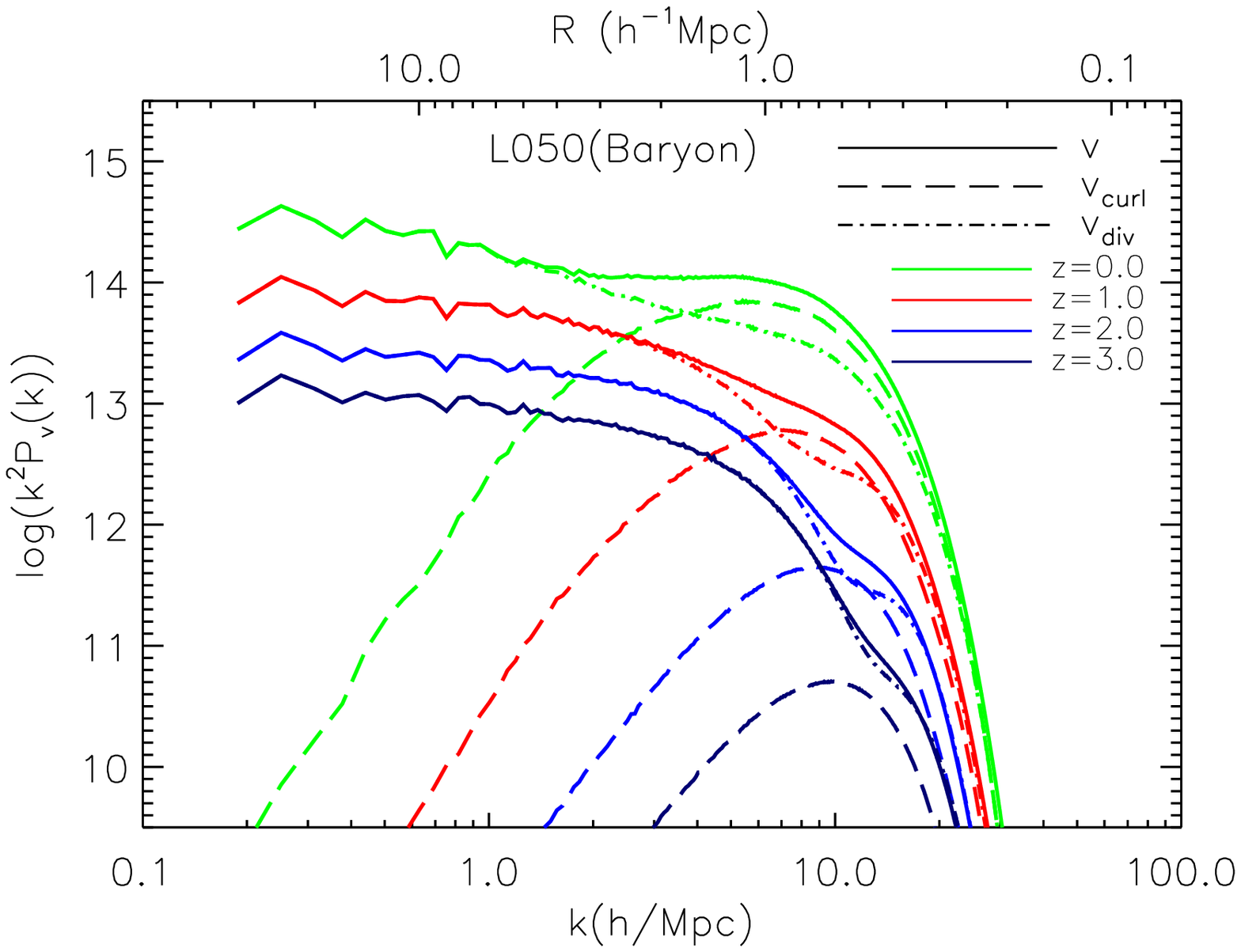}
\includegraphics[width=0.50\textwidth]{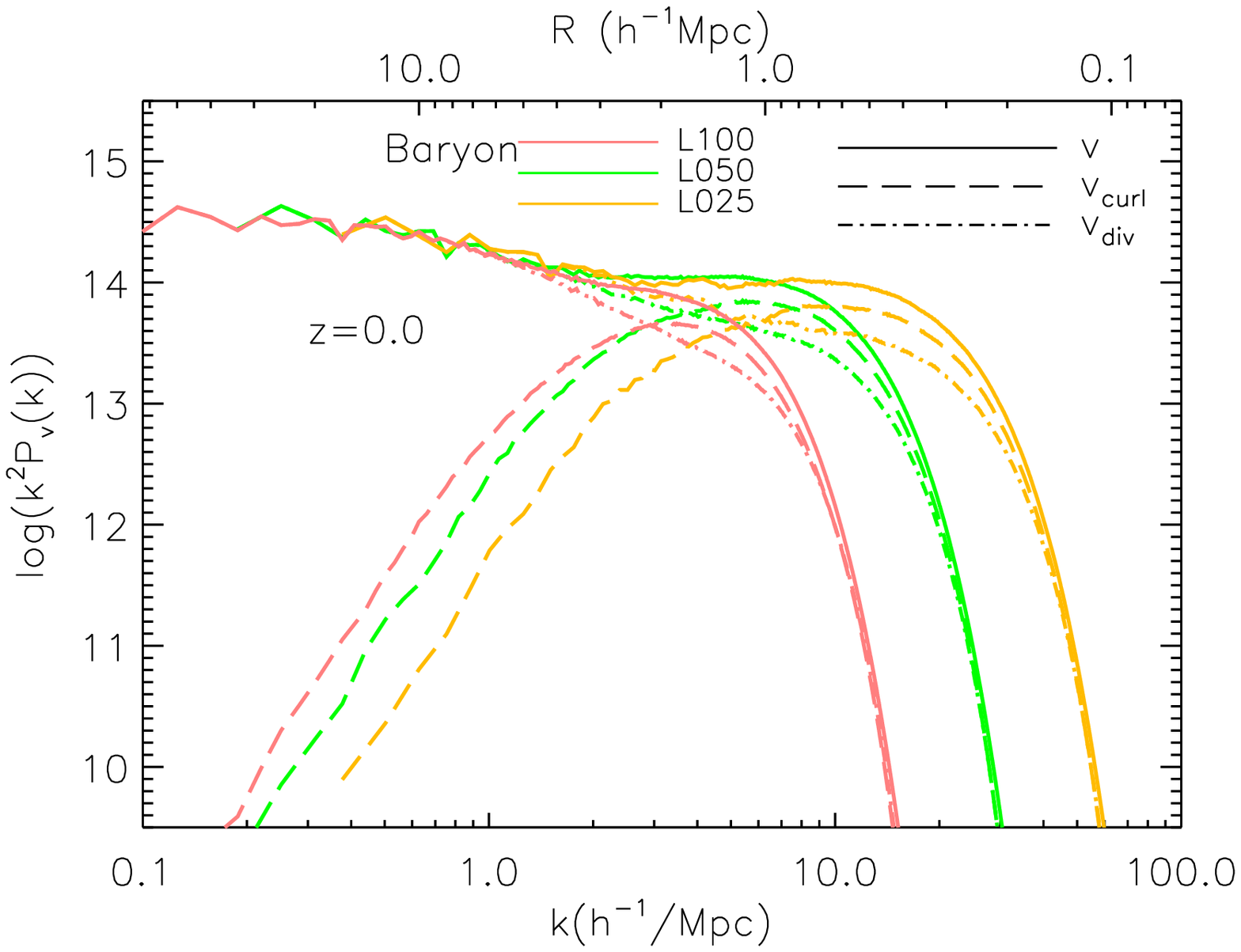}
\vspace{1.2cm}
\includegraphics[width=0.50\textwidth]{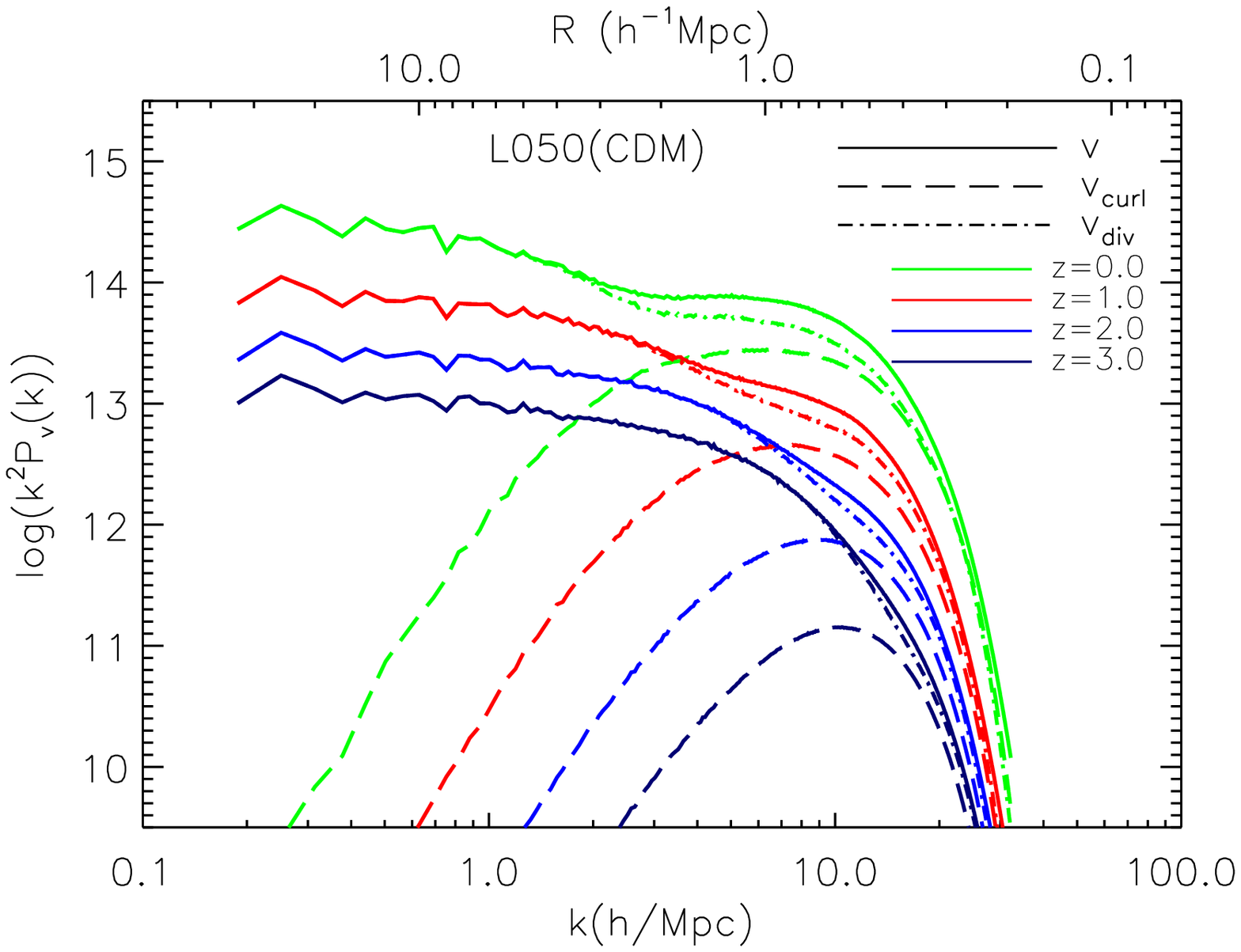}
\includegraphics[width=0.50\textwidth]{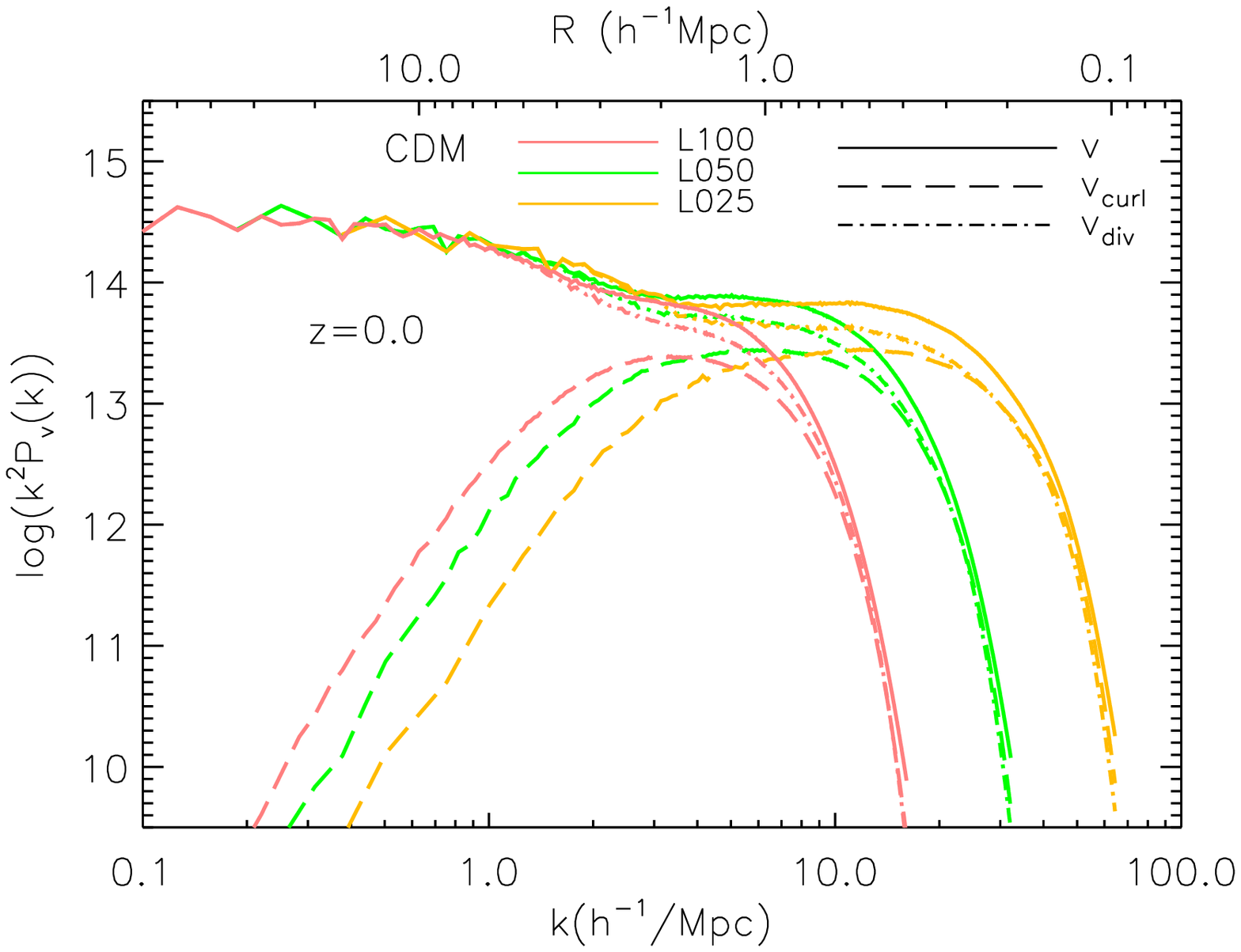}
\caption{Left: The compensated power spectrum $k^2P(k)$  of total, curl and divergence velocity in L050 at $z=3.0, 2.0,1.0,0.0$ for baryonic(Top) and cold dark matter(Bottom).Right: The compensated velocity power spectrum in three simulations at $z=0.0$}
\label{figure17}
\end{figure*}

Power spectrum of velocity at different time reflect the spacial correlation of motions associated to the evolving cosmic web. In Figure 17, we plot the compensated power spectrum $k^2P(k)$ of the total, curl and divergence velocity. For the sake of clarity, the comoving velocity rather than the proper velocity are used in this figure to seperate lines at different redshifts. On large scales, namely, in the linear regime, the power spectrum is dominated by the divergence component, in agreement with theoretical expectation and previous study (Zheng et al. 2013). The dominancy of irrotational velocity can be extended to the mildly nonlinear regime, i.e., around a few megaparsecs. Actually, the power spectrum of divergence velocity has been well developed at $>2$ Mpc by $z=3$. Combining with the exploration on mass and velocity distribution given in the above sections, we conclude that the outline of cosmic web above $\sim 2-3$ Mpc have been formed by $z=3$, when the sheets served as the frame. The power spectrum of curl velocity increase rapidly since $z=3.0$, in accordance with the growth of curl motions largely developed in filament components. At $z=0.0$, the contribution from the curl motion reaches 10\% at scale of $\sim 5$ Mpc, and can exceed the divergence velocity at scale small than $\sim 2 $ Mpc for the baryonic matter. The power spectrum of both modes drop dramatically at a few times of the resampling smooth length. 

The relative contribution of curl velocity in dark matter is weaker than baryonic matter, remaining subordinate to irrotational velocity in the highly nonlinear regime. Pichon \& Bernardeau(1999) predicted that, the curl motion will become significant at scale $\sim 3-4$ Mpc after the shell crossing in large scale caustics. The predicted scale is in good agreement with our simulations. Using a set of cold dark matter only simulations, Zheng et al.(2013) demonstrated that the power spectrum of curl motions would equal to the divergence motions at $k \sim 10 h/Mpc$, comparable with our result in L100. They also expect that the contribution of curl velocity would be twice of the divergence at sufficiently small scales, which is hard to verify for the dark matter due to numerical factors related to the discrete particle, limited resolution and smooth processes. For the baryonic matter in our simulations, the ratio of contribution of curl velocity to divergence velocity reach to a peak value of  $\sim 1.5$ at the scale $\sim 1.3, 0.8, 0.4$ Mpc in L100, L050, L025 respectively, i.e., close to the expectation in Zheng et al.(2013).

In the nonlinear regime, the power spectrum of velocity approximates a power law within particular scale range for both the baryonic and dark matter. For the baryonic matter, the upper end is about the same, $\sim 3 Mpc$ in three simulations. The lower end varies with the resolution, extended to $\sim 0.3 Mpc$ in L025. The growth of the upper end scale is sync with the evolution of vorticity in filaments presented at Figure 8. The upper end of the scale range for dark matter is small than baryonic matter. In the highly nonlinear regime, the anisotropic collapsing of gas would develop numerous strong curved shocks, mostly associated to filaments, and hence supersonic intermittent velocity field exhibiting $P(k) \propto k^{-2}$ law in the power spectrum(e.g., Porter et al. 1992; Kritsuk et al. 2007; Zhu et al. 2013). After the shell crossing in caustics, the curl motions of dark matter in filaments may have developed similar intermittent behaviour. 

\subsection{Velocity Structure Functions and Fractal Dimensions}

\begin{figure}[htbp]
\vspace{-1.0cm}
\includegraphics[width=0.45\textwidth]{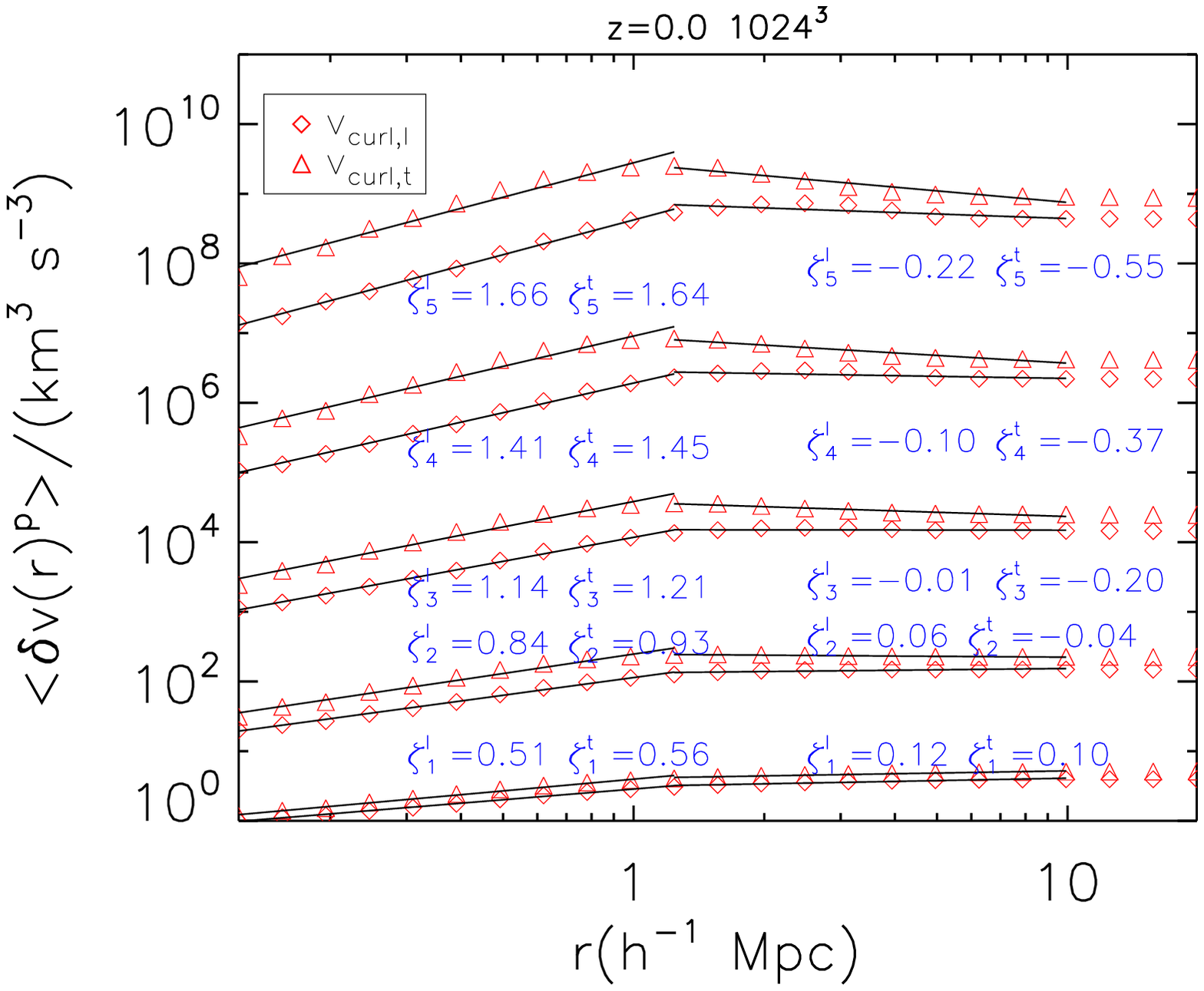}
\includegraphics[width=0.45\textwidth]{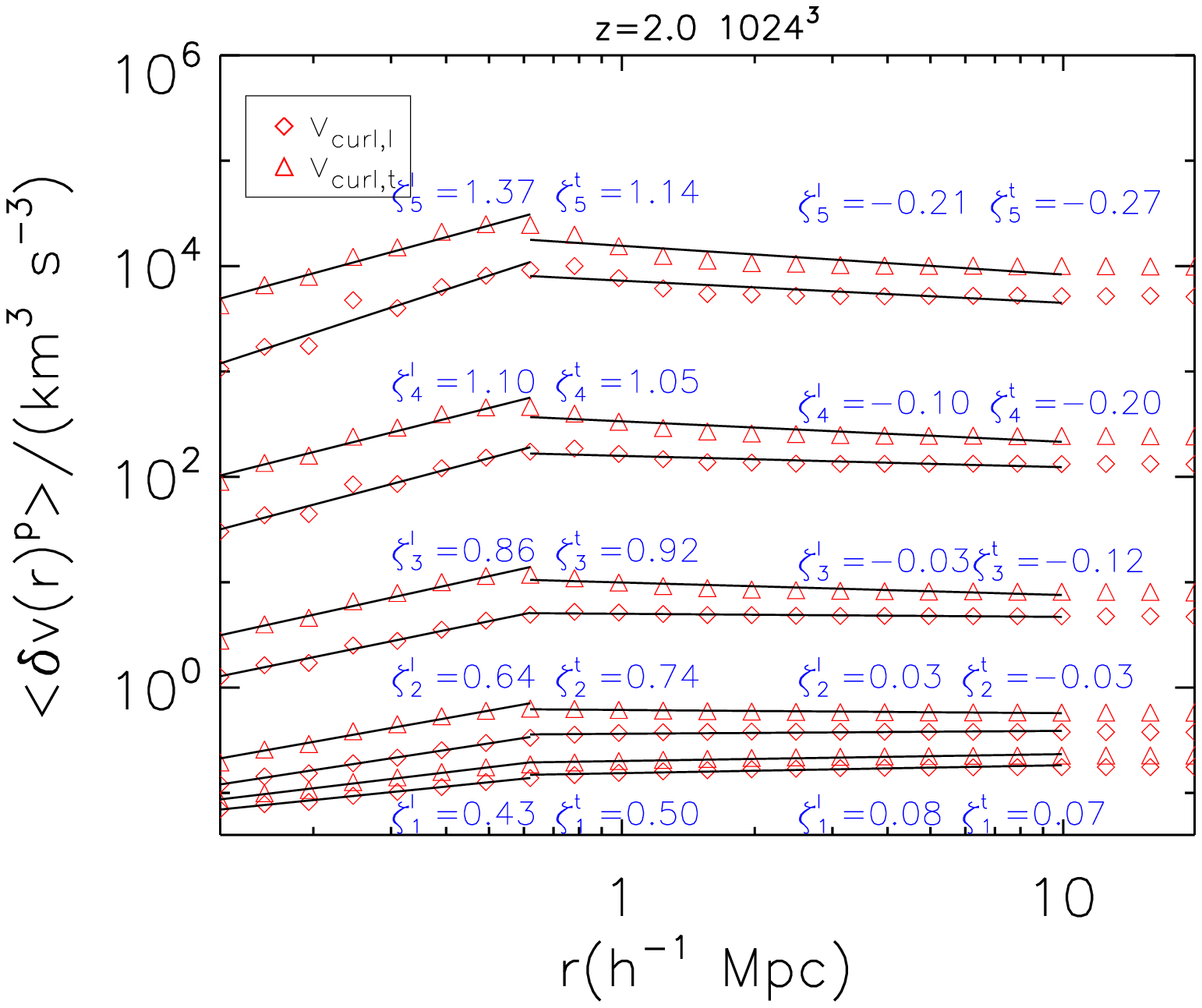}
\includegraphics[width=0.45\textwidth]{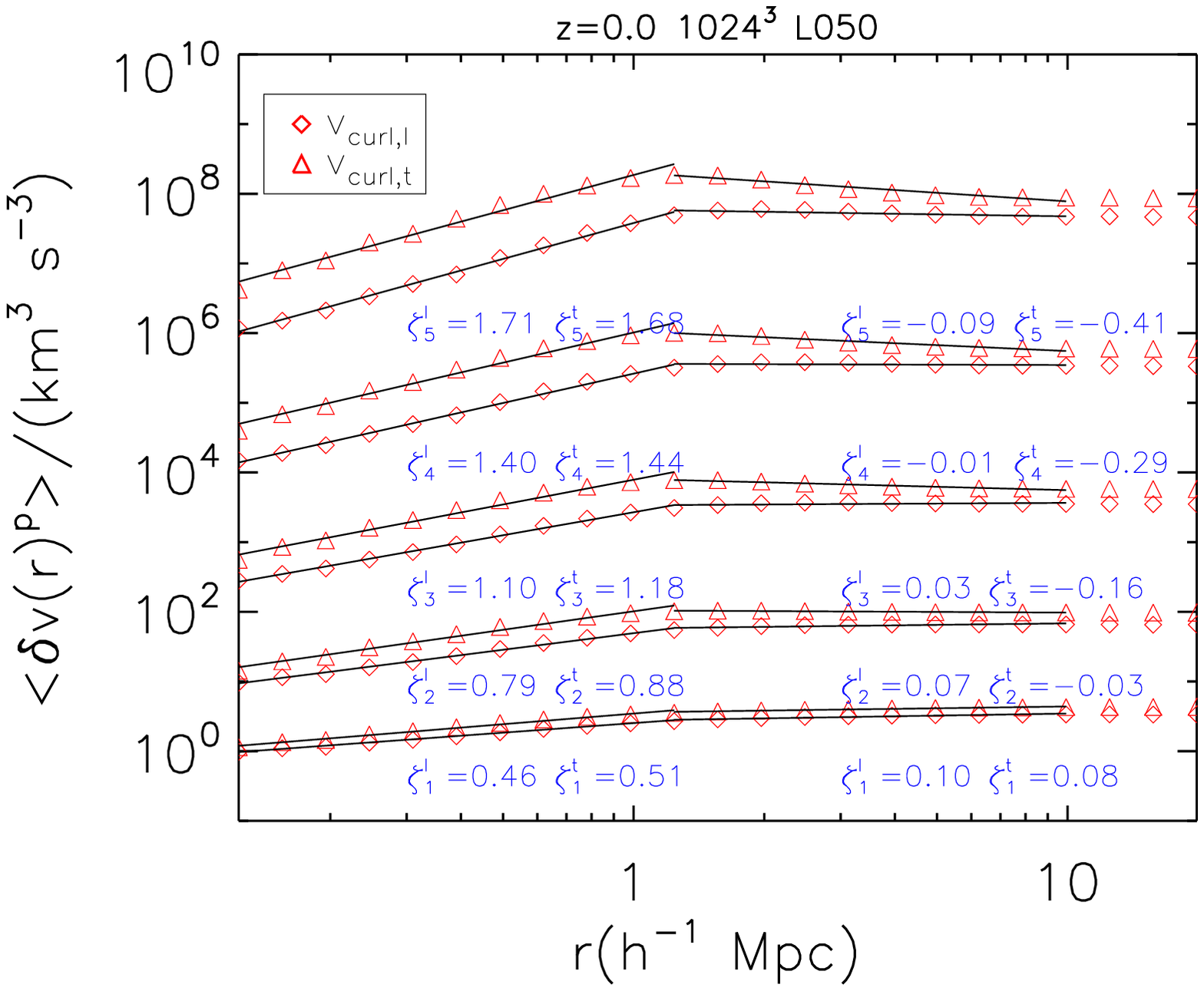}
\includegraphics[width=0.45\textwidth]{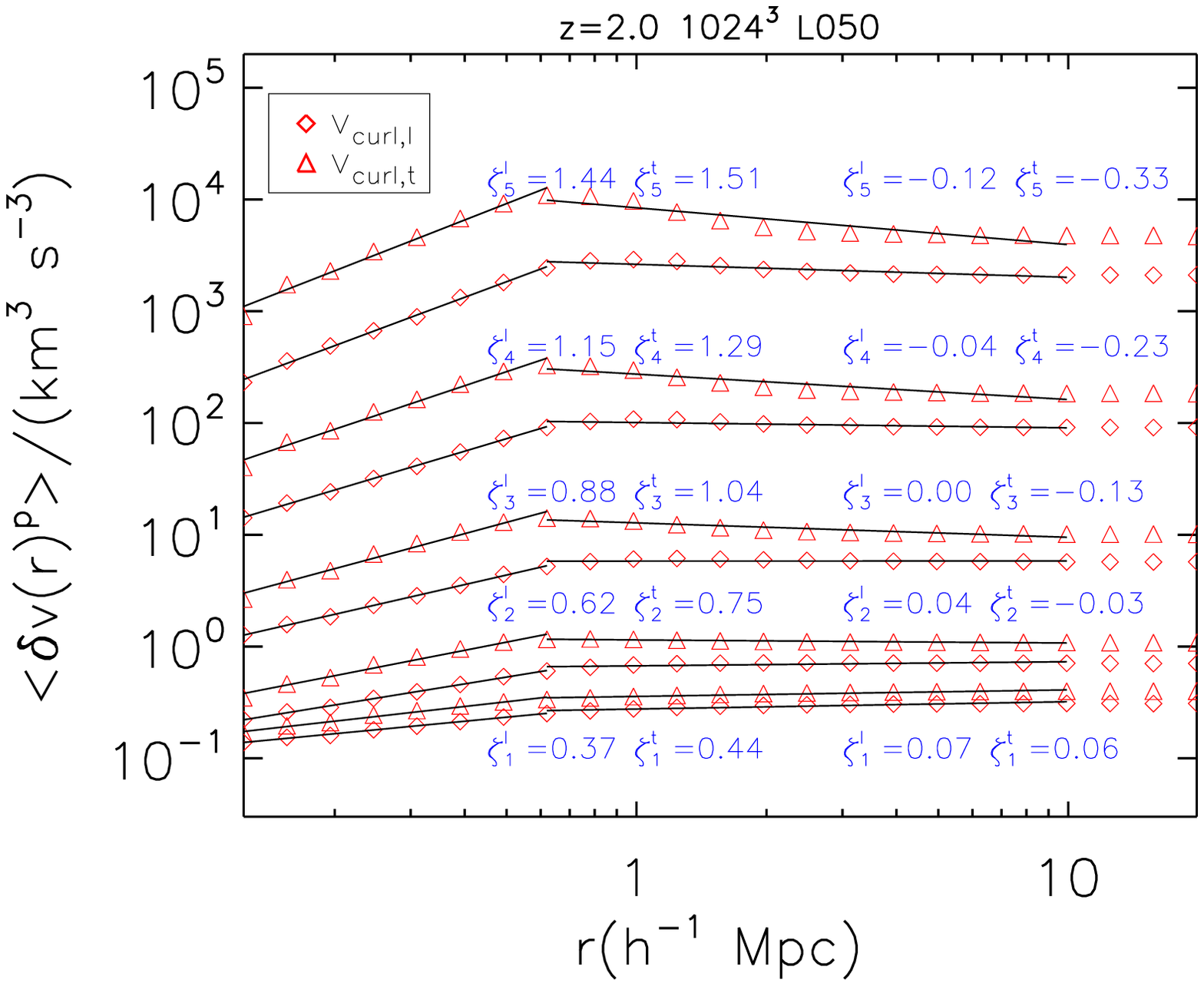}
\vspace{0.1cm}
\caption{The velocity structure functions of curl motions at $z=2.0, 0.0$ for baryonic(Top two) and cold dark matter(Bottom two) in L050.}
\label{figure18}
\end{figure}
 
The growth of magnitude and power spectrum of curl velocity indicates a close connection between matter flows and cosmic web. Given that the vorticity in filaments may play very important role in building the spins of halo and galaxy(e.g., Laigle et al. 2015), we carry a further check on the correlations between vorticity and filaments through the study of velocity structure functions and fractal spacial dimensions. The velocity structure functions can give a more insightful view on the coherence of curl motions in the sheets and filaments. In ZF15, we found that the velocity structure functions of curl motions of baryonic matter can be described by the intermittent model proposed by She \& Leveque(1994)(also Dubrullel 1994) in the supersonic scale range. The velocity structure function reads as, 
\begin{equation}
S_p(r) \ = \  < \delta v_r^p> = < |v(x)-v(x+r)|^p> \  \propto \ r^{\zeta_p},
\end{equation}
, where $p=1, 2, 3, ...$. She \& Leveque(1994) and Dubrulle(1994) suggested that the relative scaling exponents $Z_p$ would follow a universal scaling law, i.e.,  
\begin{equation}
Z_p=\frac{\zeta_p}{\zeta_3}=(1-\Delta)\frac{p}{3}+\frac{\Delta}{1-\beta}(1-\beta^{p/3}),
\end{equation}
where $\beta$ and $\Delta$ are related to the hierarchy properties of intermittent structures(More details can be found in Dubrullel 1994 and ZF15). 

\begin{figure*}[tbp]
\includegraphics[width=0.45\textwidth]{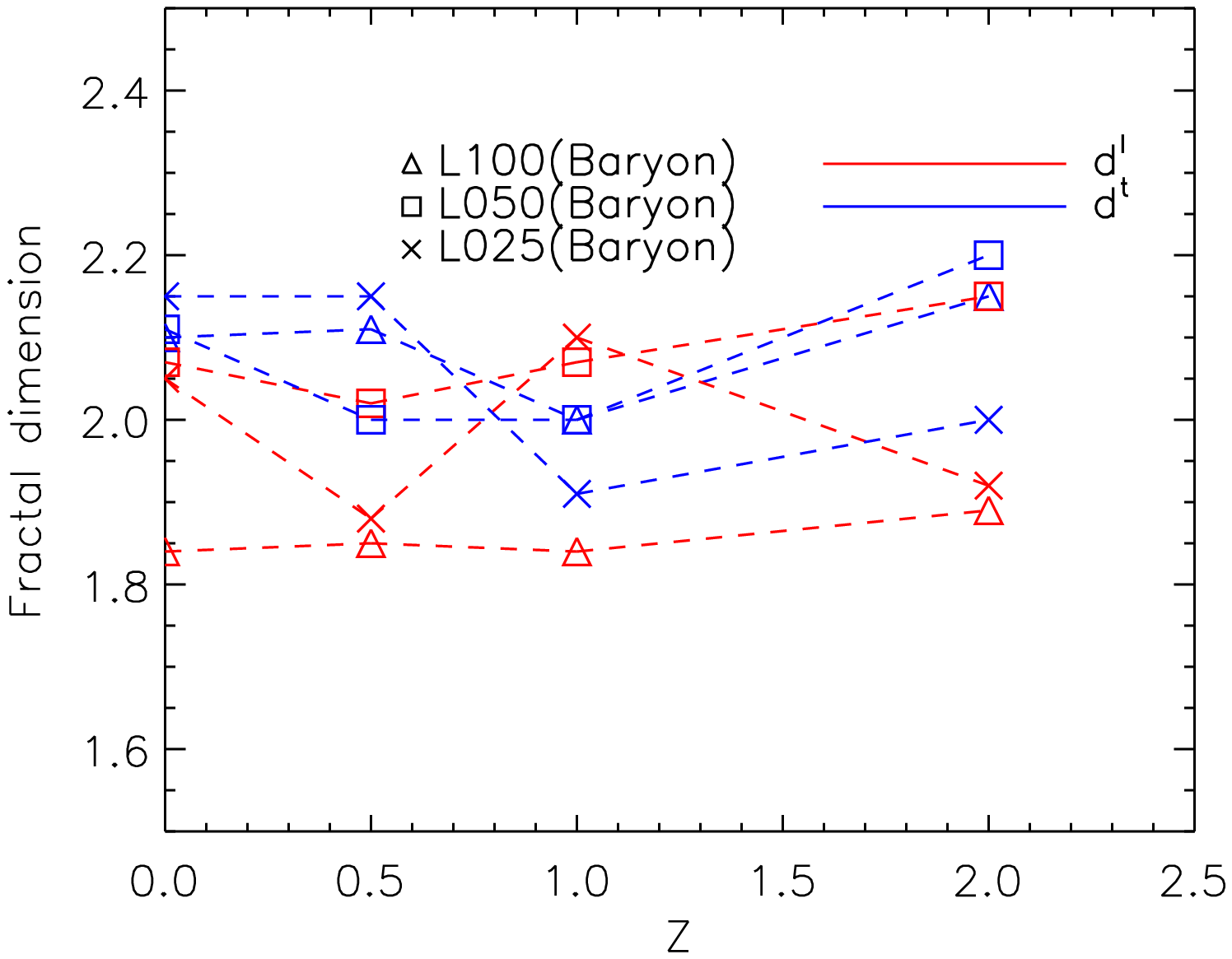}
\includegraphics[width=0.45\textwidth]{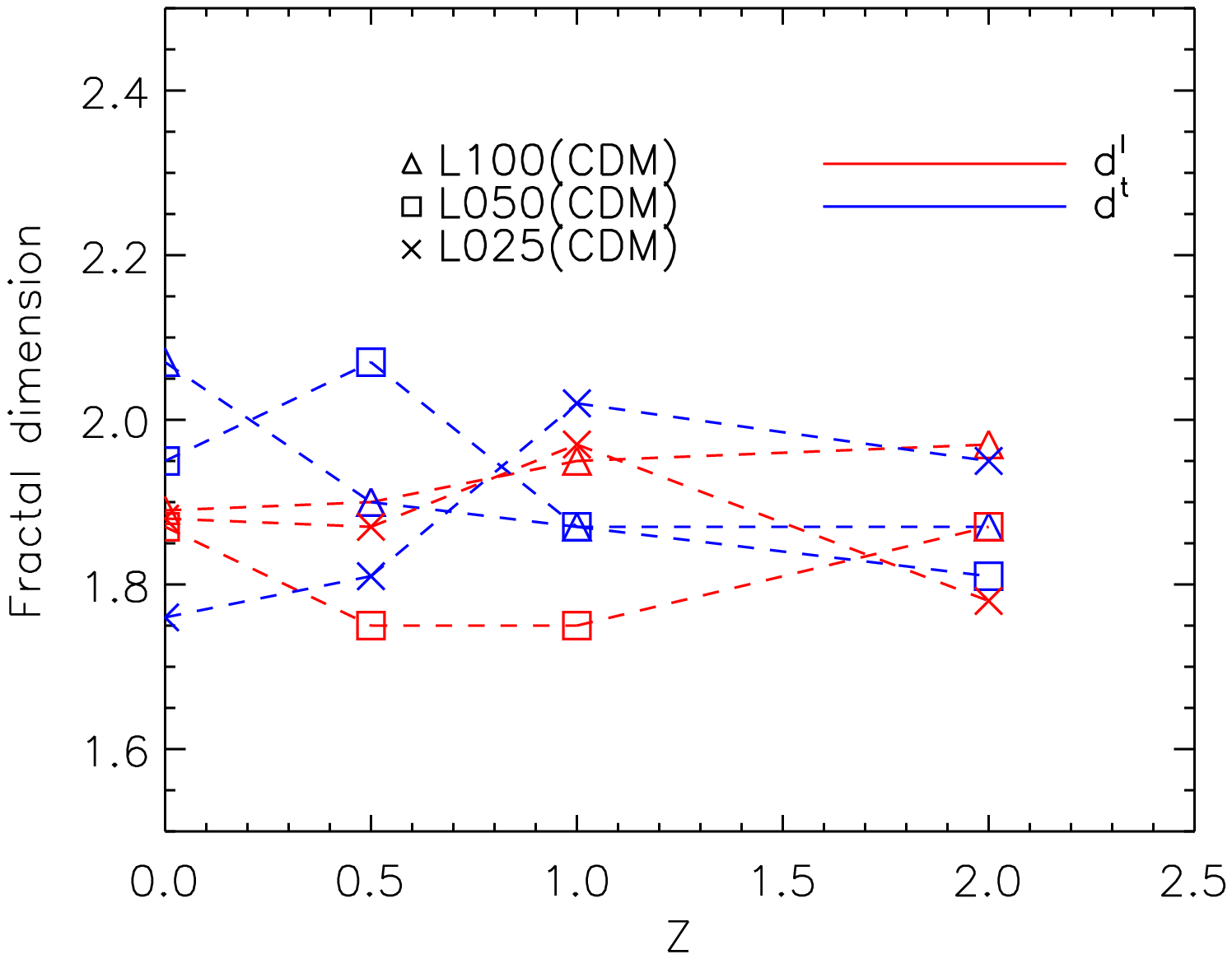}
\caption{The fraction dimensions of curl velocity derived from structure functions for baryonic(Left) and cold dark matter(Right) in three simulations.}
\label{figure19}
\end{figure*}

\begin{figure*}[htbp]
\vspace{-0.5cm}
\includegraphics[width=0.45\textwidth]{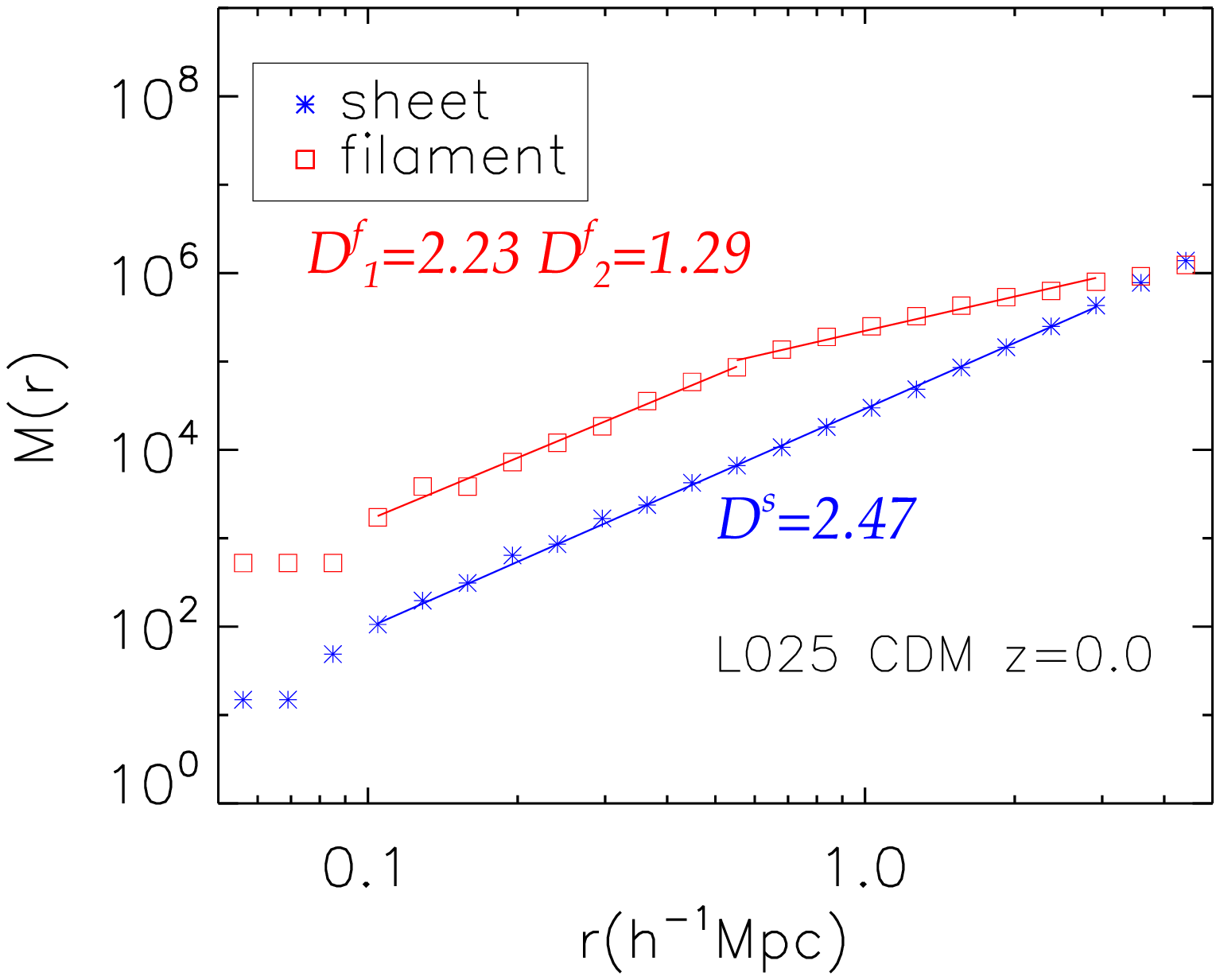}
\includegraphics[width=0.45\textwidth]{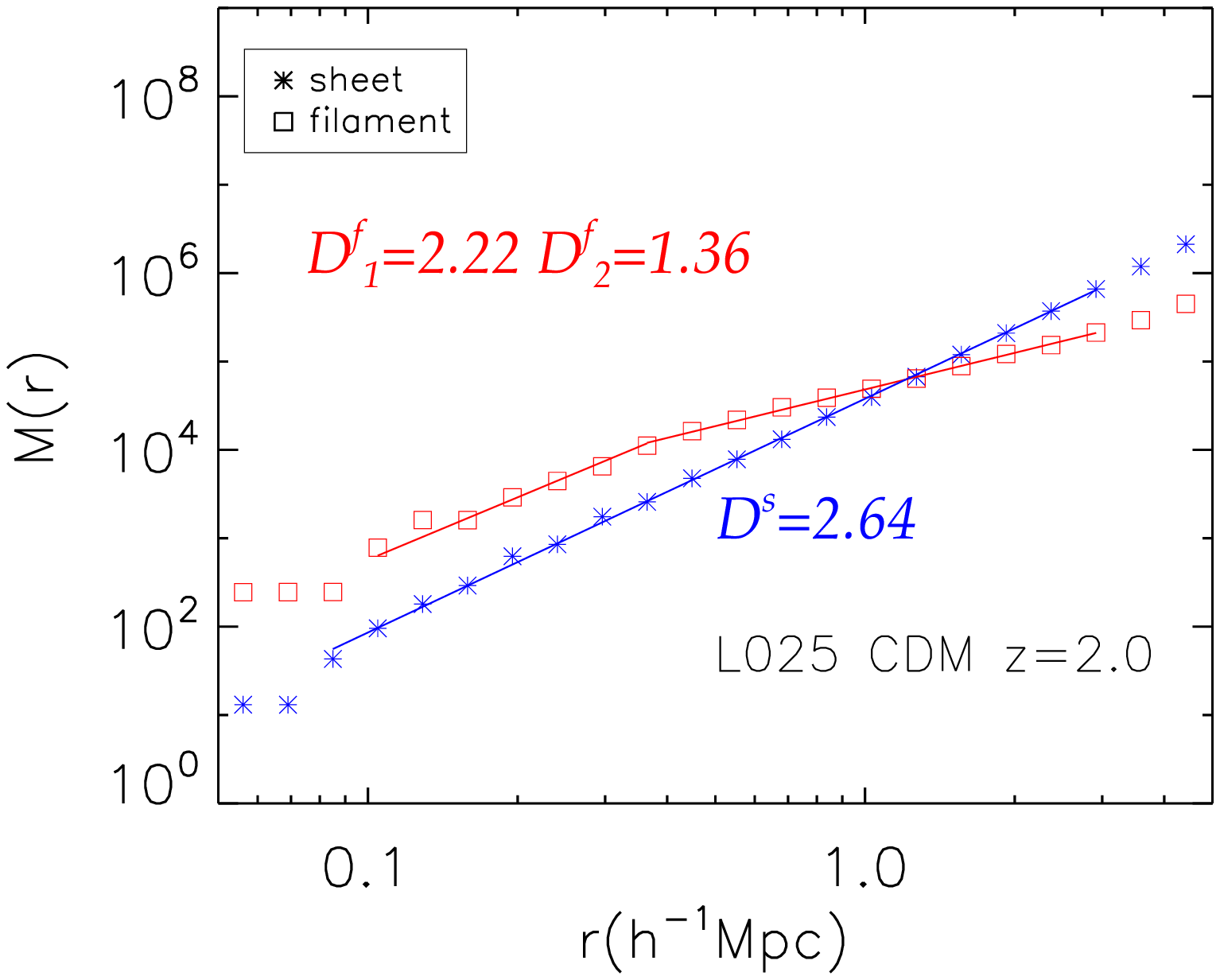}
\vspace{0.5cm}
\caption{The fractal dimension of mass distribution in filaments and sheets for cold dark matter at z =0(Left), 2(Rigt) in L025.}
\label{figure20}
\end{figure*}

We apply the same analysis to the resampled curl velocity fields of both baryonic and cold dark matter in all the three simulations. Figure 18 shows the longitudinal and transverse structure functions, $S_p^{l}$ and $S_p^{t}$, at $z=2$ and $z=0$ in L050. In agreement with ZF15, the structure functions are broken into two distinctive sections, where the breaking occurs at the scale that marks the turnover in the compensated power spectrum of curl velocity. The breaking scale is almost the same for baryonic and dark matter. In other word, the curl motions show significant coherence below the breaking scale denoted by $k_{b}$, strongly confirms the results from power spectrum study. 

The fractal dimensions of the most singular structures, derived from the curl velocity structure functions, are given in Figure 19, where $d^l$ and $d^t$ are from $S_p^{l}$ and $S_p^{t}$ respectively. According to the SL model, the most singular structures of the vortical motions have dimension of $1.85-2.15$ and $1.75-2.05$ for baryonic and dark matter respectively. The $d^l$ and $d^t$ in dark matter are smaller than the values of filament in C14. On the other hand, the fractal dimensions of filaments and sheets can be measured by box counting method(Mandelbrot 1983). Figure 20 presents the results of dark matter in L025, the values are slightly larger than the baryonic matter in ZF15. A transition of mass distribution at the scale around $k_{b}$ is also observed in the filaments. The fractal dimensions of curl motions is close to the fractal dimensions of filaments below $k_{b}$. In a word, the velocity power spectrum, structure functions, the fractal dimension of mass in filaments show analogous transition over $k_{b}$, and the value of fractal dimensions of curl motions and mass in filaments is similar. Combing the growth history of curl motions revealed in subsection \S4.2 and ZF15, we attribute that the vortical kinetic energy are primary associated to filaments for both baryonic and dark matter. The statistics properties investigated in our simulations indicate that the curl motions of cosmic matter is mainly driven by the formation of filaments, usually from further collapsing within sheets. This result is consistent with Laigle et al. (2015), in which they claimed that vorticity in large scale structures tends to be confined to, and predominantly aligned with filaments. 

\section{The relative orientation of shear tensor and vorticity}

\begin{figure}[htbp]
\hspace{-1.0cm}
\includegraphics[width=0.65\textwidth]{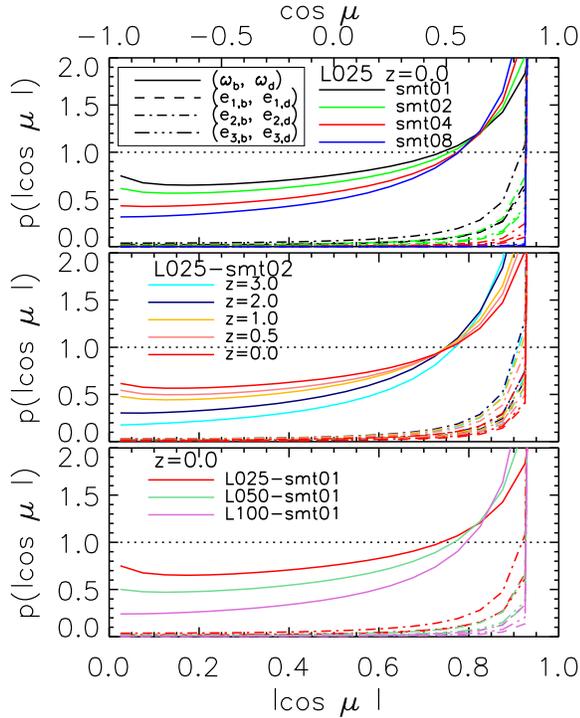}
\vspace{0.5cm}
\caption{The probability distribution $p(cos\mu)$($p(|cos\mu|)$) as a function the angle between vorticity(shear eigenvectors) of baryonic and dark matter. Top: different smooth length at $z=0$ in L025; Middle: with $R_s=2R_g$ from $z=3.0$ to $z=1.0$ in L025; Bottom: with $R_s=R_g$ at $z=0$ in three simulations.}
\label{figure21}
\end{figure}

\begin{figure}[htbp]
\hspace{-1.0cm}
\includegraphics[width=0.65\textwidth]{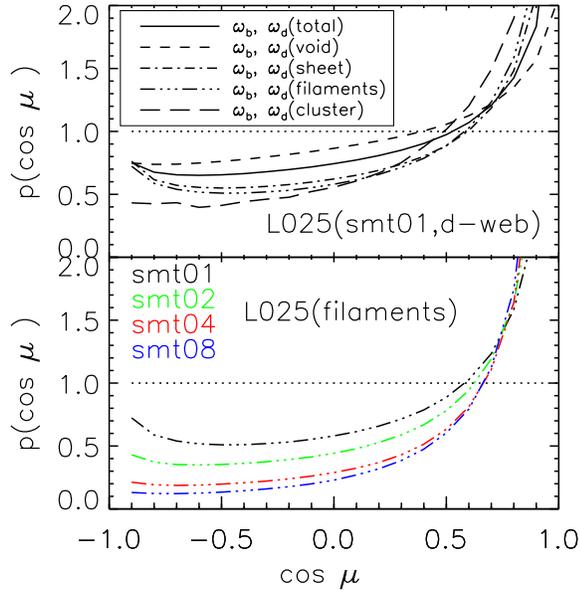}
\caption{The probability distribution $p(cos\mu)$ in different environment between vorticity of baryonic and dark matter in L025. }
\label{figure22}
\end{figure}

The ambient vortical flow around dark matter halo are suggested to play important role in shaping the halo spin and galaxies, and shows orientation preference with respect to the velocity shear (Libeskind et al 2013; Libeskind, Hoffman \& Gottlober 2014). Tentative evidence of alignment of galaxy spin axes with respect to filaments have been reported in works analysing observation samples. However, the spin axis of spiral galaxy is found to be randomly distributed with respect to the $e_1$ vector, different from the strong alignment of angular momentum of dark matter halos to $e_1$ found in N-body simulations(Temple et al. 2013; Temple \& Libeskind 2013). The limited resolution and lack of star formation in our simulation prohibit us from a direct investigation on the scale of galaxies. Nevertheless, comparison between baryonic and dark matter on the evolution of shear tensor and vorticity over scale range of tens of kpc to a couple of Mpc is feasible in our samples, which may be helpful to diagnose the cause of disparity between observations and N-body simulations. Actually, the growth history and statistics including power spectrum of vorticity reported in last section already indicate signs of discrepancy between baryonic and dark matter below a couple of Mpc.

Figure 21 presents the probability distribution $p(cos \mu )$ as a function of the angle between vorticity of baryonic and dark matter, i.e.,  $\vec{\omega}_b \cdot \vec{\omega}_d$. $p(|cos \mu |)$ is shown for the shear eigenvectors of $\vec{\bold{e}}_{b,i} \cdot \vec{\bold{e}}_{d,i}$, where $i=1,2,3$. The velocity shear of the baryonic matter are almost parallel to the dark matter, only weak misalignment can be observed for $\vec{e}_2$. Misalignment of $\vec{\omega}_b$ with respect to $\vec{\omega}_d$ is found, and became more apparent for shorter smooth length, i.e., higher vorticity. The median value of $cos(\mu)$ decrease from about $0.75$ for $R_s=0.4 Mpc$ to $0.38$ for $R_s=0.05 Mpc$ in L025. The evolution of $p(cos \mu )$ since $z=3.0$ is also demonstrated in Figure 21. The growth history of misalignment between vorticity of two matter components matches the global development of vorticity displayed in Figure 14. The increased resolution in L025 leads to higher vorticities and hence large misalignments.

\begin{figure*}[tbp]
\includegraphics[width=1.00\textwidth]{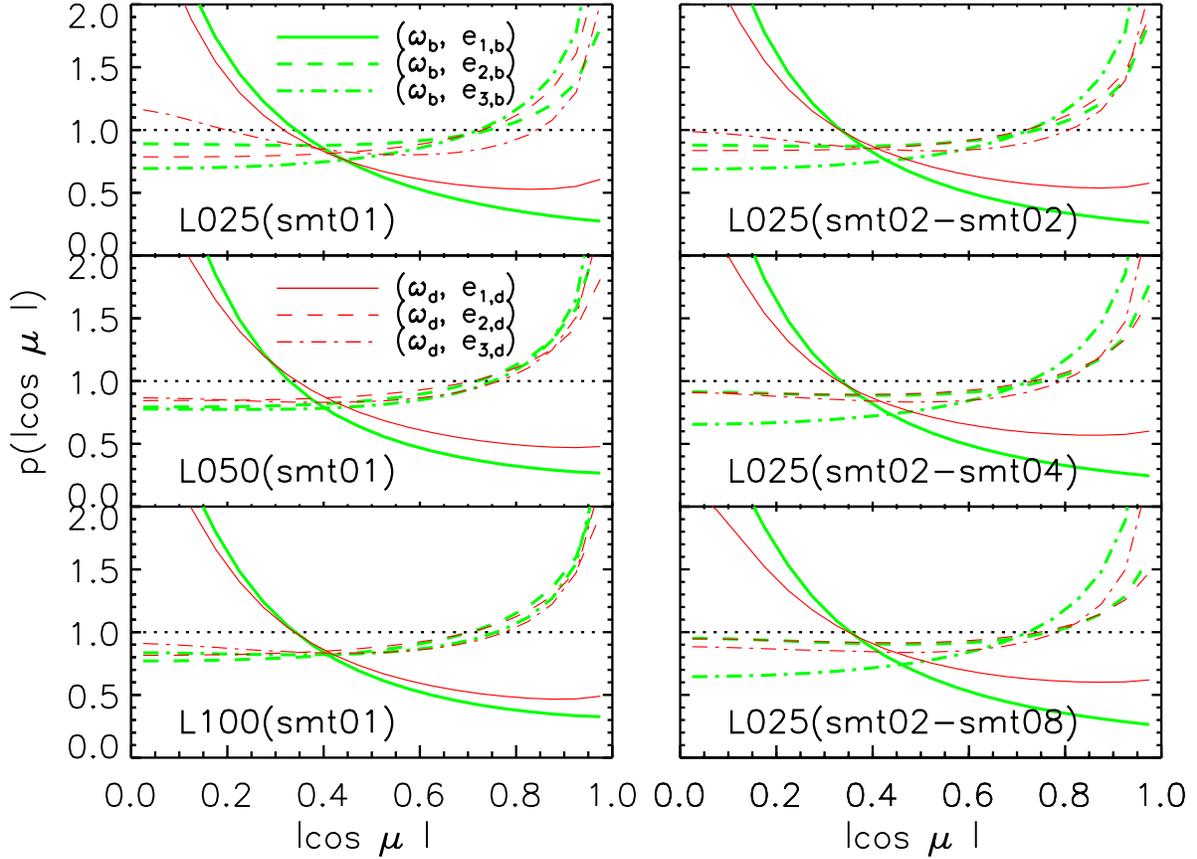}
\caption{The probability distribution $p(|cos\mu|)$ of the angle between the vorticity and the three eigenvectors of velocity shear, green and red lines represent baryonic and dark matter respectively. The smooth length $R_s$ used to calculate vorticity and shear eigenvectors are equal in the left column. In the right column, $R_s$ is set to $2R_g$ for the calculation of vorticity(denoted as smt02 before the dash), and $R_s$ for the calculation of shear tensor varies in three panels(smtxx after the dash). }
\label{figure23}
\end{figure*}

\begin{figure*}[htbp]
\includegraphics[width=1.00\textwidth]{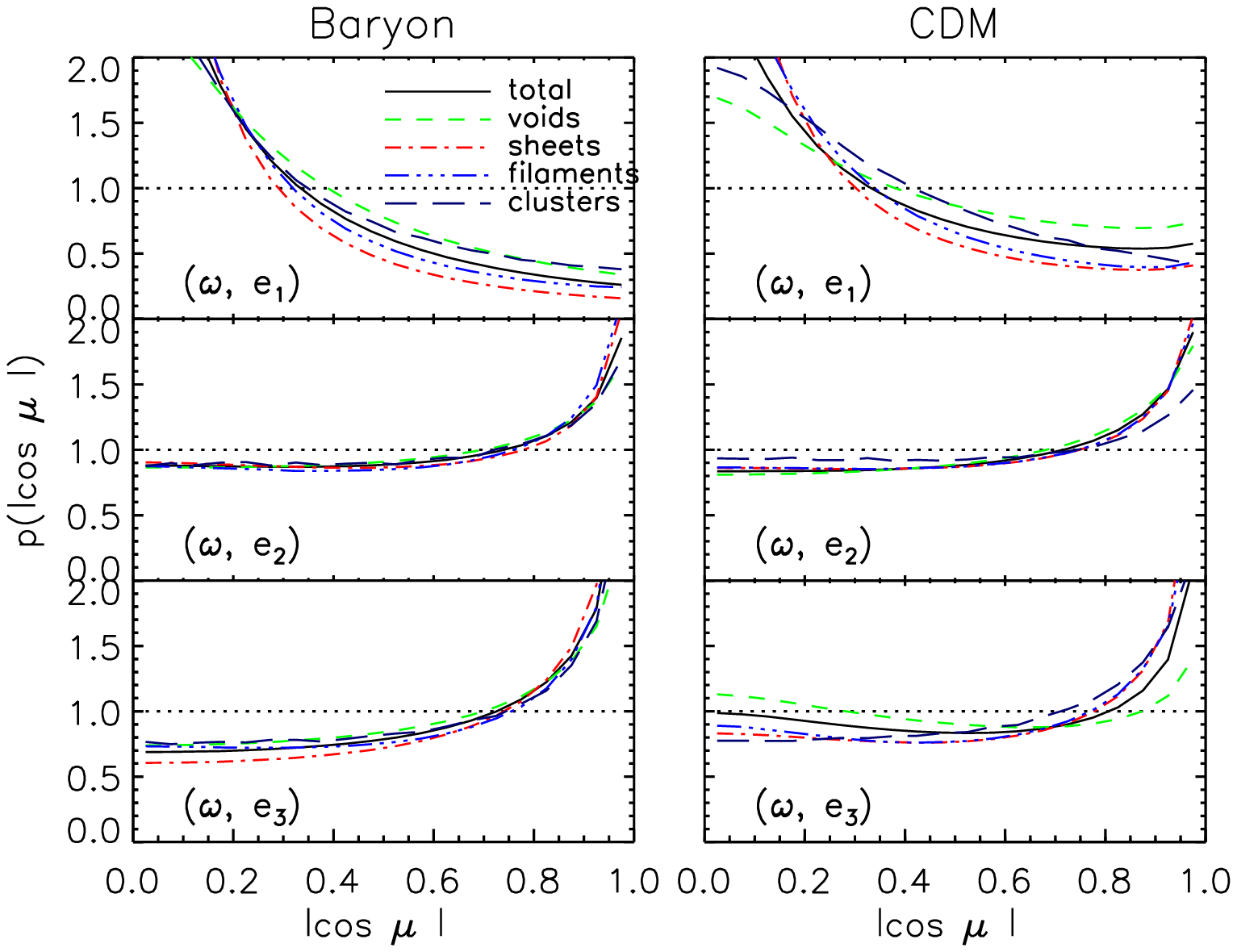}
\caption{The probability distribution $p(|cos\mu|)$ of the angle between the vorticity and the three eigenvectors of velocity shear in different web environment in L025. $R_s=2R_g$ is used to identify \textbf{d-web} and calculate vorticity and shear tensor. Left: Baryonic matter; Right: Dark matter. }
\label{figure24}
\end{figure*}

As the calculation of vorticity of dark matter suffers from poor sampling in under-dense region to some extent, the misalignment between $\vec{\omega}_b$  and $\vec{\omega}_d$ would likely be amplified in voids by numerical errors. We examine $p(cos \mu )$ in the four types of large scale structures identified by \textbf{d-web} applied in baryonic density field.  The result is plotted in Figure 22. The global misalignment is indeed raised by cells in voids, where vorticity is small. Nevertheless, the misalignment between $\vec{\omega}_b$ and $\vec{\omega}_d$ is also observable in collapsed structures. The distributions of $p(cos(\mu))$ in filaments, where the vorticity is the highest among four types of cosmic environment, against different smooth lengths are shown in Figure 22. The median value of $cos(\mu)$ in filaments are about $0.60, 0.78, 0.87, 0.9$ for $R_s=1.0, 2.0, 4.0, 8.0 R_g$ respectively. Obviously, the misalignment is well developed mostly below the scale of $0.2 Mpc$.

The orientations of vorticity with respect to velocity shear in both baryonic and dark matter are shown in Figure 23. The vorticity prefers to be perpendicular to $\vec{e}_1$ in all the three simulations. Meanwhile, $\vec{\omega}$ prefers to align with $\vec{e}_2$ and $\vec{e}_3$. These tendencies have been discussed in Libeskind et al(2013) and Libeskind, Hoffman \& Gottlober (2014). In the L025 sample, the median values of $|cos \mu|$ between $\vec{\omega}$ and $\vec{e}_1$, $\vec{e}_2$ and $\vec{e}_3$ are estimated to be $0.21, 0.57$, and $0.63$ for baryonic matter, and $0.26, 0.59$, and $0.52$ for dark matter. The latter results for dark matter is actually in good agreement with Libenskind et al(2013). The preference between $\vec{\omega}$ and $\vec{e}_1$, and $\vec{e}_3$ are relatively stronger for baryonic matter, which may partly because the vorticity of baryonic matter is developed well than dark matter.  The right column of Figure 23 shows the orientation of vorticity filtered with $R_s=97.6 kpc$ with respect to the velocity shear eigenvectors with $R_s=97.6, 195.2, 390.4 kpc$ respectively. The vorticity tends to lie in the $\vec{e}_2, \vec{e}_3$ plane, which is slightly weakened by larger $R_s$ for dark matter, while for baryonic matter any variation is hardly observed. The weaken of alignment for dark matter became more apparent for $R_s > 0.5 Mpc$ in Libeskind, Hoffman \& Gottlober (2014). The parallel alignment of $\vec{\omega}$ with $\vec{e}_2$ is quite similar in two matter components, and is also getting weaker with larger $R_s$. The median value of $|cos \mu|$ between $\vec{\omega}$ and $\vec{e}_2$ of dark matter decreases from 0.65 for $R_s=97.6kpc$ to 0.53 for $R_s=390.4kpc$ in L025, basically in agreement with Libeskind, Hoffman \& Gottlober (2014).

Figure 24 presents the alignment between $\vec{\omega}$ and $\vec{e}_i$ of baryonic and dark matter in different cosmic web environment. In consistent with Libeskind et al.(2013), the strength of perpendicular alignment between $\vec{\omega}$ and $\vec{e}_1$ decrease from sheets, to filaments, clusters, and voids. The trend is more evident for dark matter, with the median $|cos \mu|$ are about $0.17, 0.21, 0.28, 0.34$ respectively.  The parallel alignment between $\vec{\omega}$ and $\vec{e}_2$, and $\vec{e}_3$ are similar in all the environments, except that $(\vec{\omega}$, $\vec{e}_2$) in clusters and ($\vec{\omega}$, $\vec{e}_3$) in voids for dark matter. The corresponding median $|cos \mu|$ are 0.54 and 0.52. For a short summary, the alignment between vorticity and eigenvectors of shear tensor in baryonic matter field resembles dark matter, and is even moderately stronger between $\vec{\omega}$ and $\vec{e}_1, \vec{e}_3$ in the highly nonlinear regime. The tension between observation and simulation regarding the spin axis of spiral galaxy with respect to the $e_1$ vector would persist if the vorticity of stellar follows ambient gas.

\section{Discussion}

The formation and evolution of cosmic web is a long standing question. The Zel'dovich approximation suggests that the sheets/walls appear first, and then filaments form as a result of proceeding anisotropic collapse. Alternatively, the peak patch cosmic web scenario(Bond \& Myers 1996) offers an inverse sequence, i.e., clusters and their prototypes forms firstly around high density peaks on large smoothing scale, i.e., from several to tens of Mpc, then prominent filaments emerges as clusters-cluster bridges from the quadrupole matter distribution, and sheets/walls occurs at last.  In reality,  the evolution of cosmic web might be more complex. Formation of web components in the nonlinear regime involves hierarchical development(van de Weygaert \& Bond 2008). For example, the formation of prominent filaments is likely to result from the assembly of small scale filaments.

Cautun et al.(2014) found that, the configuration of prominent filaments and clusters in N-body simulations are basically consistent with the prediction of peak patch scenario. Meanwhile, the configuration of tenuous filaments and sheets, which are dominant at high redshifts, is not agree with the cluster-filament-cluster picture. Tenuous filaments and sheets of both baryonic and dark matter, identified with the tidal tensor and velocity shear tensor, also appears early in our simulations. Small scale filaments are embed by (proto)sheets at high redshifts. The layout of sheets on scales above $2-3$ Mpc may have been set up at $z=3$. In term of mass fraction, the evolution of identified d-web and v-web in our samples indicates a transition of the dominant structure from sheets to filaments taking place at $z \sim 2-3$. This interpretation, however, should be treated carefully. The transition and associated redshift are subjected to the scheme and parameters employed in web classification. Moreover, a fundamentally physical definition of filaments and sheets is still not available.

Given the classification scheme used in this work, such a transition is a natural expectation of the anisotropic collapse and hierarchical structure formation in the LCDM universe. The initial anisotropic density deformation leads to collapse in the primary axes firstly and the formation of local proto-sheets, which will assemble with each other and build the outline layouts above $2-3$ Mpc at $z>3$. Meanwhile, the divergence velocity field can be well developed over corresponding scales, and the further collapsing in secondary axes triggers the emerging of small-scale filaments within sheets. The filaments keep accreting matter from sheets and voids and surpass the sheets in term of mass fraction at some redshift. With \textbf{d-web}, and taking $\lambda_{th}=0.2-1.2$, the transition redshift is found to be around $\sim 2-3$ in our simulation samples. The vortical motions is boosted after the transition redshift.

A transition from sheets to filaments as the dominant web component of the universe in mass fraction may have an effect on the cosmic star formation history to some extent. The gas accretion in halos at different redshift might be affected by the change of dominant cosmic environment and associated velocity modes.  At redshifts larger than $\sim 2$, the overwhelmingly dominant divergence velocity may help the halo acquiring gas more rapidly. Afterwards, the rising curl motions may slow down the gas accretion. Moreover, since the accretion shocks surrounding different large scale structures also have different statistical properties(e.g. Vazza et al. 2009; Zhu et al. 2013), the thermal state of the accreted gas onto halos before and after the transition redshift can be different. This naive scenario requires a careful verification by taking account of the formation and evolutionary history of halos. Their connections, however, worth further investigating in depth, as the impact of cosmic environment on the star formation in galaxies is not well resolved yet.

The growth of filaments along with time is closely related to the role of vorticity of ambient flow field in the spins of dark matter halo and galaxies. Recent works suggest that the vorticities of the gas and dark matter share the same orientation on large scales, and prefer to be confined and aligned with filaments. Also the spins of halo and galaxy show alignment with the large scale vorticity of dark matter and gas respectively (Libeskind et al 2013; Dubois et al. 2014; Laigle et al. 2015). Laigle et al.(2015) demonstrated the emerging of vorticity in filaments due to quadripolar mult-flow, as predicted by Codis et al.(2012). The primary role of filaments in developing curl velocity revealed in this work, including the growth and distribution of vorticity, curl velocity, power spectrum and velocity structure functions, provides a solid support for results in Codis et al.(2012) and Laigle(2015).

Similar to dark matter, the vorticity of baryonic matter is found to perpendicular to the $\vec{e}_1$ vector, and parallel to $\vec{e}_2$, and $\vec{e}_3$. The alignment between $\vec{\omega}$ and $\vec{e}_1$, and $\vec{e}_3$ for baryonic matter are even stronger than dark matter, which is more significant with higher resolution. However, the spin axis of spiral galaxies in observations is found to be random with respect to $\vec{e}_1$(Temple et al. 2013; Temple \& Libeskind 2013). Hence the discrepancy between our results of baryonic matter and observation regarding $\vec{\omega}$ and $\vec{e}_1$ is likely more serious than N-body simulations. To have an in-depth probe of this discrepancy, hybrid N-body/hydrodynamic simulations with sufficient higher resolutions to track the dynamics of vorticity on galactic scales, and implemented with star formation and feedback module are urged.
 
\section{Conclusion}

Based on a set of cosmological hydrodynamical simulations, we make a comprehensive investigation on the mass and velocity evolution of baryonic and cold dark matter since $z=5$ in the four types of large scale environment, i.e., voids, sheets/walls, filaments and clusters/knots. The web classification methods on the basis of tidal tensor and velocity shear tensor have been used. We found that the baryonic and cold dark matter show similar trends regarding the evolution of mass and velocity in the cosmic web, but differences are observed. We summarise our results as folllowing:

1. Sheets/walls are found to have formed in the early stage of anisotropy collapsing processes. The large scale distribution of sheets had been formed by $z=3$, and served as the frame of cosmic web. The mass fraction in sheets was up to $\sim 40\%$ at $z=5$, and had changed slowly at $z >2$, and then decreased gradually. Filaments also emerged at an earlier time before $z=5$, and became the primary structure as measured by mass fraction after $z \sim 2$. The mass fraction in filaments have been increased from around $20\%$ at $z=5$ to $\sim 50\%$ at $z=0$. Massive clusters/knots became well developed and evident at around $z=0.5$. The mass fraction in clusters/knots was smaller than $10\%$ till $z=2.0$ and is $\sim 15\%$ at z=0.0. The exact numbers of these fractions and redshifts are dependent on the web classification scheme and threshold parameter $\lambda_{th}$, for which $\lambda_{th}=0.2-1.2$ have been investigated in this work.

2. In accordance with the formation of the frame of sheets/walls, the cosmic divergence velocity had been well developed above $2-3$ Mpc by $z=3$. The vortical motion have grown sharply along with the rapid rising of filamentary structures since $z=3.0$, and become comparable to the divergence motion under $2-3$ Mpc at $z=0$, due to strong curved shocks and shell crossing for baryonic and dark matter respectively at the boundaries of filaments. The coherent scale of vortical motion increase with time correspondingly, reach at $\sim 2$ Mpc at $z=0$, and is demonstrated by the velocity power spectrum, structure functions and fractal dimensions. The relative orientation of vorticity with respect to the three eigenvectors of velocity shear tensor for baryonic matter is similar to dark matter, i.e., vorticity prefers to perpendicular to $\vec{e}_1$ and parallel to $\vec{e}_2$ and $\vec{e}_3$, which is in consistent with the results of N-body simulations in the literature(e.g., Libeskind et al 2013; Libeskind, Hoffman \& Gottlober 2014)

3. Differences between baryonic and dark matter rise in the nonlinear regime. Mildly less baryonic matter is residing in filaments and clusters than dark matter, at a level of $\lesssim 20\%$ in relative percentage. The vorticity in baryonic matter is more significant than dark matter. Below a couple of Mpc, velocity power spectrum of curl velocity can dominate over divergence velocity in baryonic matter, but the divergence mode keeps dominant in dark matter. In addition, the alignment between vorticity and $\vec{e}_1$, and $\vec{e}_3$ are moderately stronger in baryonic matter. These differences may be underestimated due to the limited resolution and lack of star formation in our simulation.

In short, our simulations indicate that the rapid growth of vortical motions is in sync with the emerging and rising of filaments within sheets since $z \sim 3$ . The over dense regions in the universe may have transited from sheets dominated to filaments dominated at about $z \sim 2$, which may affect the gas supply for galaxy formation and evolution. The primary role of filaments in developing curl velocity since $z \sim 3$ is also expected to have important impact on the spins of halos and galaxies. Simulations with higher resolution and implemented with the star formation and feedback could bring more vivid view, and provide a direct comparison with observation to verify our understanding of the cosmic web.

\begin{acknowledgements}
Acknowledgements: 
The authors thank the anonymous referee for very insightful and helpful comments to improve the manuscript. The simulations were run at Supercomputing Center of the Chinese Academy of Sciences, and Sun Yat-Sen University. WSZ is supported by the National Natural Science Foundation of China(NSFC) under grants 11203012, 11673077 and the Fundamental Research Funds for the Central Universities. FLL is supported under the NSFC grants 11273060, 91230115 and 11333008, and State Key Development Program for Basic Research of China (No. 2013CB834900 and 2015CB857000).  
\end{acknowledgements}


\begin{references}

\reference{}Andrae, R., \& Jahnke, K. 2011, MNRAS, 418, 2014 

\reference{} Aragon-Calvo, M.A., 2007,  Morphology and dynamics of the cosmic web. Ph.D. thesis,
Groningen University, the Netherlands

\reference{} Aragón-Calvo, M. A.; Jones, B. J. T.; van de Weygaert, R.; van der Hulst, J. M., 2007a, A\&A, 474, 315

\reference{} Aragon-Calvo M. A., van de Weygaert R., Jones B. J. T., van der Hulst J. M., 2007b, ApJ, 655

\reference{} Aragon-Calvo M. A., van de Weygaert R., Jones B. J. T.,2010, MNRAS, 408, 2163

\reference{} Bond J. R., Kofman L., Pogosyan D., 1996, Nature, 380, 603

\reference{} Bond J. R., Myers S. T., 1996, ApJS, 103, 1

\reference{} Bond N. A., Strauss M. A., Cen R., 2010a, MNRAS, 406, 1609

\reference{} Bond N. A., Strauss M. A., Cen R., 2010b, MNRAS, 409, 156


\reference{} Cautun M., van de Weygaert R., Jones B. J. T., 2013, MNRAS, 429, 1286

\reference{} Cautun, M., van de Weygaert, R., Jones, B. J. T., \& Frenk, C. S., 2014, MNRAS, 441, 2923

\reference{} Colberg J. M., Krughoff K. S., Connolly A. J., 2005, MNRAS, 359, 272

\reference{} Colless M. et al., 2003, preprint arXiv:astro-ph/0306581

\reference{} Codis S., Pichon C., Devriendt J., Slyz A., Pogosyan D., Dubois Y., Sousbie T., 2012, MNRAS, 427, 3320

\reference{} de Lapparent V., Geller M. J., Huchra J. P., 1986, ApJ, 302, L1

\reference{} Dong, X. C., Lin, W. P., Kang, X., et al., 2014, ApJL, 791, L33

\reference{} Dubois, Y., Pichon, C., Weler, C., Le Borgne, E., Devriendt, J., et al., 2014, MNRAS, 444, 1453

\reference{} Dubrulle, B., 1994, Phys. Rev. Lett., 73, 959

\reference{} Eisenstein D. J., Loeb A., 1995, ApJ, 439, 520

\reference{} Feng, L.L., Shu, C.-W., \& Zhang, M.P. 2004, \apj, 612, 1

\reference{} Forero-Romero J. E., Hoffman Y., Gottl¨ober S., Klypin A., Yepes G., 2009, MNRAS, 396, 1815

\reference{} Gheller, C.; Vazza, F.; Favre, J.; Brüggen, M., 2015, MNRAS, 453, 1164

\reference{} Hahn O., Carollo C. M., Porciani C., Dekel A., 2007a, MNRAS, 381, 41

\reference{} Hahn O., Porciani C., Carollo C. M., Dekel A., 2007b, MNRAS, 375, 489

\reference{} Hahn O., Teyssier R., Carollo C. M., 2010, MNRAS, 405,274

\reference{} Hoffman, Y., Metuki, O., Yepes, G., Gottlober, S., Forero-Romero, J. E., Libeskind, N. I., \& Knebe, A. 2012, MNRAS, 425, 2049

\reference{} Icke V., 1973, A\&A, 27, 1

\reference{} Jones, B. J. T., van de Weygaert, R., \& Aragon-Calvo, M. A. 2010, MNRAS, 408, 897

\reference{} Komatsu, E., et al. 2009, ApJS, 180, 330

\reference{} Kritsuk, A.G., Norman, M.L., Padoan, P., \& Wagner, R.,  2007, \apj, 665, 416

\reference{} Laigle, C., Pichon, C., Codis, S., Dubois, Y., Le Borgne, D., et al., 2015, MNRAS, 446, 2744

\reference{} Libeskind N. I., Hoffman Y., Steinmetz M., Gottlober S.,
Knebe A., \& Hess S., 2013, ApJL., 766, L15

\reference{} Libeskind N. I., Hoffman Y., Gottlober, S., 2014,  MNRAS, 441, 1974

\reference{} Mandelbrot, B. B. 1974, J. Fluid Mech., 62, 331 
 
\reference{} Matarrese, S., \& Mohayaee, R. 2002, MNRAS, 329, 37     

\reference{} Mehmet, A., Aaron, S. G. R., Simon, D., Peder, N., Ivan, B., et al., 2014, MNRAS, 438, 177

\reference{} Pichon, C., \&  Bernardeau, F., 1999, A\&A, 343,663

\reference{} Porter, D. H., Pouquet, A., \& Woodward, P. R. 1992, ThCFD, 4, 13

\reference{} Pueblas, S.,  \& Scoccimarro, R., , 2009, Phys. Rev. D 80, 043504

\reference{} Shandarin S. F., Zel'dovich Ya. B., 1989, Rev. Mod. Phys., 61, 185

\reference{} She, Z.S., \& Leveque, E., 1994, Phys. Rev. Lett., 72, 336

\reference{} She, Z. S., \& Waymire, E. C. 1995, Phys. Rev. Lett., 74, 262
	
\reference{} Shull, J. M., Smith, B. D., Danforth, C. W., 2012, \apj, 759, 23

\reference{} Sousbie, T., 2011, MNRAS, 414, 350

\reference{} Tegmark M. et al., 2004, ApJ, 606, 702
	
\reference{} Tejos, N., Prochaska, J. X., Crighton, N. H. M., Morris, S. L., Werk, J. K., et al., 2016, MNRAS, 455, 2662

\reference{} Tempel, E., Libeskind, N., 2013, ApJL, 775, L42
	
\reference{} Tempel, E.; Stoica, R. S.; Martínez, V. J.; Liivamagi, L. J.; Castellan, G.; Saar, E., 2014, MNRAS, 438,3465

\reference{} van de Weygaert R., Bond J. R., 2008, in Lecture Notes in Physics, Berlin Springer Verlag, Vol. 740, A Pan-
Chromatic View of Clusters of Galaxies and the Large-Scale Structure, Plionis M., Lopez-Cruz O., Hughes D.,eds., p. 335

\reference{} Vazza, F., Brunetti, G., Gheller, C., 2009, MNRAS, 395, 1333

\reference{} White S. D. M., Silk J., 1979, ApJ, 231, 1

\reference{} Wang, X., Szalay, A., Aragon-Calvo, M. A., et al., 2014, \apj, 793, 58

\reference{} Wang, Y. O., Lin, W. P., Kang, X., et al. 2014, \apj, 786, 8

\reference{} Zel'dovich Y. B., 1970, A\&A, 5, 84

\reference{} Zhang, Y. C., Yang, X. H., Faltenbacher, A., et al, 2009, \apj, 706, 747

\reference{} Zhang, Y. C., Yang, X. H., Wang, H. Y., et al. 2015, \apj, 798, 17
	
\reference{} Zheng, Y., Zhang, P. J., Jing, Y. P., Lin, W. P., \& Pan, J., Physical Review D, vol. 88, Issue 10, id. 103510

\reference{} Zhu, W.S., Feng, L.L., \& Fang, L.Z., \apj, 2010, 712, 1

\reference{} Zhu, W.S., Feng, L.L.,  Xia, Y. H., Shu, C. W., Gu, Q. S., \& Fang, L.Z.,\apj, 2013, 777, 48

\reference{} Zhu, W.S., Feng, L.L., \apj, 2015, 811,94

\end{references}
\end{document}